\documentclass[aps,prd,preprint,a4paper,showpacs,nofootinbib,
superscriptaddress]{revtex4-1}
\usepackage{bm}
\usepackage{indentfirst}
\usepackage{amsmath}
\usepackage{graphicx}
\usepackage{float}
\usepackage{amssymb}
\usepackage{subfigure}
\usepackage{amssymb}
\usepackage{hyperref}
\usepackage{epstopdf}
\usepackage[section]{placeins}

\usepackage[utf8]{inputenc}
\hypersetup{
    colorlinks=true,
    linkcolor=red,
    citecolor=blue,
}
\usepackage{color}
\begin{document}

\title{Echoes from phantom wormholes}
\author{Hang Liu}
\email{hangliu@sjtu.edu.cn}
\affiliation{School of Physics and Astronomy,
Shanghai Jiao Tong University, Shanghai 200240, China}
\affiliation{Center for Gravitation and Cosmology, College of Physical Science and Technology, Yangzhou University, Yangzhou 225009, China}
\author{Peng Liu}
\email{phylp@jnu.edu.cn}
\affiliation{Department of Physics and Siyuan Laboratory, Jinan University, Guangzhou 510632, China}
\author{Yunqi Liu}
\email{yunqiliu@yzu.edu.cn}
\affiliation{Center for Gravitation and Cosmology, College of Physical Science and Technology, Yangzhou University, Yangzhou 225009, China}
\author{Bin Wang}
\email{wang\_b@sjtu.edu.cn}
\affiliation{Center for Gravitation and Cosmology, College of Physical Science and Technology, Yangzhou University, Yangzhou 225009, China}
\affiliation{School of Aeronautics and Astronautics, Shanghai Jiao Tong University, Shanghai 200240, China}
\author{Jian-Pin Wu}
\email{Corresponding author, jianpinwu@yzu.edu.cn}
\affiliation{Center for Gravitation and Cosmology, College of Physical Science and Technology, Yangzhou University, Yangzhou 225009, China}

\begin{abstract}

We study the time evolution of the test scalar and electromagnetic fields perturbations in configurations of phantom wormholes  surrounded by dark energy with state parameter $\omega< -1$. We observe obvious signals of echoes reflecting wormholes properties and disclose the physical reasons behind such phenomena. In particular, we find that the dark energy equation of state has a clear imprint in echoes in wave perturbations. When $\omega$ approaches the phantom divide $\omega=-1$ from below, the delay time of echoes becomes longer. The echo of gravitational wave is likely to be detected in the near future, the signature of the dark energy equation of state in the echo spectrum can serve as a local measurement of the dark energy.

\end{abstract}

\maketitle

\section{Introduction}

In the universe, there is accumulating evidence that classical black holes enveloped by event horizons can be developed from gravitational collapse.  There have been observational arguments, sometimes very strong ones, in favour of the existence of the event horizon \cite{Akiyama:2019bqs}, but it was shown that they cannot prove it \cite{Abramowicz:2002vt}. It was also claimed in \cite{Cardoso:2016rao} that the probe of initial gravitational waves (GWs) ringdown signals can not be regarded as the evidence of the existence of event horizons, as it suggested that the horizonless objects which are compact enough to guarantee a photon sphere can mimic the black holes in the sense that they can generate almost the identical  initial ringdown signals of GWs  to that of black holes.

On the other hand, there are also strong theoretical motivations and rooms for different kind of alternatives to black holes, motivated from the quantum gravity effects or the resolutions to the black hole information paradox \cite{Almheiri:2013hfa}. In a broad sense, these horizonless compact objects other than a neutron star are referred to as exotic compact objects (ECOs) \cite{Cardoso:2019rvt}. We can design ECO models to mimic all observational properties of black holes with arbitrary accuracy, such that we can compare the characteristics of black hole with that of ECOs and to find a classification for different models \cite{Cardoso:2019rvt}.
In this context, a number of ECOs have been proposed,
including wormholes \cite{Cardoso:2016rao,Bueno:2017hyj}, gravastars \cite{Mazur:2004fk,Visser:2003ge,Wang:2018cum}, boson stars \cite{Schunck:2003kk}, anisotropic stars \cite{Carloni:2017bck,Isayev:2018hqx,Raposo:2018rjn}, quasiblack holes \cite{Lemos:2003gx,Lemos:2008cv},
dark stars \cite{Kawai:2013mda,Baccetti:2017oas,Baccetti:2016lsb,Baccetti:2018qrp,Giddings:1992hh,Unruh:2017uaw},
fuzzballs \cite{Lunin:2001jy,Lunin:2002qf,Mathur:2005zp,Mathur:2008nj,Mathur:2012jk}, firewalls \cite{Almheiri:2013hfa,Mathur:2012jk,Almheiri:2012rt},  the Planckian correction from the dispersion relation of gravitational field \cite{Oshita:2018fqu,Oshita:2019sat}. For more detailed introduction to ECOs, please refer to Ref.\cite{Cardoso:2019rvt} and the references therein.

Given the  theoretical importance of ECOs, it is of great significance to explore the evidence of the existence of such objects.  The recent direct observation of GWs \cite{Abbott:2016blz,TheLIGOScientific:2017qsa} provides us a brand new method in astronomical observations, and it suggests the ending of the era of a single electromagnetic wave channel observation and  announces the dawn of multi-messenger astronomy with GW as a new probe to study our universe. Hence GWs spectroscopy will play an increasingly important
role as more and more events with large signal-to-noise ratio (SNR) are detected. It is expected to employ GWs spectroscopy to detect and  distinguish different ECOs \cite{Cardoso:2016rao,Giudice:2016zpa,Macedo:2013jja} with the detection of GWs with large SNR. Apart from using GWs spectroscopy to study ECOs, the recently proposed GWs echoes can be served as  a new characteristic of ECOs \cite{Cardoso:2016oxy}. Several groups have claimed that the potential evidence of GW echoes in the LIGO/Virgo data have already been found \cite{Cardoso:2016rao,Cardoso:2016oxy} (see \cite{Abedi:2020ujo,Cardoso:2019rvt} for a review). The GWs echoes are produced due to the centrifugal barrier for the ECOs which can reflect the incident GWs, or the quantum corrections to conventional black holes event horizons \cite{Cardoso:2016rao,Cardoso:2016oxy,Foit:2016uxn,Bueno:2017hyj,Price:2017cjr,Nakano:2017fvh,Mark:2017dnq,Burgess:2018pmm,Konoplya:2018yrp, Testa:2018bzd,Wang:2018mlp,Cardoso:2019apo,Wang:2019rcf,Ghersi:2019trn,Li:2019kwa,Wang:2019szm}which renders the event horizon reflective to incident waves.  Therefore, the GW echoes play a key role in exploring physics near ECOs and also pave a path towards quantum gravity. On the other hand, the electromagnetic counterpart \cite{Graham:2020gwr,Mastrogiovanni:2020gua,Soares-Santos:2017lru,Valenti2017} to the GW events are also an important probe to study our universe. Therefore, apart from the GW echoes, the possible  future detection of  echoes of electromagnetic field waves may serve as another important prob in the exploration of fundamental physics.

Wormholes are a class of important and simple horizonless objects which connect  two distinct universes or two distant regions of the same universe by throat of wormholes \cite{Morris:1988cz}, and they are also considered as a kind of ECOs as they can be constructed to have any arbitrary
compactness and usually the exotic matter is required to build such models \cite{Cardoso:2019rvt}. Different wormhole spacetimes can have very different properties. Since people are
interested in wormhole that mimic black holes, some simple but useful wormhole models are usually discussed, such as wormhole constructed by gluing two Schwarzschild black hole spacetimes \cite{Cardoso:2016oxy,Cardoso:2016rao,Dai:2019mse,Mark:2017dnq,Wang:2018mlp}, and Damour-Solodukhin Wormhole \cite{Damour:2007ap,Bueno:2017hyj,Volkel:2018hwb}.
We can detect the wormholes by gravitational lensing as it is known to be one of the most important tools to test the predictions of general relativity \cite{Cramer:1994qj,Safonova:2001vz,Nandi:2006ds,Nakajima:2012pu,Dey:2008kn,Bhattacharya:2010zzb,Tsukamoto:2012xs,Kuhfittig:2013hva,Tsukamoto:2016qro,Jusufi:2017mav,Ovgun:2018fnk,Nedkova:2013msa,Amir:2018szm}.
Recently, the authors in \cite{Dai:2019mse} interestingly proposed that the objects in our universe can be influenced by the objects in other universe (or other region of the same universe) through wormholes and thus provided a new method to detect wormholes.

Note that the wormholes are ECOs such that it is natural to apply GWs echoes to detect wormholes and reveal the characteristics of such ECOs. In \cite{Cardoso:2016rao,Cardoso:2016oxy}, the GW echoes are first disclosed by studying the simple spherically symmetric wormholes.
The GW echo from a wormhole model with a slowly evolving throat radius is also studied \cite{Wang:2018mlp}. It gives rise to a nonconstant delay time of echoes. Also, the GW echoes are regarded as a probe to characterize the black hole–wormhole ``transition" \cite{Bronnikov:2019sbx,Churilova:2019cyt,Bronnikov:2019sbx}.
Since the object after merge would rotate rapidly, several groups have studied the signals of GW echoes from Kerr-like wormholes \cite{Micchi:2019yze,Bueno:2017hyj}. The studies indicate that rotation leads to two key characteristics: the break of the degeneracy of the quasinormal frequencies and the emergence of the modes of the late-time instabilities. In addition, some preliminary attempts have been implemented to establish templates of echoes from wormhole models \cite{Mark:2017dnq,Testa:2018bzd,Bueno:2017hyj}.

In this paper, we intend to study the GW echoes from a phantom wormhole proposed in \cite{Lobo:2012qq}.
This wormhole geometry are supported by phantom dark energy with equation of state $w< -1$ \cite{Caldwell:1999ew}, which violates the null energy condition. The wormhole solutions in \cite{Lobo:2012qq} are asymptotically flat and there is no need to surgically paste the interior wormhole geometry to an exterior vacuum spacetime. They are contrary to other phantom wormhole solutions which are not asymptotically flat \cite{Sushkov:2005kj,Lobo:2005us,Lobo:2005yv,PhysRevD.72.061303,Jamil:2009vn} or for which the interior wormhole metric is glued to a vacuum exterior spacetime at a junction interface \cite{Lemos:2003jb,Lobo:2004id,Lemos:2004vs,Lobo:2005zu,Lemos:2008aj,Garcia:2011aa}. From this point of view, this new wormhole solutions are more natural.

It is very interesting to study wormholes in the background filled with phantom dark energy.
As we all know, the astrophysical compact objects are believed to be influenced by their environment, for example, through the accretion disks. In Ref. \cite{He:2009jd}, the authors studied perturbations around black holes absorbing dark energy and  they observed distinct perturbation behaviors when black holes swallow  different
kinds of dark energy. This provides the possibility of disclosing whether the dark energy is of quintessence type or phantom type. In Ref. \cite{MersiniHoughton:2008aw}, the authors proposed that  gravitational radiation from binary systems of supermassive black holes can provide information of local properties of dark energy. Its feasibility was further discussed in  \cite{Enander:2009pq}. These attempts provide ideas of possible local detection of dark energy in astronomical observations \cite{He:2009jd,MersiniHoughton:2008aw}. Since the echoes of gravitational wave may be detected in the near future in the  third-generation ground based GW detectors or space based GW detectors, examining the influences of the phantom dark energy parameter on the echo behaviors is interesting, which is possibly a new way to  probe dark energy locally in phantom wormhole backgrounds.

It is necessary to point out that, for simplicity, in this paper we would like to discuss echoes of  scalar field and electromagnetic field waves which are regarded as a proxy for gravitational perturbations instead of directly dealing with gravitational perturbations in our calculation of echoes. This simplified treatment is justified by the fact that the perturbation of  scalar fields usually behave qualitatively similarly to gravitational perturbations. Therefore, we believe that the main features of GW echoes can be reflected by echoes of scalar fields. On the other hand, the discoveries of the electromagnetic  counterparts to GW events \cite{Graham:2020gwr,Soares-Santos:2017lru,Valenti2017} also motivate us to investigate echoes of electromagnetic fields, for the reason that both the GW echoes and electromagnetic field echoes could be observed at the same time for one single GW event.

Our paper is organized as what follows. In Section \ref{section2}, we present a brief introduction of the  phantom wormhole models and introduce the numeric method to calculate the waveforms of GW echoes. The echo signals from phantom wormhole model \uppercase\expandafter{\romannumeral 1} and \uppercase\expandafter{\romannumeral 2} are worked out and analyzed in Section \ref{section3}  and  Section \ref{section4}, respectively. In Section \ref{section5}, we calculate the quasinormal modes (QNMs) frequencies in these two wormhole backgrounds. The final remarks are given in Section \ref{section5}. Throughout this paper, we set $G=c=1$.

\section{wormhole models and methods}
\label{section2}
In this section, we shall present a brief review on the asymptotically flat phantom wormholes. And then, we write down the evolution equations of a test scalar field and an electromagnetic field in the tortoise coordinate. Next, we introduce the numerical method of time-domain profile to solve the equation of motion (EOM).
\subsection{Asymptotically flat phantom wormholes}
A new asymptotically flat phantom wormhole solution is constructed in \cite{Lobo:2012qq}, which is given by
\begin{eqnarray}
&&
ds^2=-A(r)dt^2+B(r)dr^2+r^2(d\theta^2+\sin^2\theta d\phi^2)\,,
\
\\
&&
B(r)=\frac{1}{1-\frac{b(r)}{r}}\,.
\end{eqnarray}
The function $b(r)$ above is the shape function which describes the shape of the wormhole.
The wormhole throat connecting two asymptotic regions is located at a minimum radial coordinate $r_0$ satisfying $b(r_0)=r_0$. The so-called flaring-out condition gives $(b(r)-b'(r)r)/2b^2(r)>0$, which results in $b'(r_0)<1$ at the wormhole throat. To have a well-defined wormhole, we require $B(r)>0$ for $r\geq r_0$, which leads to $b(r)<r$. In addition, to have an asymptotically flat geometry, we shall impose the following conditions $A(r)\rightarrow 1$ and $b(r)/r\rightarrow 0$ at $r\rightarrow\infty$.

Considering phantom energy as the content of the spherically symmetric spacetime, there are several strategies to solve the Einstein field equations. Here,
we focus on two specific wormhole solutions in \cite{Lobo:2012qq}, which is obtained by two specific strategies. The first one, denoted by Wormhole Model \uppercase\expandafter{\romannumeral 1}, can be constructed by specifying a specific shape function. The other one, denoted by Wormhole Model \uppercase\expandafter{\romannumeral 2}, is established by specifying two equations of state relating the tangential pressure and the energy density. For more details for finding the phantom  wormhole geometries, we can refer to Ref. \cite{Lobo:2012qq}.
\subsubsection{Wormhole Model  \uppercase\expandafter{\romannumeral 1}}
We construct Wormhole Model  \uppercase\expandafter{\romannumeral 1} by constructing a specific shape function, which is
\begin{equation}
\frac{b(r)}{r_0}=a\left(\frac{r}{r_0}\right)^{\alpha}+C.\label{eq1}
\end{equation}
$\alpha$, $a$ and $C$ are dimensionless constants. Asymptotically flat condition $b(r)/r\rightarrow 0$ at $r\rightarrow \infty$ gives rise to $\alpha<1$. At the throat $r=r_0$, the condition $b(r_0)/r_0=1$ gives rise to the relation $C=1-a$.
Positive energy density imposes the condition $a\alpha>0$ and the flaring-out condition $b'(r_0)<1$ forces $a\alpha$ to satisfy $a\alpha<1$. Therefore, we have the restrictions of parameters as
\begin{equation}
\alpha<1,  \quad 0<a\alpha<1.
\end{equation}

After the shape function is specified, we shall work out the redshift function $A(r)$, which is determined by the following ordinary differential equation \cite{Lobo:2012qq}
\begin{equation}
\frac{A'(r)}{A(r)}=\frac{r_0}{r^2}\times\frac{1+a\left[\left(\frac{r}{r_0}\right)^{\alpha}(1+\omega\alpha)-1\right]}{1-\frac{r_0}{r}\left\{1+
a\left[\left(\frac{r}{r_0}\right)^{\alpha}-1\right]\right\}}.
\end{equation}
The exact solution for the above differential equation only exists for some specific model parameters. Here we only consider the specific case of $\alpha=1/2$,
which yields
\begin{equation}
A(r)=\kappa \left(1+\frac{1-a}{\sqrt{r/r_0}}\right)^2,
\end{equation}
where $\kappa$ is a constant of integration. For simplicity, we take $\kappa=1$. Then, the line element can be wrote down as
\begin{equation}
ds^2=-\left(1+\frac{1-a}{\sqrt{r/r_0}}\right)^2dt^2+\frac{dr^2}{1-\frac{a}{\sqrt{r/r_0}}-\frac{1-a}{r/r_0}}+r^2(d\theta^2+\sin^2\theta d\phi^2).
\end{equation}
Note that for this case, we have $\omega=-2/a$ and $0 < a < 2$.

\subsubsection{Wormhole Model \uppercase\expandafter{\romannumeral 2}}
We build Model \uppercase\expandafter{\romannumeral 2} by specifying two equations of state relating the tangential pressure and the energy density.
For this model, we take the following form of the red shift function
\begin{equation}
A(r)=1+\gamma \left(\frac{r_0}{r}\right)^{\alpha},
\end{equation}
with $\alpha>0$ and $A(r)\rightarrow 1$ in the limit of $r\rightarrow\infty$.
$\gamma$ is a dimensionless constant.
In this paper, we only focus on the case of $\alpha=1$, for which the shape function is given by
\begin{equation}
b(r)=-\gamma r_0+(r+\gamma r_0)^{-1/\omega}[r_0(1+\gamma)]^{(1+\omega)/\omega}\,.
\end{equation}
To have a well-defined wormhole spacetime without even horizon, we require $A(r)>0$, which leads to $\gamma>-1$. It is easy to check $b'(r_0)=-1/\omega<1$ for $\omega<-1$,
which satisfies flaring-out condition.
Without loss of generality, we shall take $r_0=1$ in the numerical calculation through this paper.

\subsection{The methods}
We only focus on the evolution of a test scalar field and an electromagnetic field as a proxy for gravitational perturbations in this paper. It can be expected to capture the main properties of GW echoes. In this part, we first derive the radial  equations of motion (EOMs) for scalar and electromagnetic field perturbation, and then we introduce the  method by which the time-domain profile of perturbations are obtained in present paper.

\subsubsection{Radial Wave Equations of Scalar Field}

We write down the EOMs for massless scalar field and electromagnetic field as what follows
\begin{eqnarray}
&&
\frac{1}{\sqrt{-g}}\partial_{\mu}(\sqrt{-g}g^{\mu\nu}\partial_{\nu}\phi(t,r,\theta,\phi))=0.\label{eq5}
\end{eqnarray}
We perform the separation of the variables for scalar field $\Phi(t,r,\theta,\phi)$ as follows
\begin{equation}
\phi(t,r,\theta,\phi)=\sum_{l,m}\frac{\Psi(t,r)}{r}Y_{lm}(\theta,\phi)\label{eq4},
\end{equation}
where $Y_{lm}(\theta,\phi)$ is the spherical harmonics and $l$ and $m$ stand for angular number and azimuthal number, respectively.
Substituting Eq. \eqref{eq4} into Eq. \eqref{eq5}, we obtain following radial equation
\begin{equation}
\begin{split}
-\partial_t^2\Psi(t,r)+\frac{A}{B}\partial_r^2\Psi(t,r)+\frac{BA'-AB'}{2B^2}\partial_r\Psi(t,r)
+\frac{A(rB'-2l(l+1)B^2)-rBA'}{2r^2B^2}\Psi(t,r)=0,\label{eq6}
\end{split}
\end{equation}
where a prime denotes a derivative with respect to areal radius $r$. With the introduction of tortoise coordinate $r_\ast$ defined by
\begin{equation}
dr_\ast=\sqrt{\frac{B(r)}{A(r)}}dr,
\end{equation}
the equation \eqref{eq6} can be transformed to the well-known wave equation
\begin{equation}
-\frac{\partial^2\Psi(t, r)}{\partial t^2}+\frac{\partial^2\Psi(t,r)}{\partial r^2_\ast}-V_s(r)\Psi(t, r)=0.
\end{equation}
The  effective potential of scalar field perturbation $V_s(r)$ is given by
\begin{equation}
V_s(r)=A(r)\frac{l(l+1)}{r^2}+\frac{1}{2r}\frac{d}{dr}\frac{A(r)}{B(r)}.\label{eqsc}
\end{equation}
Further more, we take the form of time dependence of $\Psi(t,r)$ as $\Psi(t,r)=e^{-i\omega t}\Phi(r)$, and then we get the wave equation in frequency domain
\begin{equation}
\frac{d^2\Phi(r)}{dr^2_\ast}+(\omega^2-V_s(r))\Phi(r)=0,\label{eq2}
\end{equation}
where $\omega$ indicates the frequency of the perturbation. The $\omega$ becomes complex and can be understood as frequencies of QNMs when only outgoing waves are required at infinity $r_\ast\rightarrow\pm\infty$.

\subsubsection{Radial Wave Equations of Electromagnetic  Field}
It is known that the  electromagnetic field perturbation can be classified into odd (axial) perturbation and even (polar) perturbation based the behavior of the perturbation field under the angular space inversion transformation $(\theta,\varphi)\rightarrow(\pi-\theta,\pi+\varphi)$. Under this transformation, odd perturbation changes the sign $(-1)^l$ while even perturbation changes the sign $(-1)^{l+1}$.
Usually, the effective potential of the two kind perturbation do not coincide, as what have reported in \cite{Toshmatov:2018tyo,Toshmatov:2018ell}. However, in the wormhole background in our consideration, we will show that the effective potential of the odd and even perturbations coincide with each other and hence the same property of these two perturbations are expected.

The EOM of electromagnetic field is given by
\begin{eqnarray}
\frac{1}{\sqrt{-g}}\partial_{\mu}(F_{\rho\sigma}g^{\rho\nu}g^{\sigma\mu}\sqrt{-g})=0,\label{eq7}
\end{eqnarray}
where $F_{\rho\sigma}=\partial_\rho A_\sigma-\partial_\sigma A_\rho$ is the electromagnetic field tensor and $A_{\sigma}$ is the related gauge potential. The odd electromagnetic field perturbation can be expressed as
\begin{equation}
A_\mu^{odd}(t,r,\theta,\varphi)=\sum_{l,m}\left(\left[
                   \begin{array}{c}
                     0 \\
                     0 \\
                     \frac{\Psi^{lm}(t,r)\partial_{\varphi}Y_{lm}(\theta,\varphi)}{\sin\theta} \\
                     -\Psi^{lm}(t,r)\sin\theta\partial_\theta Y_{lm}(\theta,\varphi) \\
                   \end{array}
                 \right]\right).
\end{equation}
Then we get the following nonvanishing covariant components of electromagnetic field tensor
\begin{subequations}\label{eq8}
\begin{align}
&F_{t\theta}=\frac{1}{\sin\theta}\partial_t\Psi^{lm}\partial_\varphi Y_{lm},\\
&F_{t\varphi}=-\sin\theta\partial_t\Psi^{lm}\partial_\theta Y_{lm},\\
&F_{r\theta}=\frac{1}{\sin\theta}\partial_r\Psi^{lm}\partial_\varphi Y_{lm},\\
&F_{r\varphi}=-\sin\theta\partial_r\Psi^{lm}\partial_\theta Y_{lm},\\
&F_{\theta\varphi}=-\Psi^{lm}\left[\partial_\theta(\sin\theta\partial_\theta Y_{lm})+\frac{1}{\sin\theta}\partial_\varphi^2Y_{lm}\right]=l(l+1)\sin\theta\Psi^{lm}Y_{lm}.
\end{align}
\end{subequations}
Substituting Eq. \eqref{eq8} into Eq. \eqref{eq7}, we obtain the  radial wave equation
\begin{equation}
-\frac{l(l+1)A}{r^2}\Psi^{lm}+\frac{BA'-AB'}{2B^2}\partial_r\Psi^{lm}+\frac{A}{B}\partial_r^2\Psi^{lm}-\partial_t^2\Psi^{lm}=0,\label{eq9}
\end{equation}
and one should not confuse the metric function $A(r)$ with the gauge potential $A_{\mu}$. In the tortoise coordinate defined by $dr_\ast=\sqrt{B(r)/A(r)}dr$, Eq. \eqref{eq9} can be rewritten as

\begin{equation}
-\frac{\partial^2\Psi^{lm}(t, r)}{\partial t^2}+\frac{\partial^2\Psi^{lm}(t,r)}{\partial r^2_\ast}-V_{em}^{odd}(r)\Psi^{lm}(t, r)=0,
\end{equation}
where $V_{em}^{odd}(r)$ denotes the effective potential of odd electromagnetic perturbation and takes the form \cite{Bronnikov:2019sbx,Zinhailo:2018ska}

\begin{align}
V_{em}^{odd}(r)=A(r)\frac{l(l+1)}{r^2}\label{eqem}.
\end{align}

Now we turn to discuss the even electromagnetic field perturbation which is given in the form
\begin{equation}
A_\mu^{even}(t,r,\theta,\varphi)=\sum_{l,m}\left(\left[
                   \begin{array}{c}
                     d^{lm}(t,r)Y_{lm}(\theta,\varphi) \\
                      h^{lm}(t,r)Y_{lm}(\theta,\varphi)\\
                      k^{lm}(t,r)\partial_\theta Y_{lm}(\theta,\varphi) \\
                    k^{lm}(t,r)\partial_\varphi Y_{lm}(\theta,\varphi) \\
                   \end{array}
                 \right]\right).
\end{equation}
The nonvanishing components of the electromagnetic field tensor are given as
\begin{subequations}\label{eq12}
\begin{align}
&F_{tr}=(\partial_t h^{lm}-\partial_r d^{lm})Y_{lm},\\
&F_{t\theta}=(\partial_t k^{lm}-d^{lm})\partial_\theta Y_{km},\\
&F_{t\varphi}=(\partial_t k^{lm}-d^{lm})\partial_\varphi Y_{km},\\
&F_{r\theta}=(\partial_r k^{lm}-h^{lm})\partial_\theta Y_{lm},\\
&F_{r\varphi}=(\partial_r k^{lm}-h^{lm})\partial_\varphi Y_{lm}.
\end{align}
\end{subequations}
By employing Eq. \eqref{eq7} and Eq. \eqref{eq12}, one can get following equations
\begin{align}
\begin{split}
\left(\frac{2rA}{H}-\frac{r^2AH'}{2H^2}\right)\partial_r d^{lm}+\frac{r^2A}{H}\partial_r^2 d^{lm}&+\left(\frac{r^2AH'}{2H^2}-\frac{2rA}{H}\right)\partial_t h^{lm}-\frac{r^2A}{H}\partial_t\partial_r h^{lm}\\
&+l(l+1)(\partial_t k^{lm}-d^{lm})=0,\label{eq10}
\end{split}
\end{align}
\begin{equation}
-\frac{r^2}{A}\partial_t\partial_r d^{lm}+\frac{r^2}{A}\partial_t^2 h^{lm}+l(l+1)(h^{lm}-\partial_r k^{lm})=0,\label{eq11}
\end{equation}
where for convenience, we have defined $H(r)=A(r)B(r)$. By differentiating Eq. \eqref{eq10} and Eq. \eqref{eq11} with respect to $r$ and $t$, respectively, and then we define new variable
\begin{equation}
\Psi^{lm}(t,r)=\frac{r^2}{\sqrt{H}}(\partial_t h^{lm}-\partial_r d^{lm}),
\end{equation}
and introduce tortoise coordinate $dr_\ast=\sqrt{B(r)/A(r)}dr$, finally we arrive at the wave equation
\begin{gather}
-\frac{\partial^2\Psi^{lm}(t, r)}{\partial t^2}+\frac{\partial^2\Psi^{lm}(t,r)}{\partial r^2_\ast}-V_{em}^{even}(r)\Psi^{lm}(t, r)=0,\\
V_{em}^{even}(r)=A(r)\frac{l(l+1)}{r^2}.\label{eqem}
\end{gather}
It is found that the effective potential $V_{em}^{even}$ of even electromagnetic field perturbation coincides with $V_{em}^{odd}$ of odd one, and henceforth we omit index $odd$ and $even$  due to the identical effective potential for both parity.
Considering time dependence of $\Psi(t,r)=e^{-i\omega t}\Phi(r)$, and then we get the wave equation in frequency domain for both odd and even perturbation
\begin{equation}
\frac{d^2\Phi(r)}{dr^2_\ast}+(\omega^2-V_{em}(r))\Phi(r)=0.
\end{equation}

Regarding the effective potential, the centrifugal barrier $\frac{l(l+1)}{r^2}$
in the first term of $V_s(r)$ and $V_{em}(r)$ comes from the decomposition
of spherical harmonic function. The redshift factor $A(r)$ in this term accounts for
relativistic effects near the throat of wormhole.

\subsubsection{Time-domain integration}
We explore the signals of the GW echo by studying the time-domain profiles of the scalar and electromagnetic fields, which can be
obtained by directly integrating the time-dependent differential equation.
To achieve this goal, we recall the EOM shared by scalar and electromagnetic  fields with the following wavelike form
\begin{equation}
-\frac{\partial^2\Psi(t, r_\ast)}{\partial t^2}+\frac{\partial^2\Psi(t,r_\ast)}{\partial r^2_\ast}-V(r(r_\ast))\Psi(t,r_\ast)=0.\label{eq3}
\end{equation}
In general, it is almost impossible to obtain an analytical solution
to the above wave equation with the given effective potential $V(r(r_\ast))$.
So we turn to resort to the numerical method.

To this end, we define $\Psi(t,r_\ast)=\Psi(i\Delta t, j\Delta r_\ast)=\Psi_{i,j}$, $V(r(r_\ast))=V(j\Delta r_\ast)=V_j$ such that we can discretize Eq. (\ref{eq3}) as
\begin{equation}
\begin{split}
-\frac{(\Psi_{i+1,j}-2\Psi_{i,j}+\Psi_{i-1,j})}{\Delta t^2}+\frac{(\Psi_{i,j+1}-2\Psi_{i,j}+\Psi_{i,j-1})}{\Delta r^2_\ast}-V_{j}\Psi_{i,j}+O(\Delta t^2)+O(\Delta r^2_\ast)=0.
\end{split}
\end{equation}
The detailed discretization scheme can be found in \cite{Zhu:2014sya}.
Considering the initial Gaussian distribution $\Psi(t=0,r_\ast)=\mathrm{exp}[-\frac{(r_\ast-\bar{a})^2}{2b^2}]$ and $\Psi(t<0,r_\ast)=0$, the discretized equation controlling time evolution of  field is derived as \cite{Zhu:2014sya}
\begin{equation}
\Psi_{i+1,j}=-\Psi_{i-1,j}+\frac{\Delta t^2}{\Delta r^2_\ast}(\Psi_{i,j+1}+\Psi_{i,j-1})+(2-2\frac{\Delta t^2}{\Delta r^2_{\ast}}-\Delta t^2V_{j})\Psi_{i,j}.
\end{equation}
The parameters are taken as $b=3$, $\Delta t/\Delta r_\ast=0.5$ (we take $\Delta t=0.1$ and $\Delta r_\ast=0.2$) in our discussion, and we will properly choose the values of $\bar{a}$ in different cases. Note that  the von Neumann stability conditions usually require that $\Delta t/\Delta r_\ast<1$, so the results for a grid size smaller than  $\Delta t/\Delta r_\ast=0.5$ made in our choice will remain unchanged if smaller size of $\Delta t$ and $\Delta r_\ast$ are taken (e.g., $\Delta t/\Delta r_\ast=0.1$, we need $\Delta t =0.02$ and $\Delta r_\ast=0.2$, and much more time is required to run the code). Actually, a reliable numerical result depends not only on the value of $\Delta t/\Delta r_\ast$, but also on the respective value of $\Delta t$ and $\Delta r_\ast$. Usually, smaller size of $\Delta t$ and $\Delta r_\ast$ leads to more accurate results.

\section{Echoes signals in Wormhole Model  \uppercase\expandafter{\romannumeral 1} background}
\label{section3}
In this section we focus on  the time-domain profile of the scalar field and electromagnetic field in Wormhole Model  \uppercase\expandafter{\romannumeral 1} background. We investigate the signals of echo in this model and further explore the effects of wormhole parameters on the behaviors of echo. Next, we shall study the echoes from scalar field and electromagnetic field, respectively. Also, a brief comment on the similarities and differences of both fields is also presented.
\subsection{Echoes of scalar field waves}
\begin{figure}
\centering
\includegraphics[height=2.in,width=3.in]{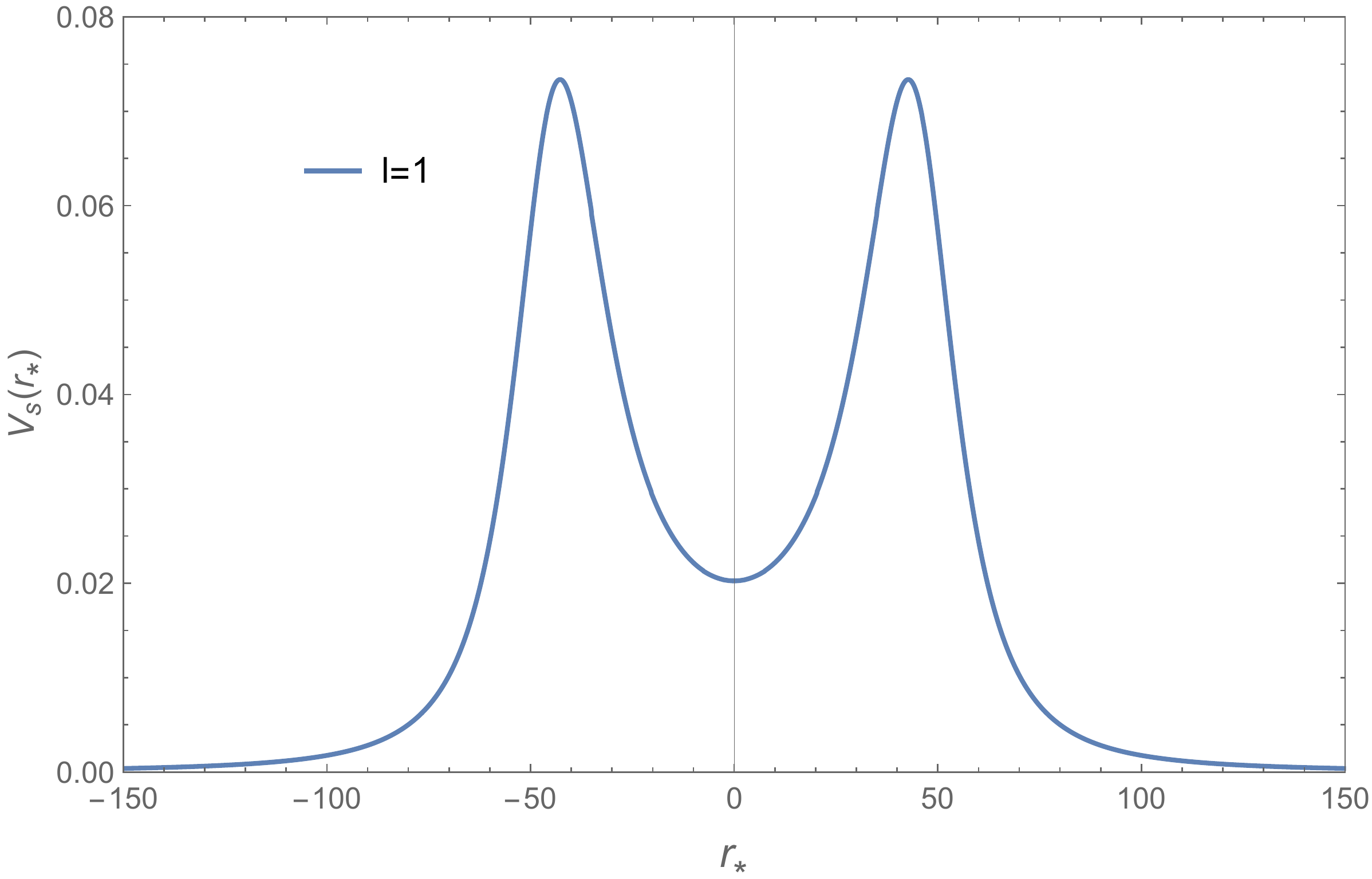}~~
\includegraphics[height=2in,width=3in]{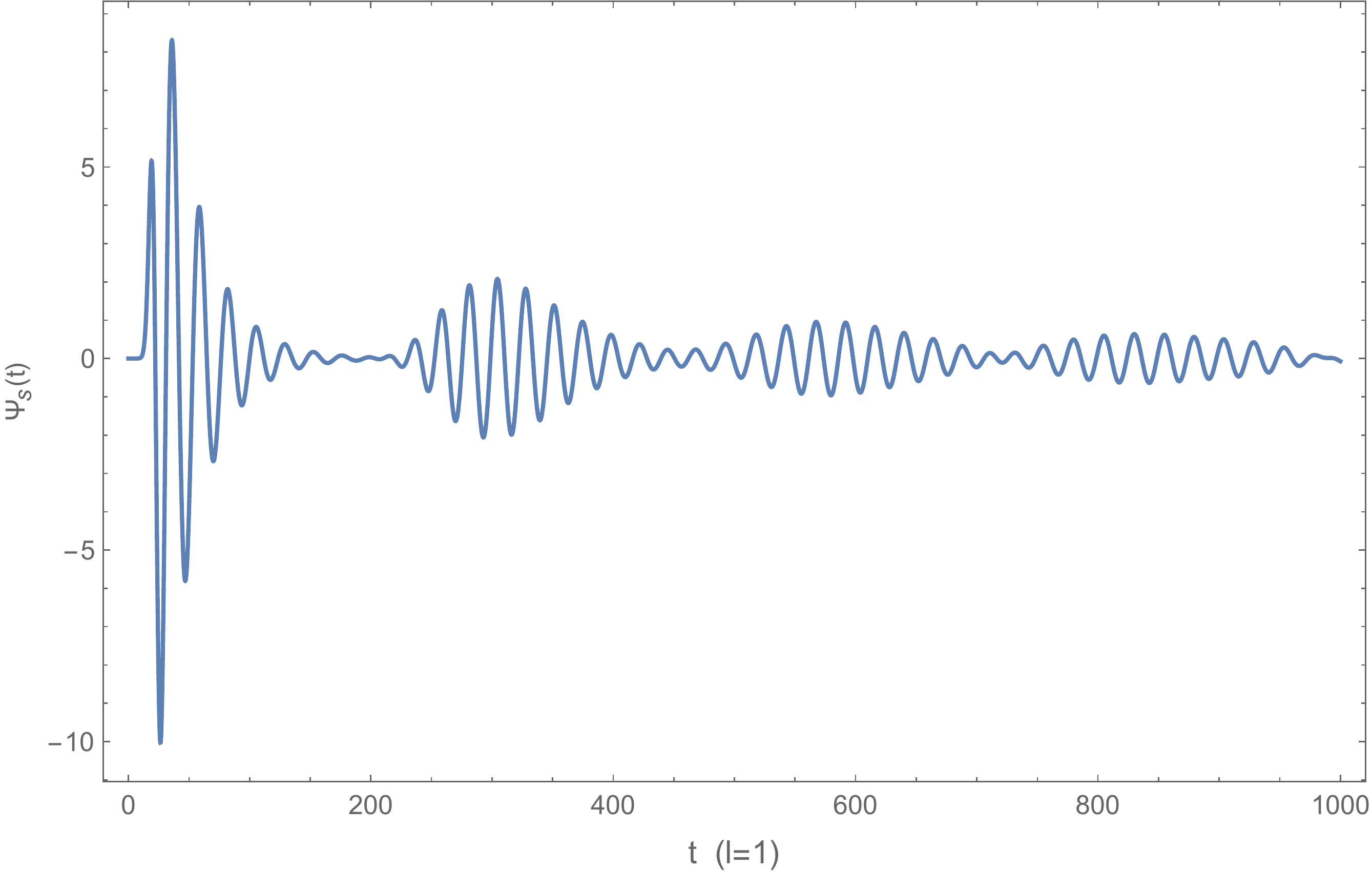}

\includegraphics[height=2.in,width=3.in]{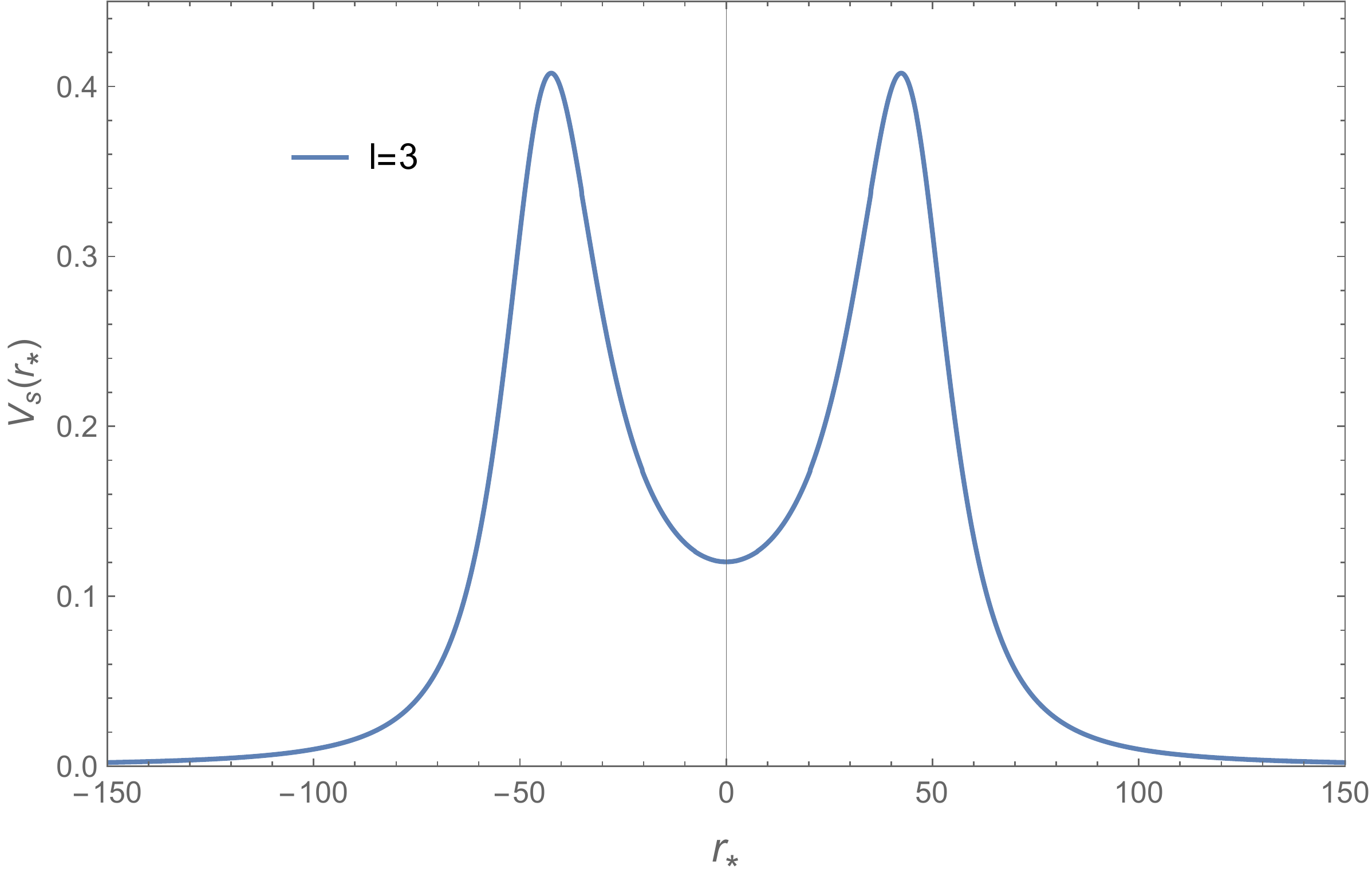}
\includegraphics[height=2in,width=3in]{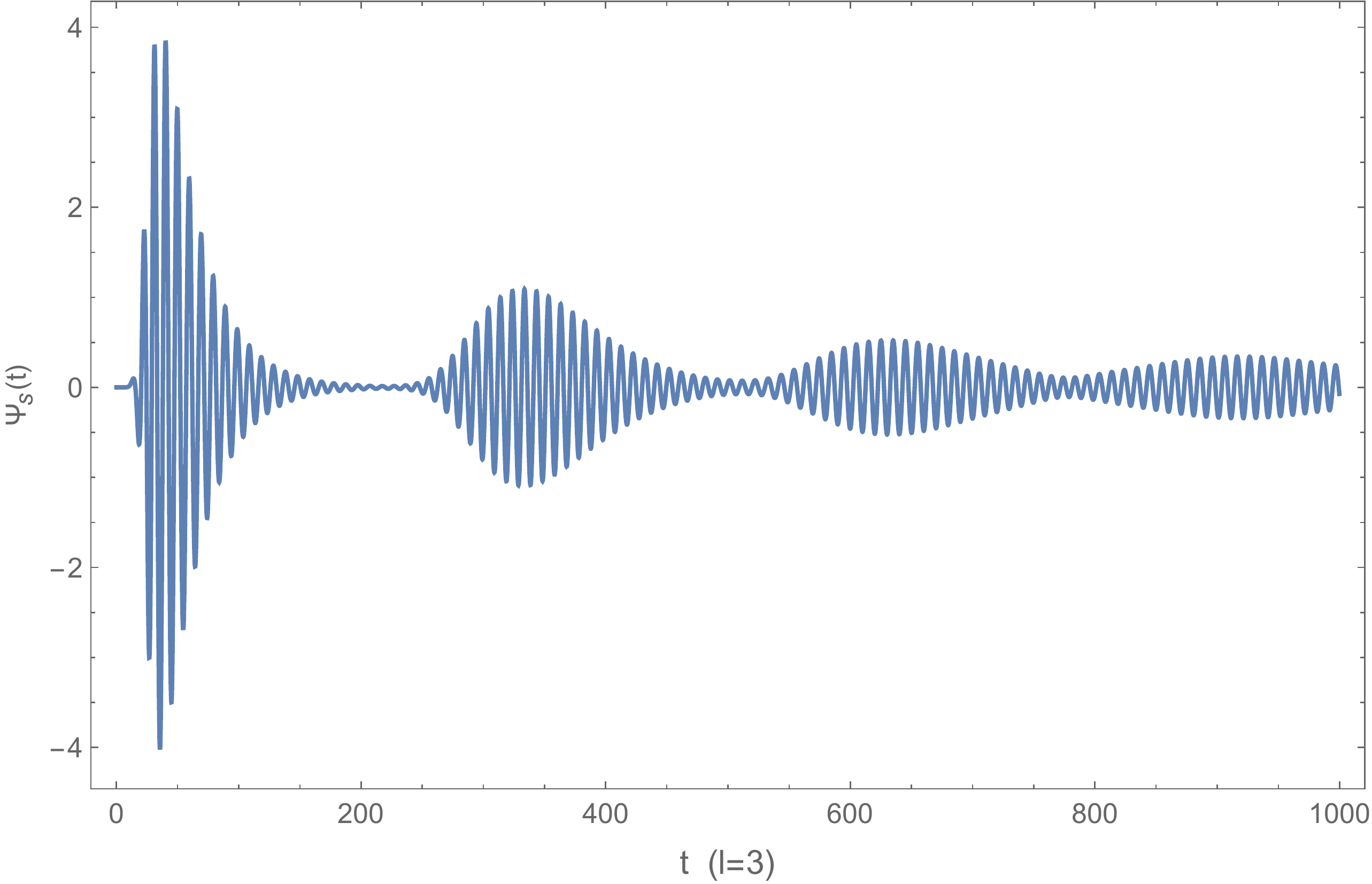}

\includegraphics[height=2.in,width=3.in]{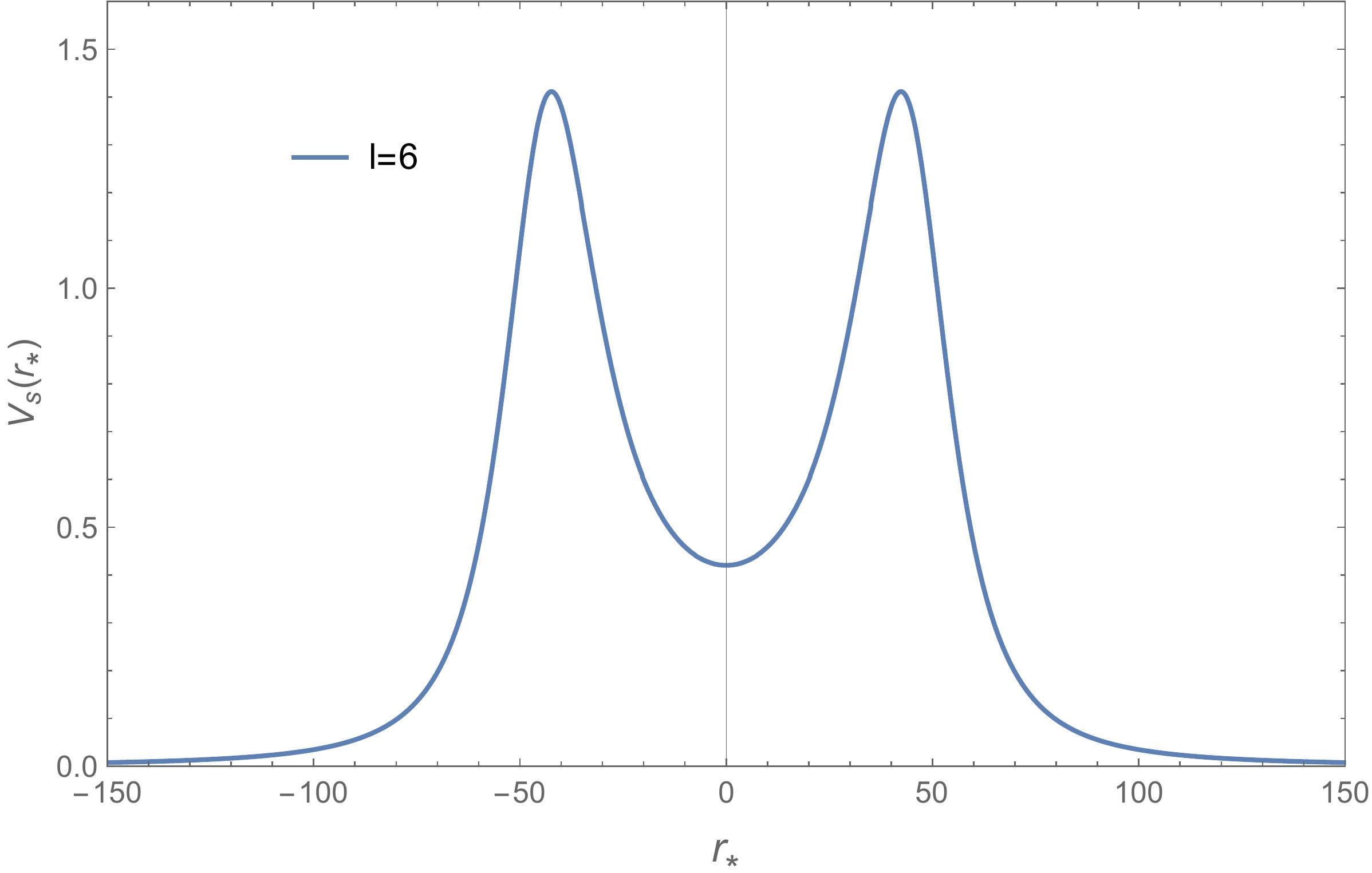}
\includegraphics[height=2in,width=3in]{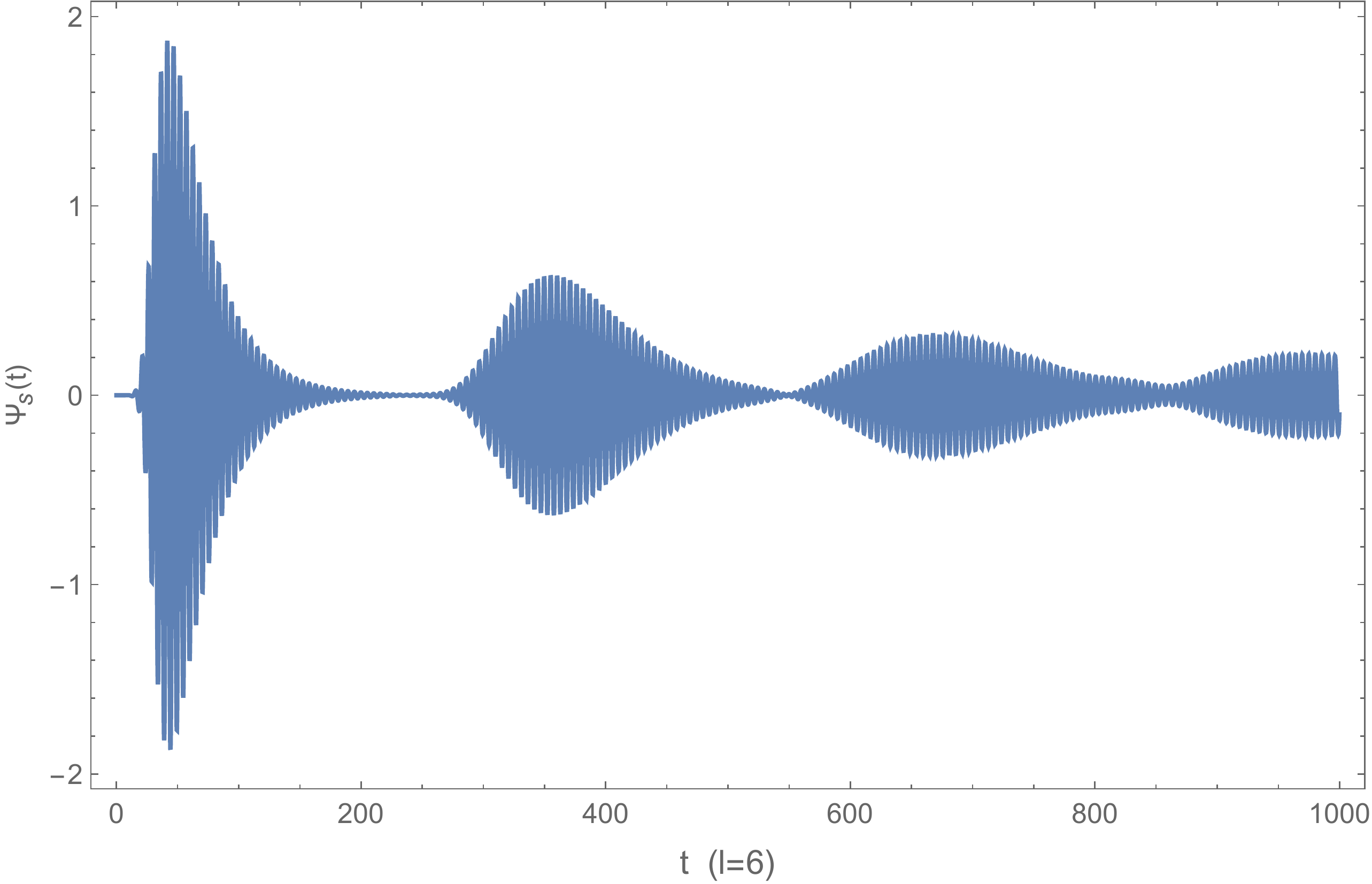}
\caption{The effective potential behavior  in coordinate $r_\ast$  and time evolution of  scalar field for different angular number  $l$, and we take parameter $a=1.9$ in this figure.}\label{fig2}
\end{figure}

Before going to explore the effects of wormhole parameters on the echoes, we want to understand the dependence of the echoes on the angular number $l$. The echoes appear thanks to a potential well as the consequence of the structure of the wormhole geometry. Therefore, the shape of the effective potential is important in understanding the properties of the echo from wormhole geometry. To this end, we show the effective potential in the left column in Fig. \ref{fig2}. Correspondingly, the time-evolution of scalar field is shown in the right column in Fig. \ref{fig2}. Note that to ensure the existence of potential well which is the necessary condition of the occurrence of echoes, the parameter $a$ is required to be $a\leq 2$ such that the value of $\omega=-2/a$ can only approach to $\omega=-1$ from below.
We will first fix $a=1.9$ and treat the angular number $l$ free to explore the influence of the angular number on echoes.

From the left column in Fig. \ref{fig2}, we indeed observe the double barriers and so thus the corresponding  echo  signals  in the time-domain profile of the scalar field (right column in Fig. \ref{fig2}). Next, we summarize the properties of the dependence of the echoes on the angular number $l$ as what follows.
\begin{itemize}
  \item Higher angular number $l$ leads to echo signals with smaller amplitudes. This observation can be understood by the potential well. From left column in Fig. \ref{fig2}, we see that with the increase of $l$, the potential grows quickly. The higher potential barriers will make it more difficult for the scalar waves to escape from the potential well to generate echoes. Also, it is instructive to explore the relation between the amplitude of the  first echoes signals and the amplitude of the initial burst for different $l$. We describe this relation by  the ratio of echoes amplitude to initial burst amplitude. Intriguingly, we find that for $l=1$, $l=3$ and $l=6$, the ratio is $0.2501$, $0.2832$ and $0.3378$, which is growing with the increase of $l$.
  \item The time delay between two successive echoes depend very weakly on the angular number $l$. It is because the width of the potential is only slightly changed with the increase of $l$.
  \item Since the higher $l$ is related to higher oscillation frequency of the waves, the waves with higher $l$  oscillate more rapidly. We can also understand this phenomenon intuitively by noting that the higher potential barriers arising from the higher angular number only allow the waves with high enough energy to escape from it, and hence waves with higher frequencies can be observed.
\end{itemize}
\begin{figure}
\centering
\includegraphics[height=2.2in,width=3.2in]{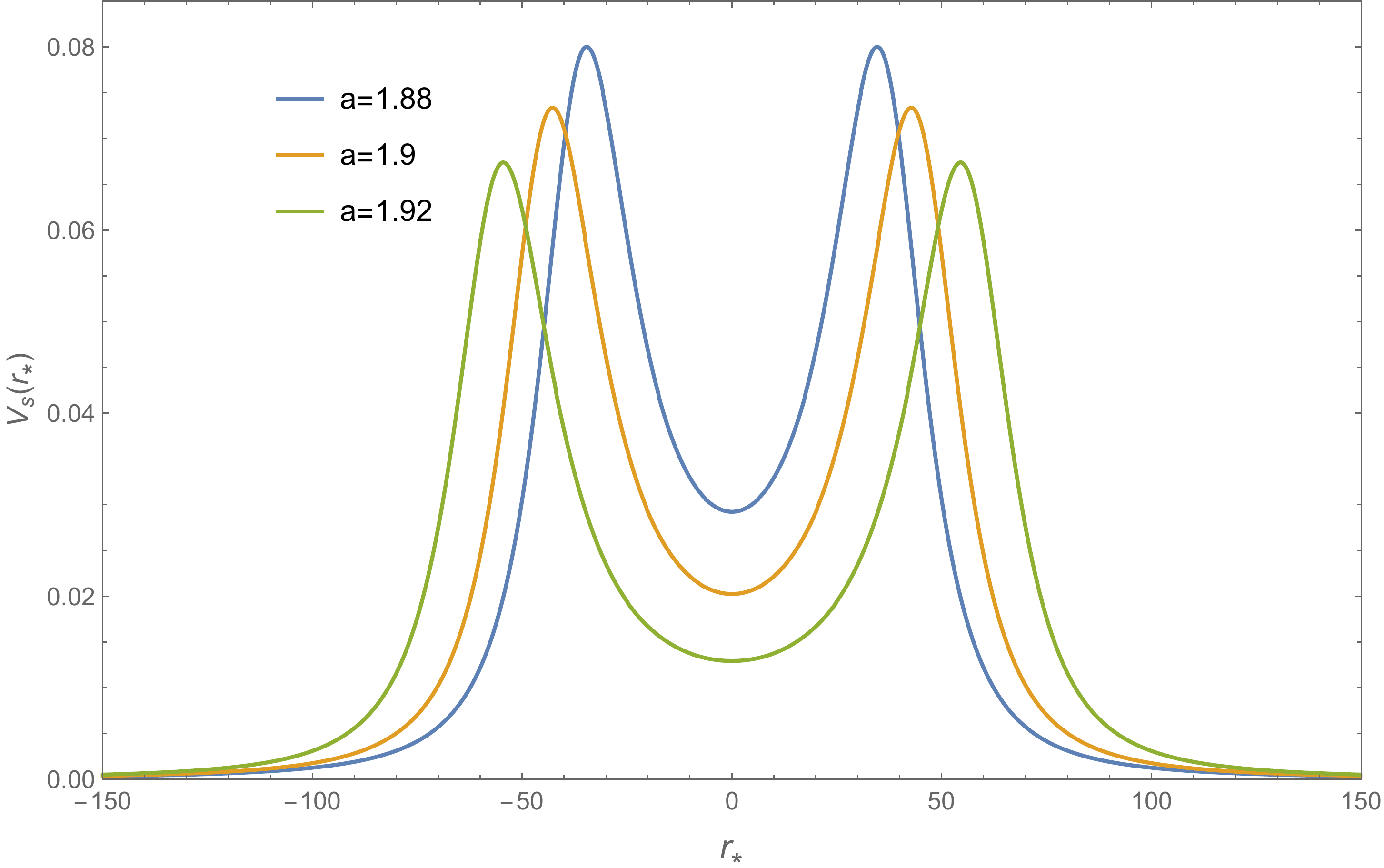}
\includegraphics[height=2.2in,width=3.2in]{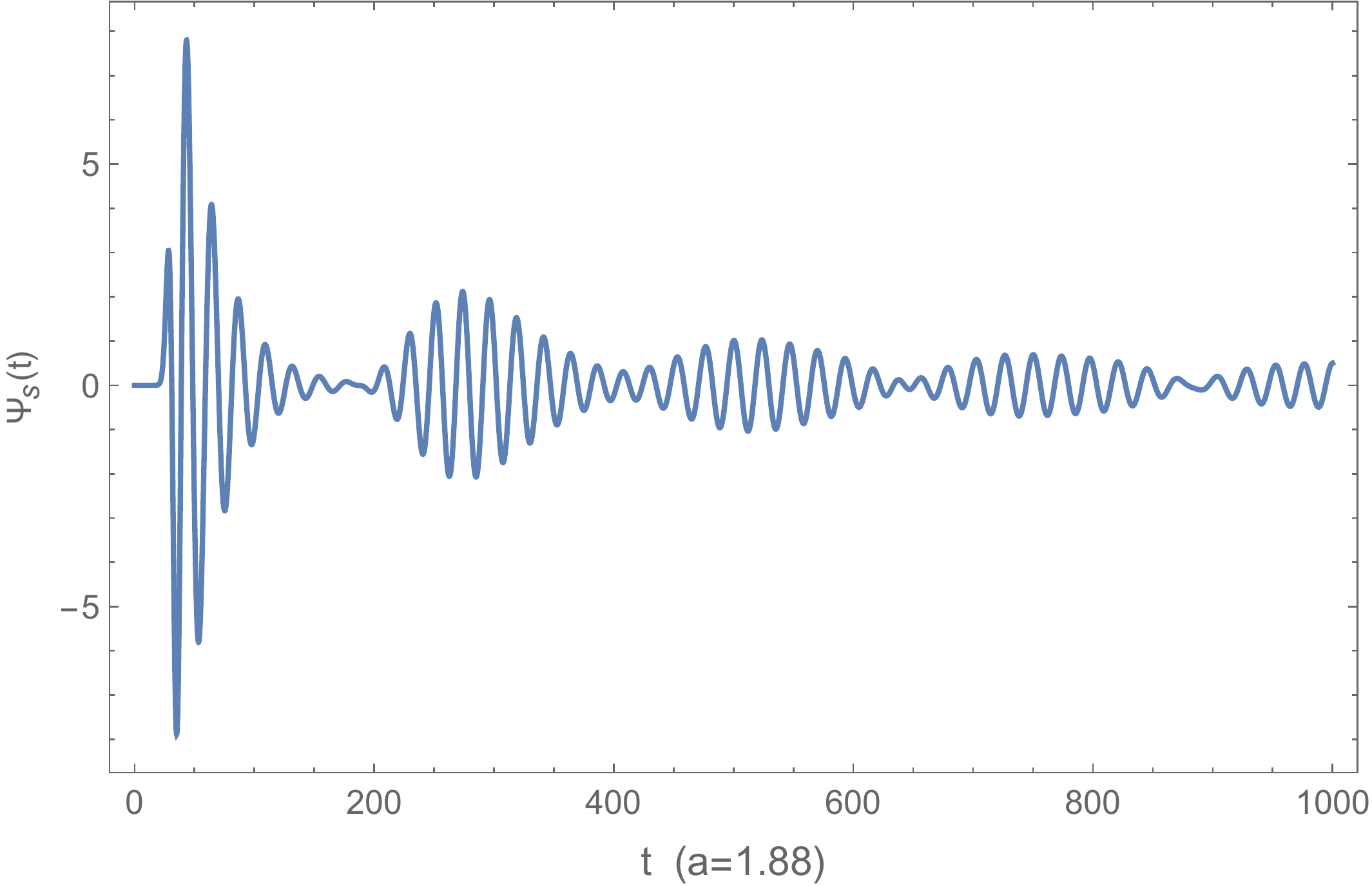}
\includegraphics[height=2.2in,width=3.2in]{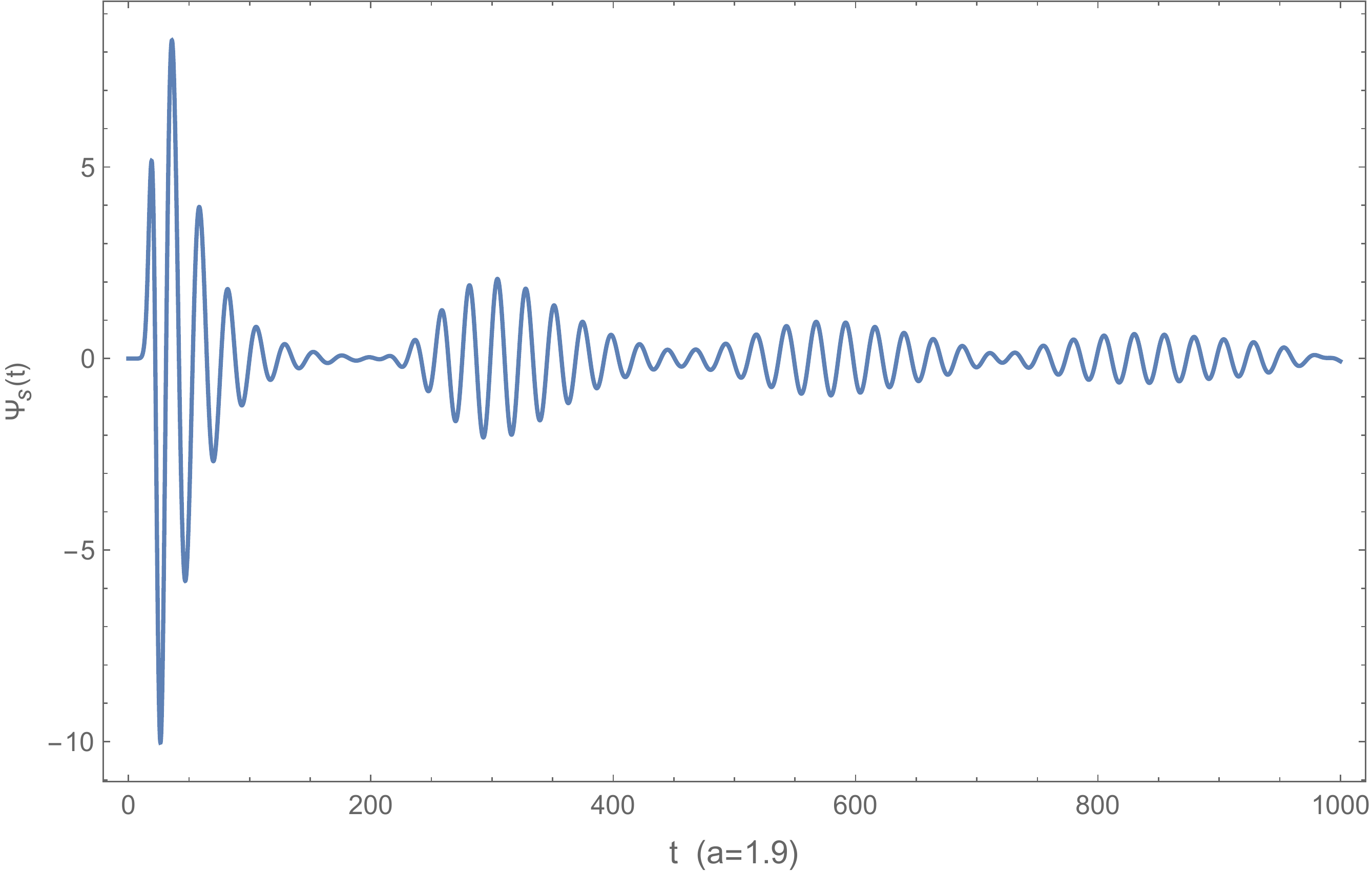}
\includegraphics[height=2.2in,width=3.2in]{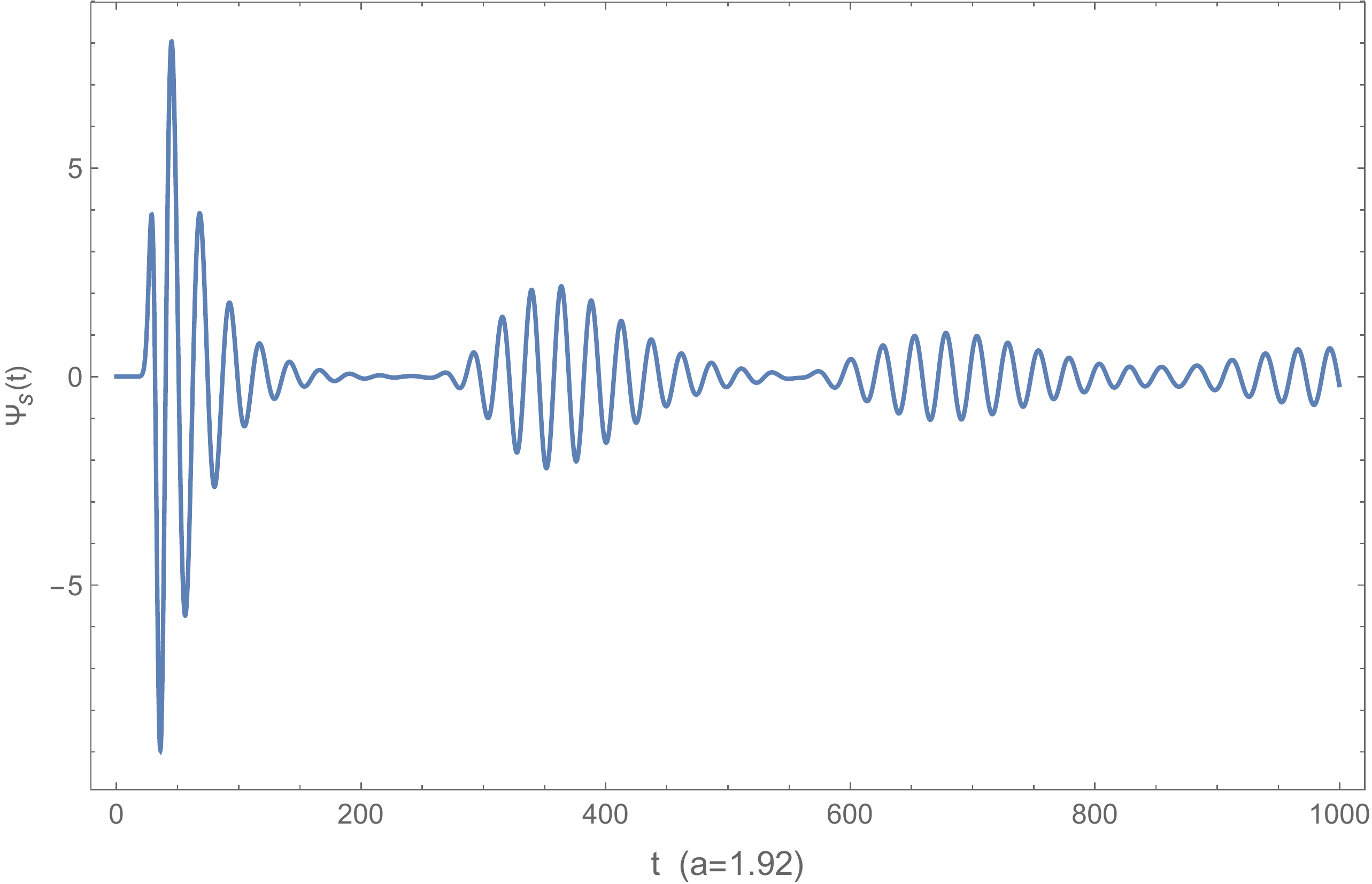}
\caption{The left one plot in the top row is  effective potential behavior of scalar field in coordinate $r_\ast$ for different parameter $a$ and the other plots are the time-domain profile of the scalar field corresponding to each potential.  We take angular number $l=1$ in this figure.}\label{fig1}
\end{figure}

Now, we fix the angular number $l=1$ and turn to focus on the effect from phantom matter on the echoes.
Fig. \ref{fig1} shows the effective potential and time evolution of scalar field for different phantom wormhole parameter $a$,
which relates the state of equation of phantom matter as $\omega=-2/a$. One can see that the model parameter $a$ has a strong effect on the time delay between echoes. As $a$ increases from small values and approaches to the upper limit $2$, the time delay becomes longer. This is because the potential becomes wider with the increase of $a$ (left plot in the top row in Fig. \ref{fig1}).
It indicates that the time delay becomes longer as the equation of state $\omega$ approaches $-1$ from the phantom state. Therefore, once the signals of echo is detected in future observations, it is possible to constrain the state parameter $\omega$ of phantom matter. In addition, we also note that the model parameter $a$ has very weak effect on the amplitudes of the echo. It can be attributed to the fact that the height of the potential barriers are almost unchanged for different model parameter $a$ (left plot in the top row in Fig. \ref{fig1}).

\subsection{Echoes of electromagnetic field waves}

\begin{figure}
\centering
\includegraphics[height=2in,width=3in]{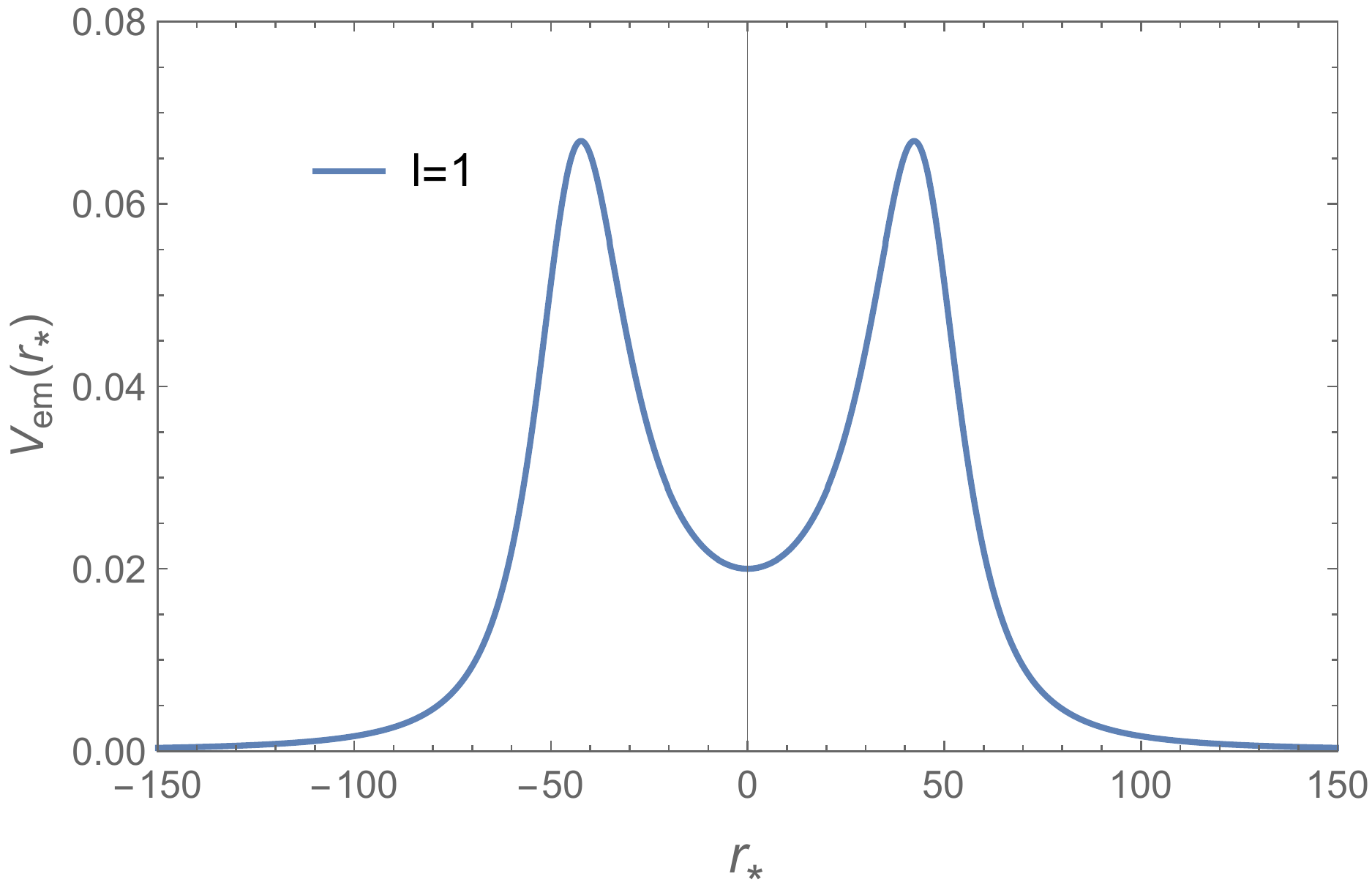}~~
\includegraphics[height=2in,width=3in]{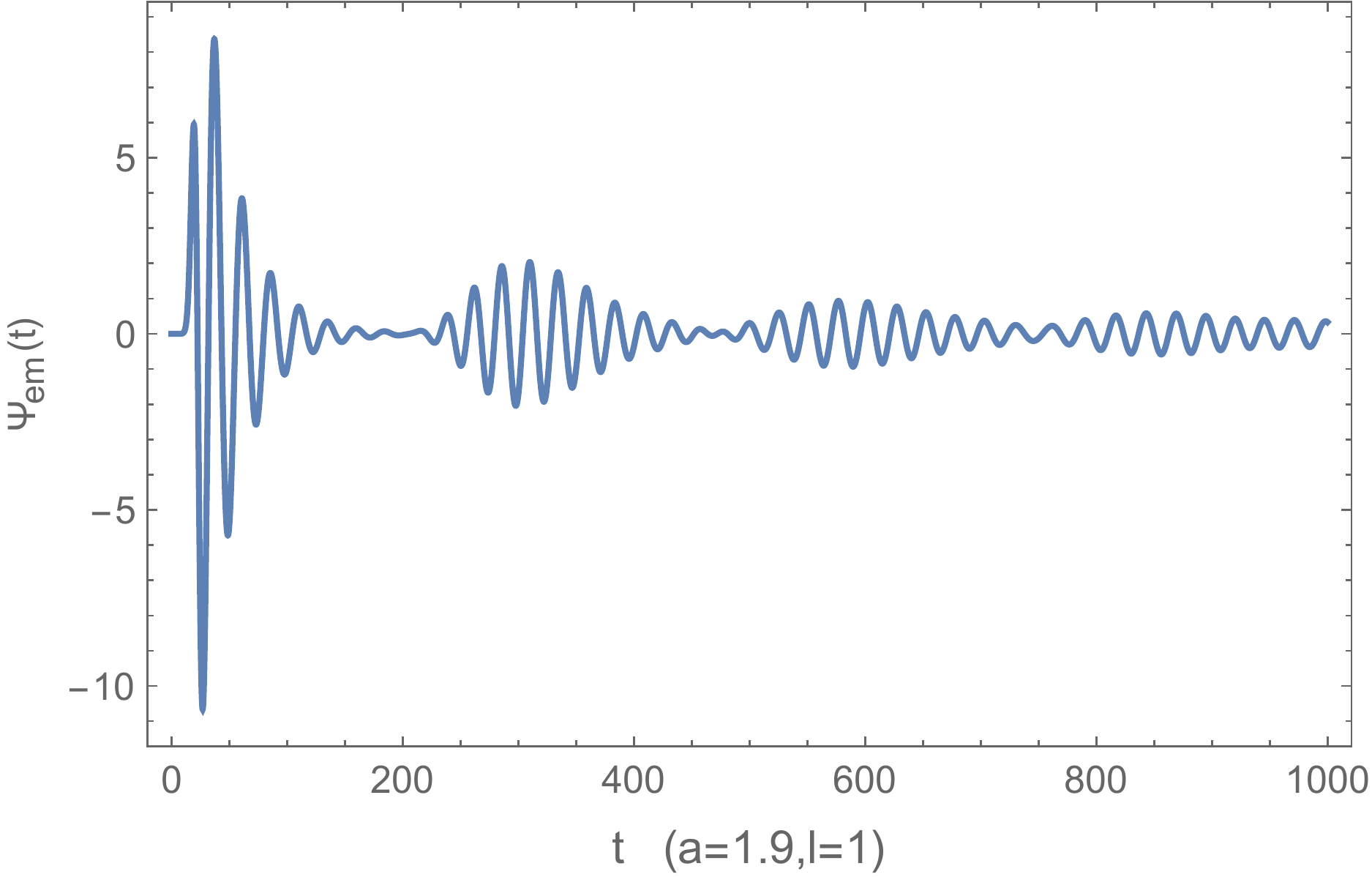}
\includegraphics[height=2in,width=3in]{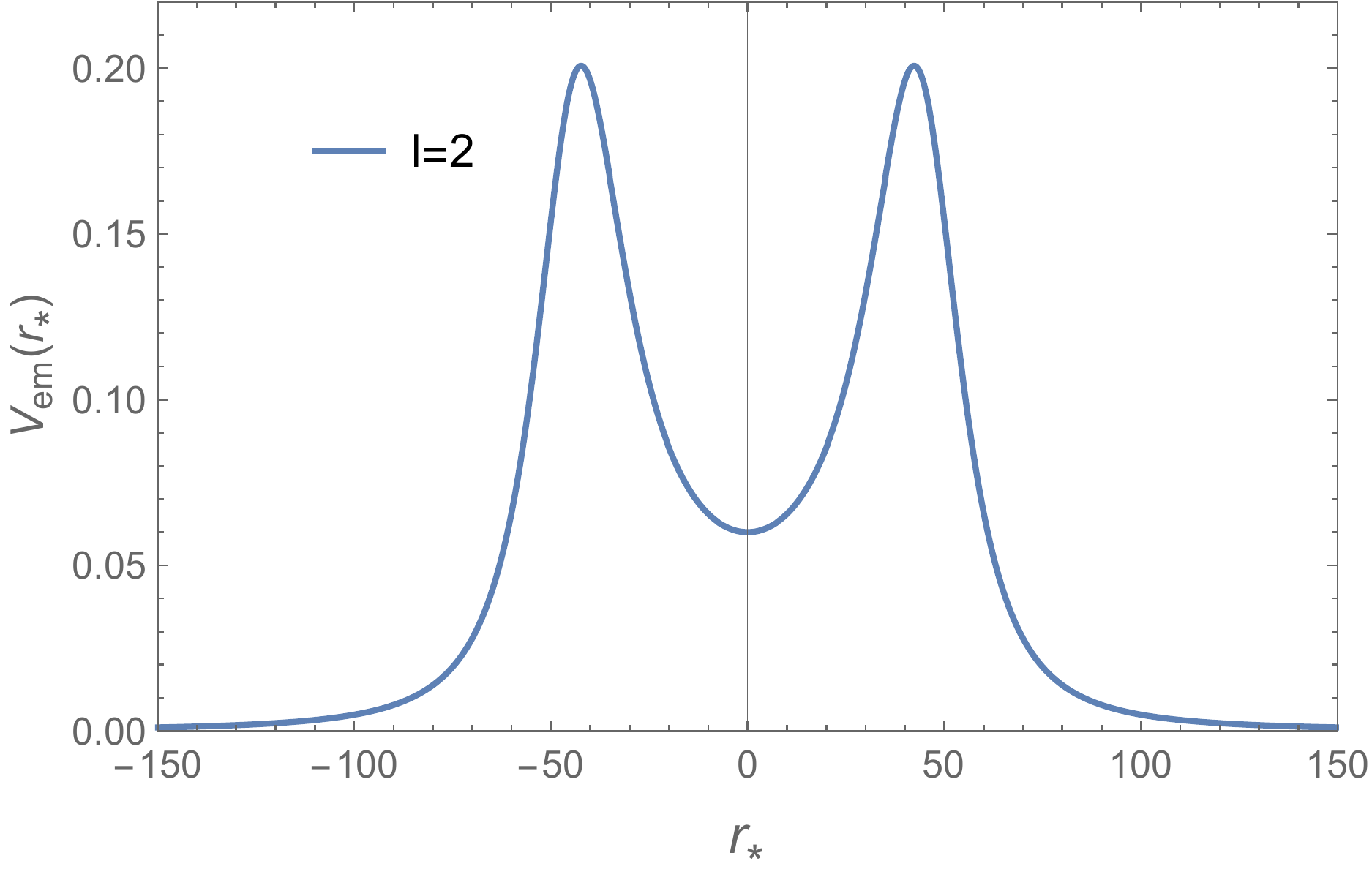}
\includegraphics[height=2in,width=3in]{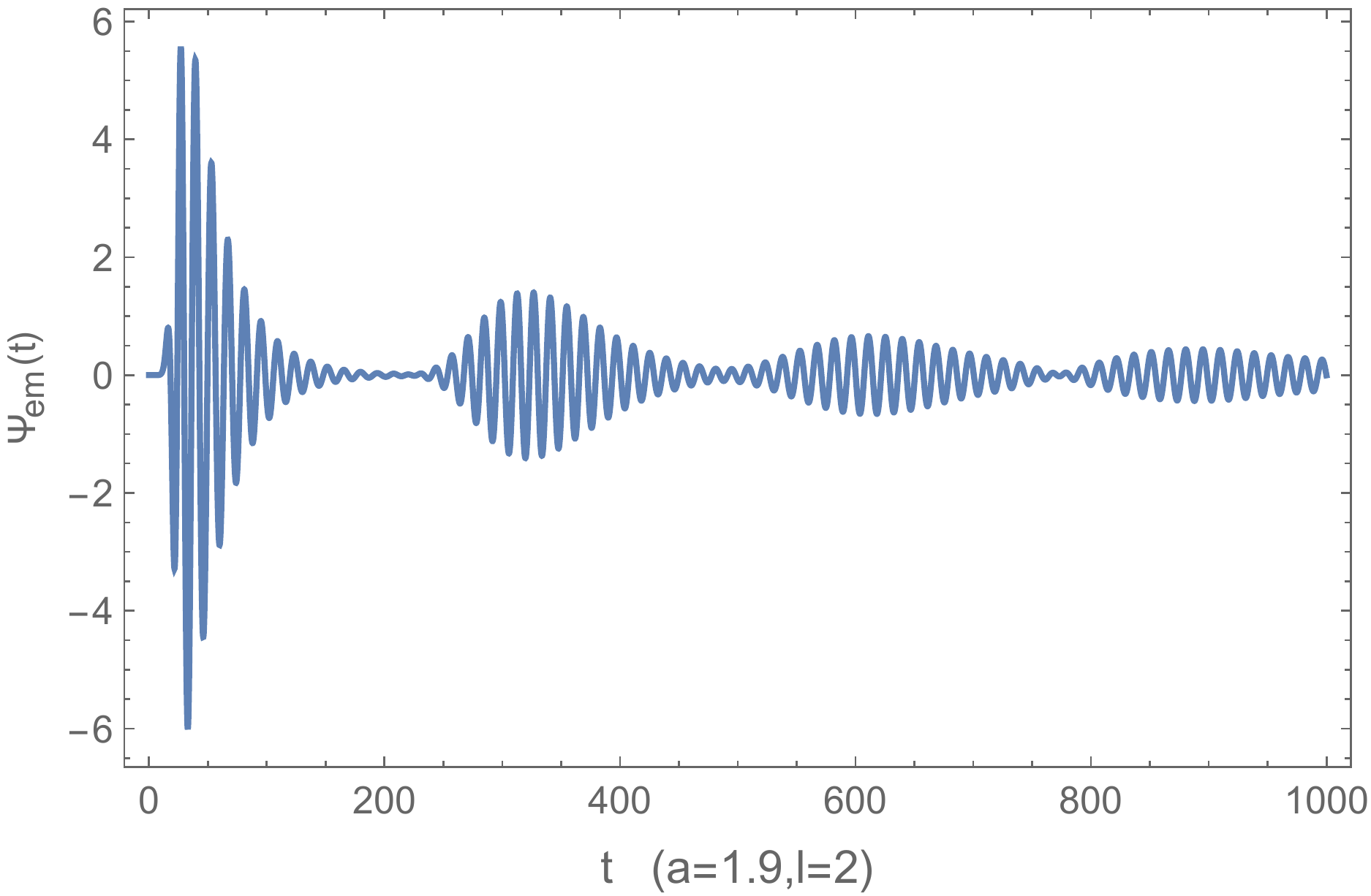}
\includegraphics[height=2in,width=3in]{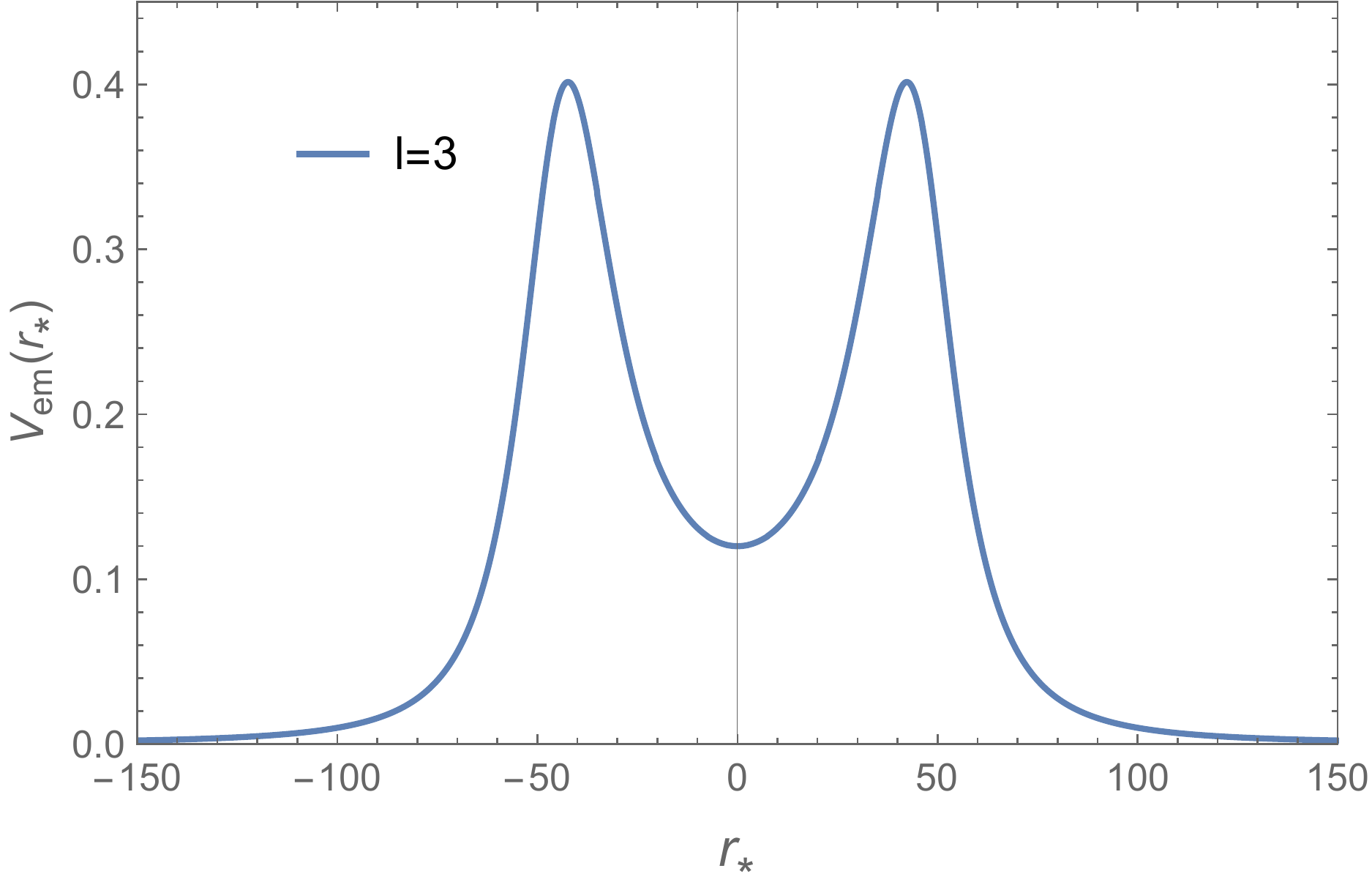}
\includegraphics[height=2in,width=3in]{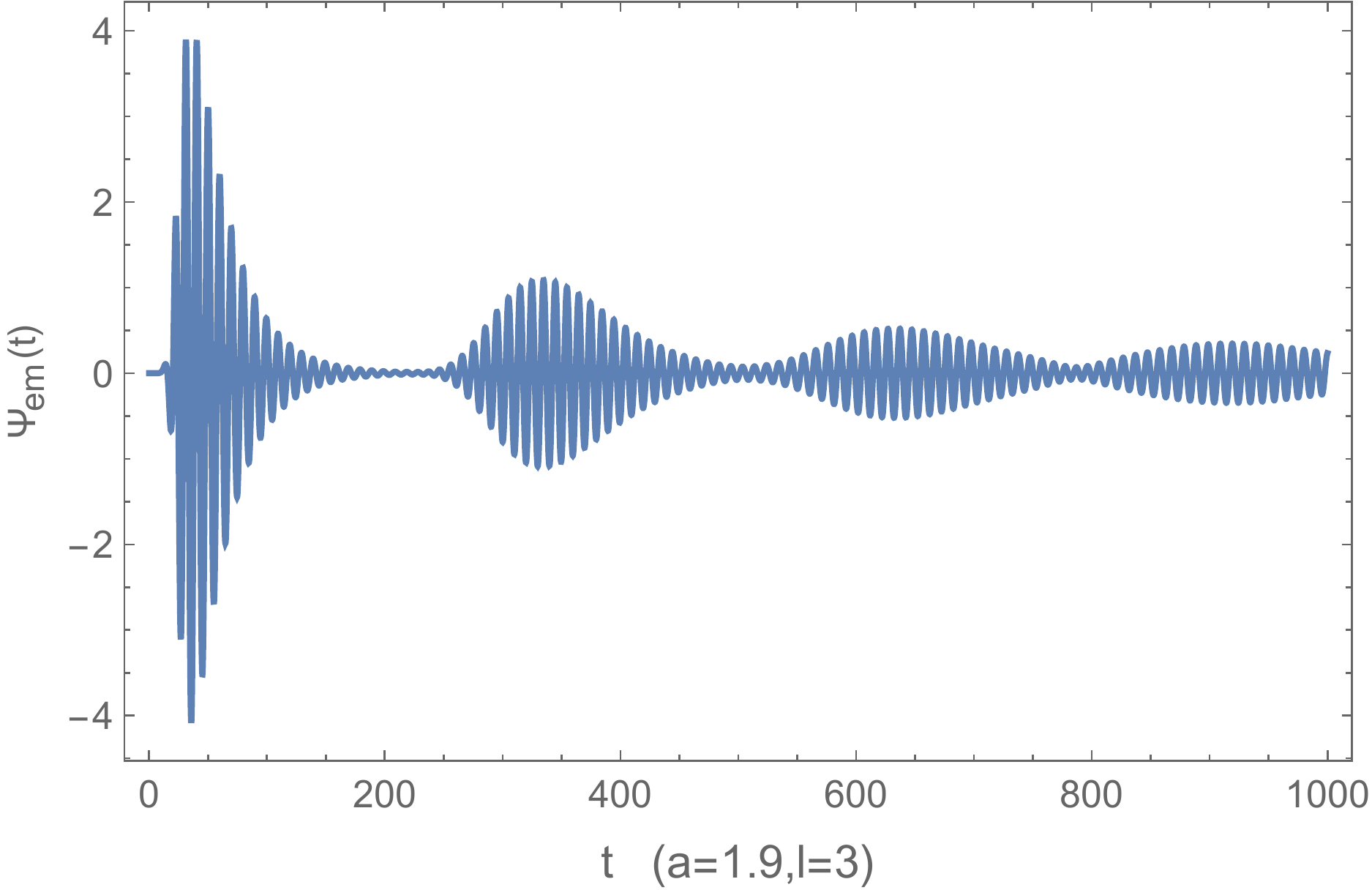}
\includegraphics[height=2in,width=3in]{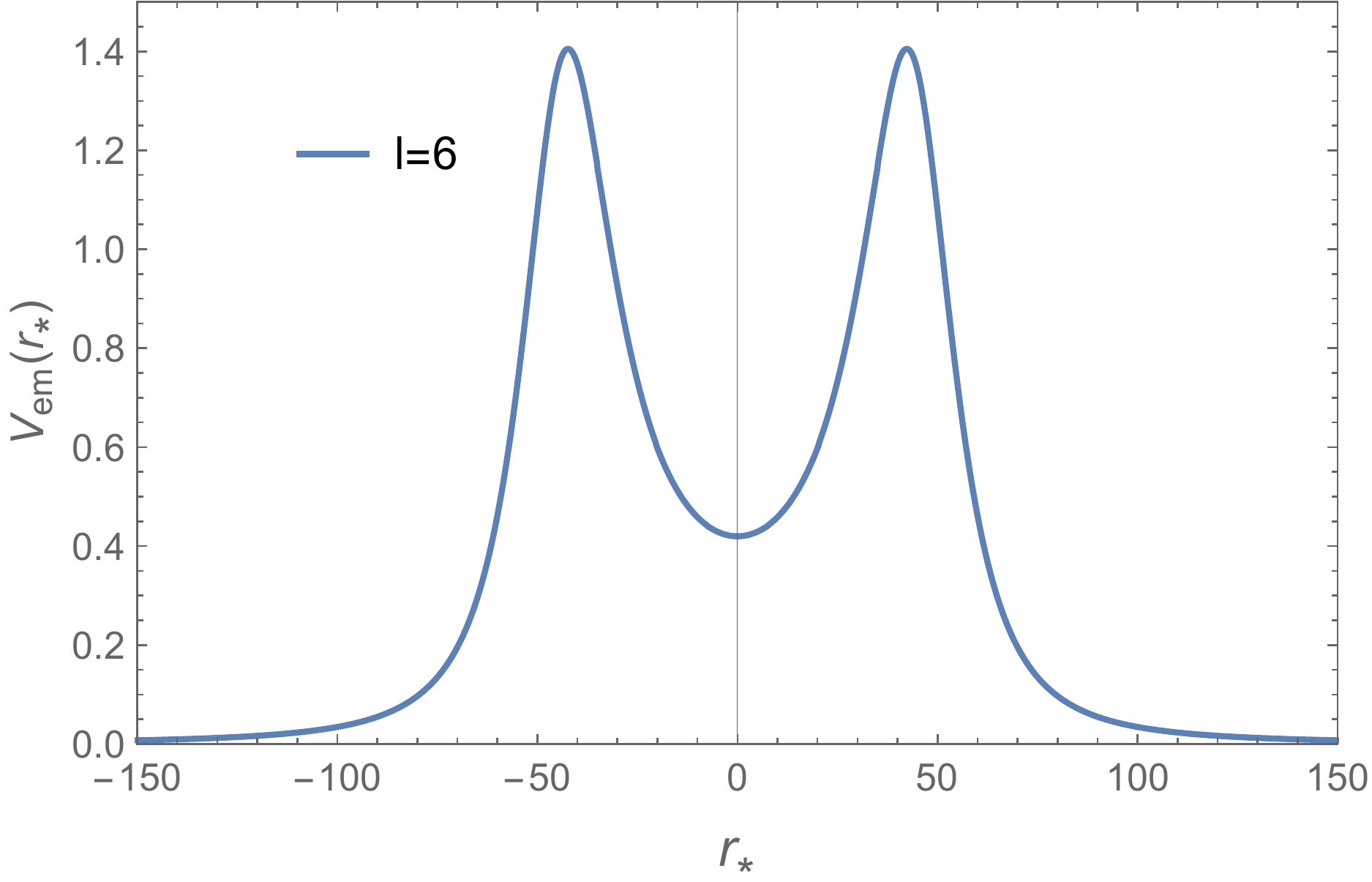}
\includegraphics[height=2in,width=3in]{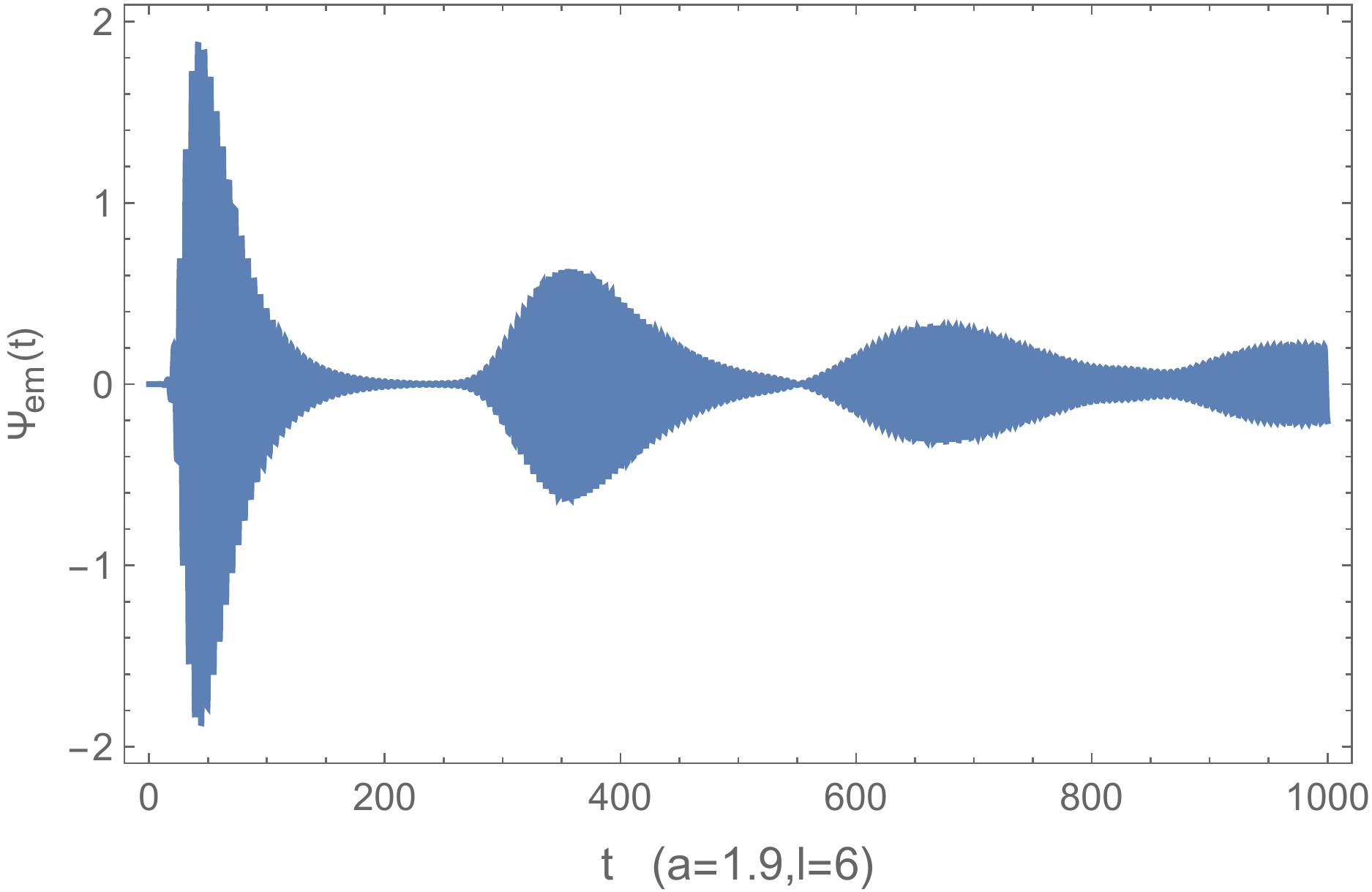}
\caption{The effective potential behavior in tortoise coordinate $r_\ast$ and time-evolution  of electromagnetic field  for different angular number  $l$, and we take parameter $a=1.9$ in this figure.}\label{fig5}
\end{figure}
\begin{figure}
\centering
\includegraphics[height=2in,width=3in]{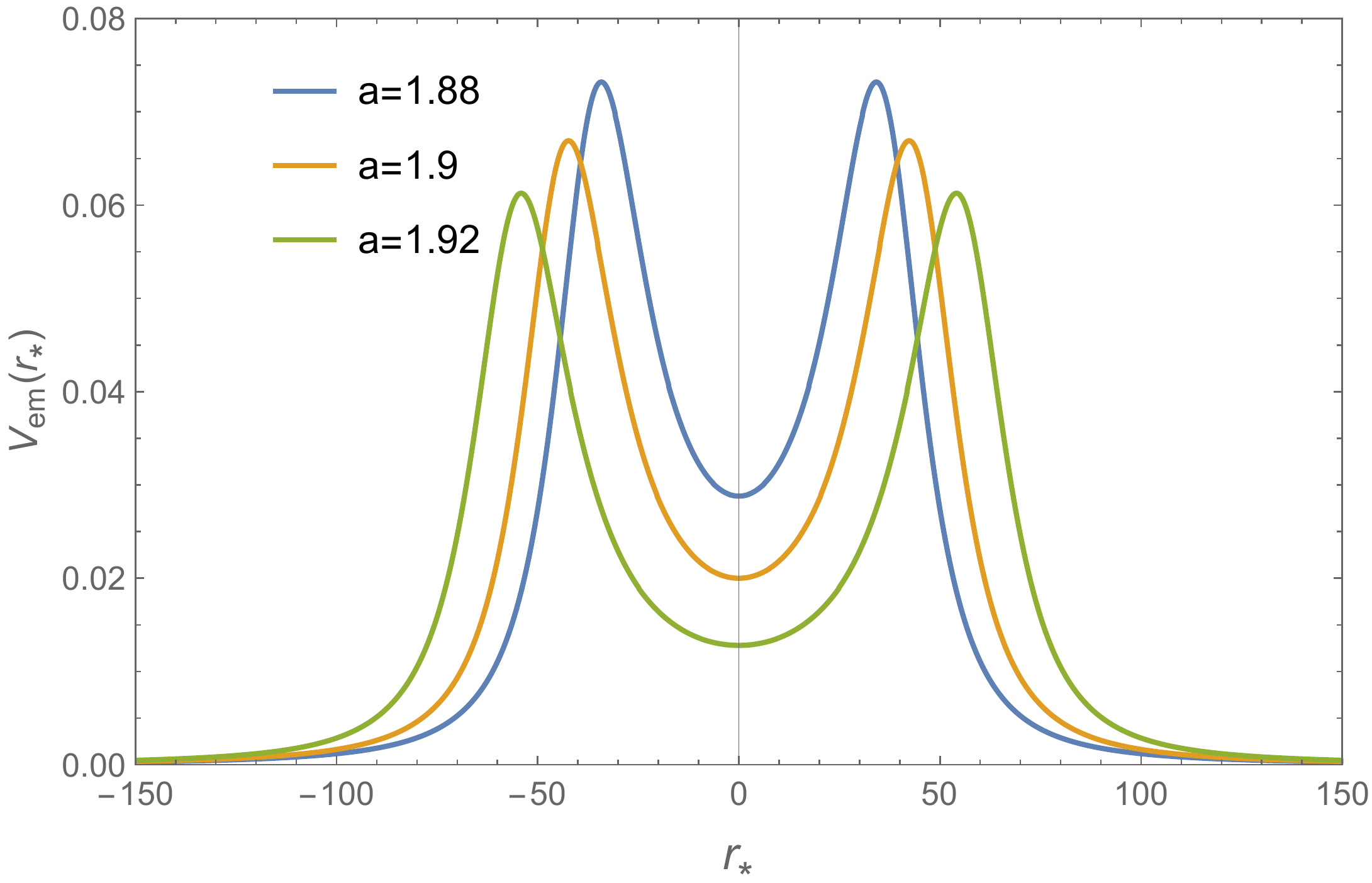}
\includegraphics[height=2in,width=3in]{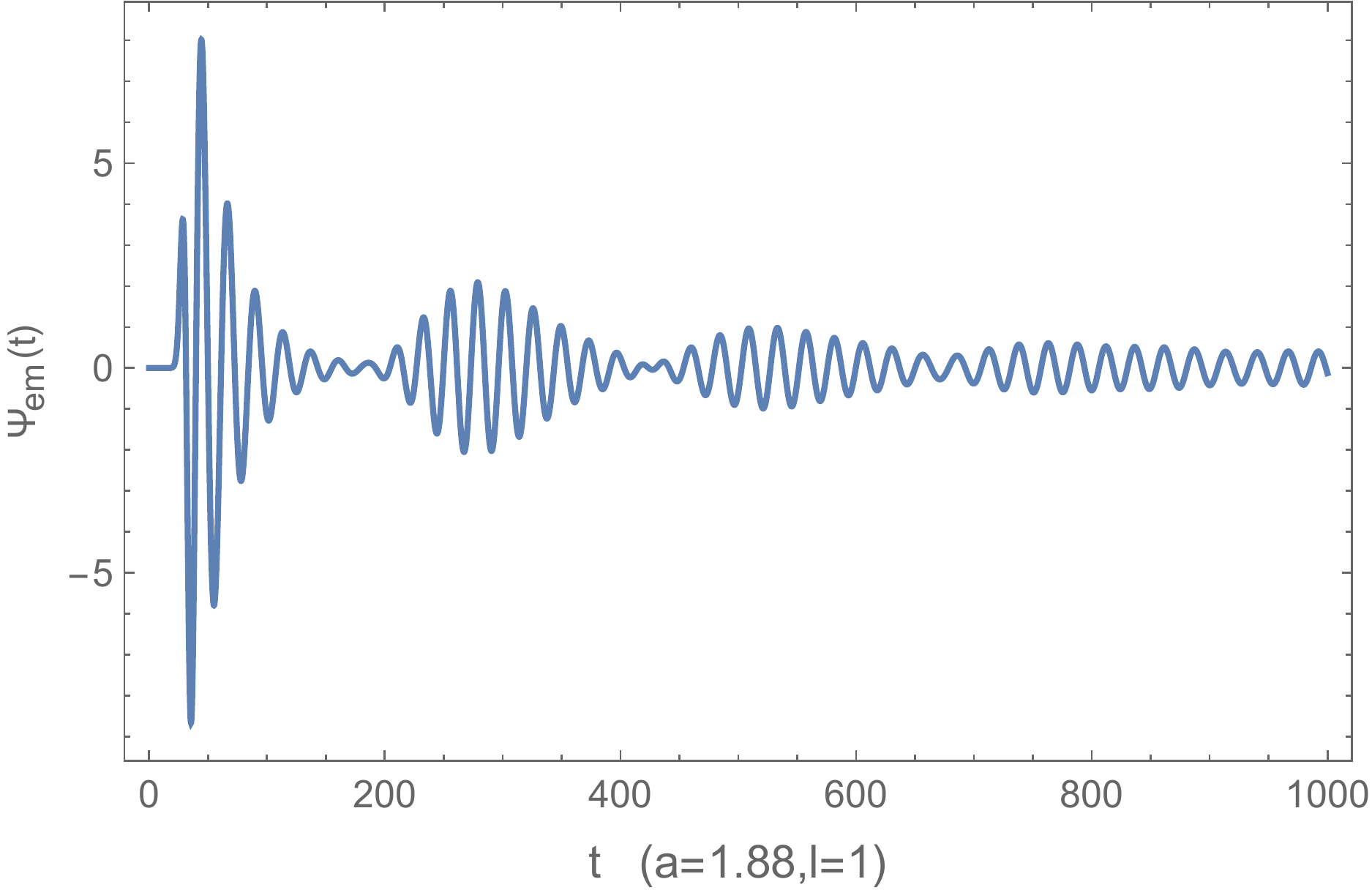}
\includegraphics[height=2in,width=3in]{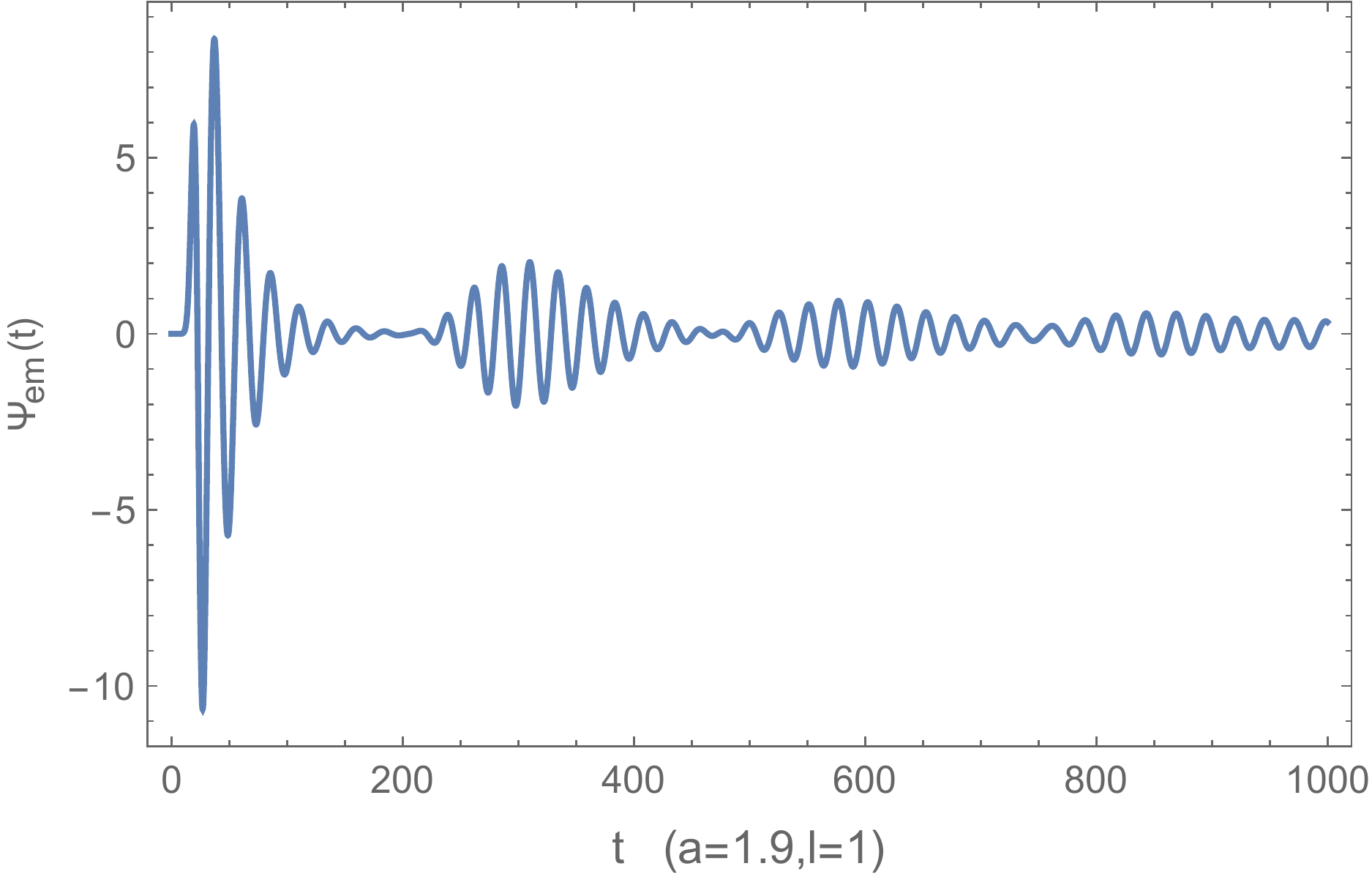}
\includegraphics[height=2in,width=3in]{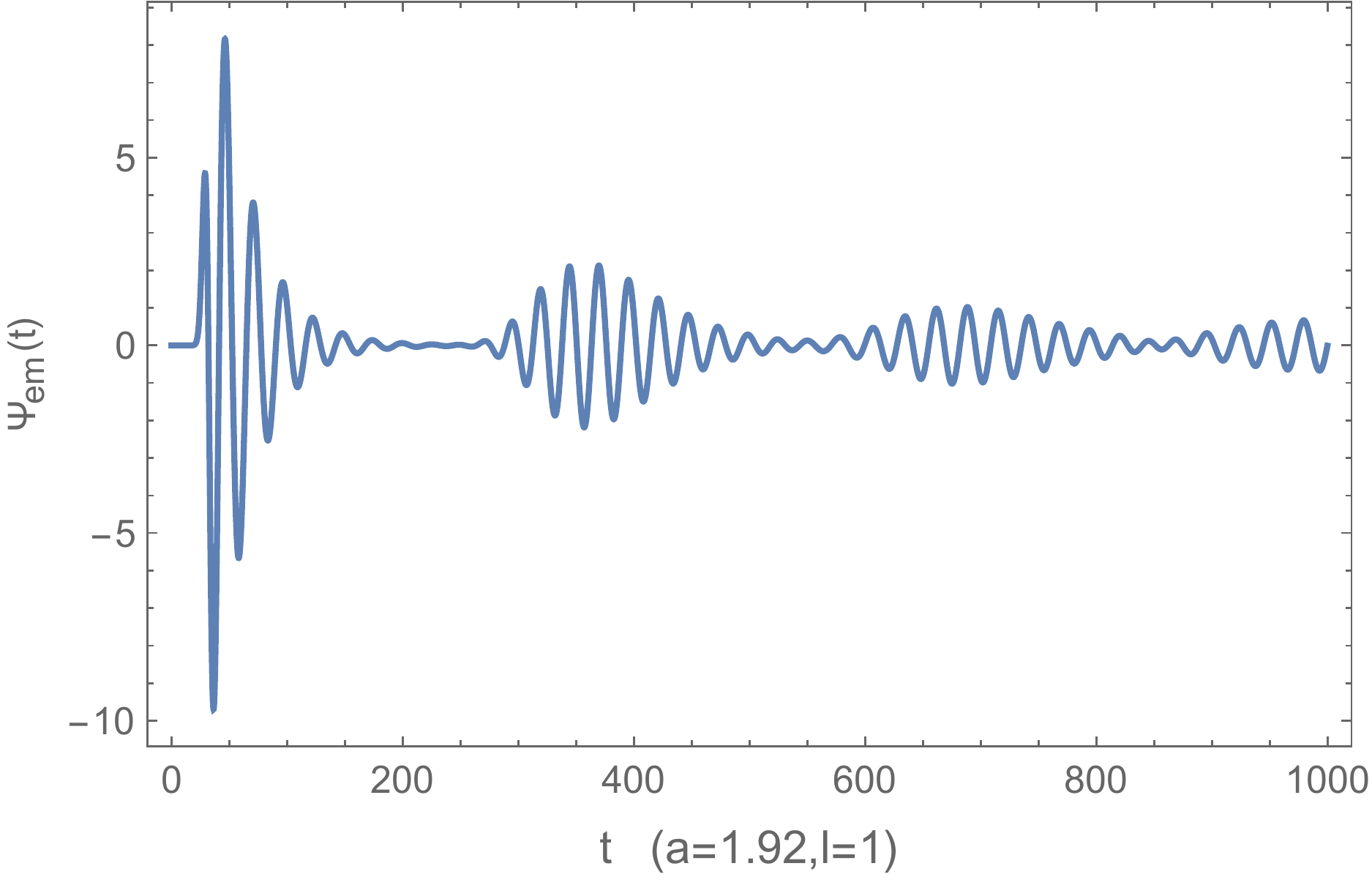}
\caption{The effective potential  in coordinate $r_\ast$ for different parameter $a$ and the time-evolution  of electromagnetic field. We take angular number $l=1$ in this figure.}\label{fig4}
\end{figure}

In this subsection we briefly discuss the properties of echoes from probe electromagnetic field by comparing the similarities and differences with that from scalar field studied in the above subsection. The left column in Fig. \ref{fig5} shows the effective potential of electromagnetic field with different angular number $l$ for fixed $a=1.9$.
We see that as $l$ increases, the potential grows quickly but the width of the potential well is only slightly changed.
This variation characteristic, even the shape and size of the potential, is almost the same as that of the scalar field observed in the left column  in Fig. \ref{fig2}.
The same thing also happens when we fix $l$ and change $a$ (see the top left in Fig. \ref{fig4}).
This observation indicates that the role the last term in the effective potential of the scalar field $V_s(r)$ playing is almost negligible,
though we cannot still give a well interpretation.

Then we present the time-domain profile of electromagnetic field with different angular number $l$ in the right column in Fig. \ref{fig5}.
Just as expected, the main characteristics of the time-evolution of electromagnetic field are almost same as that of the scala field exhibited in the above section.
The signal of echo can be clearly observed. The higher angular number $l$ results in echoes signals with smaller amplitudes.
As the angular number $l$ is turned up, the amplitudes of the echoes are dramatically decreased
and the waves oscillate more rapidly. All these are the universal properties of the time-evolution of the test field affected by the angular number $l$.

Also, Fig. \ref{fig4} exhibits the time-evolution profile of electromagnetic field with different parameter $a$.
We can clearly see that a longer time delay between echoes for a bigger $a$ when it approaches $2$ from below, which is the same as that of the scalar field.
Therefore, the time delay of the echoes from electromagnetic field also becomes longer as the equation of state $\omega$ increases from the phantom state to approach $\omega\sim -1$.

\section{Echoes signals in Wormhole Model  \uppercase\expandafter{\romannumeral 2}}
\label{section4}

\begin{figure}
\centering
\includegraphics[height=2.in,width=3.in]{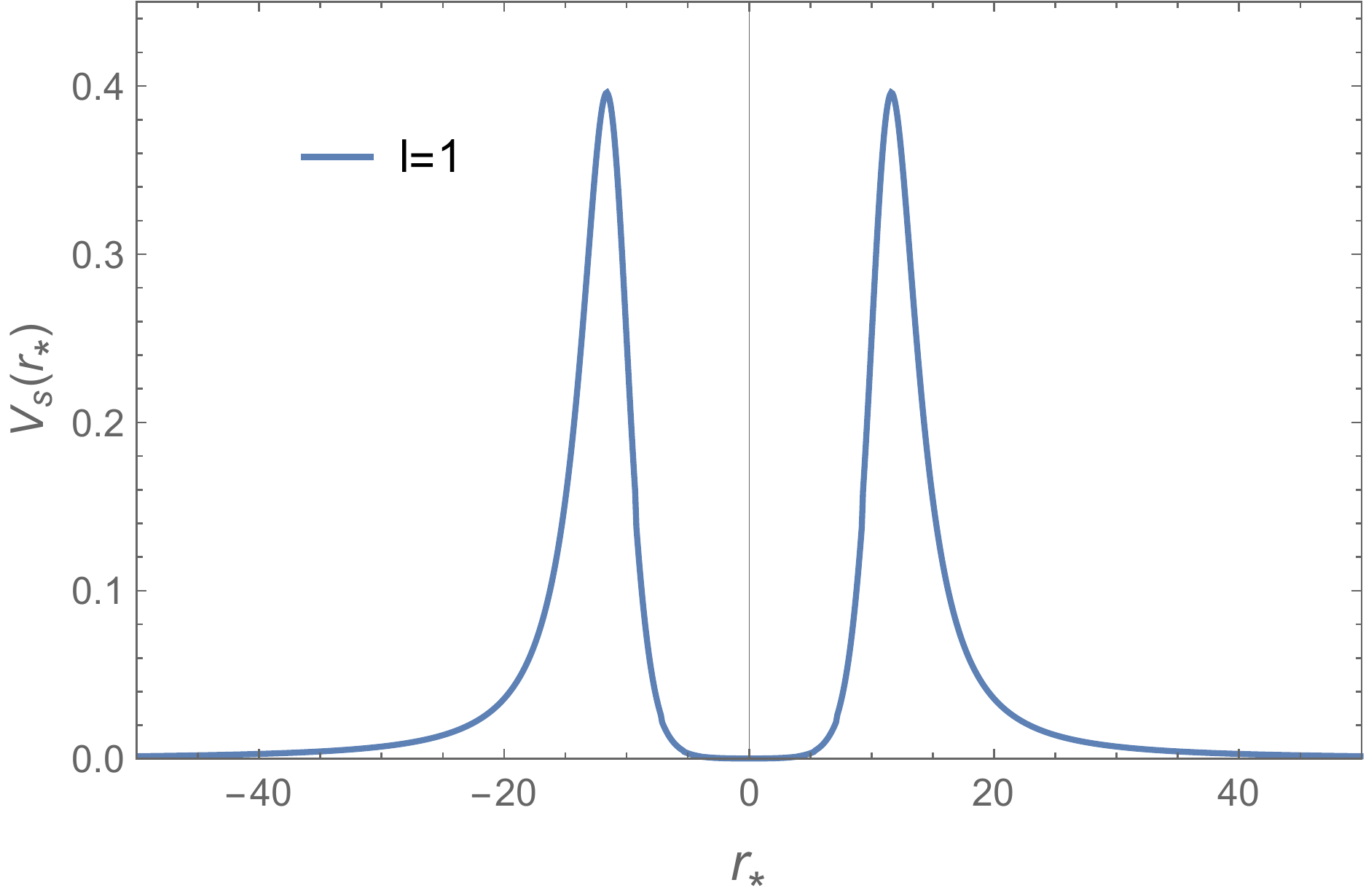}~~
\includegraphics[height=2.in,width=3.in]{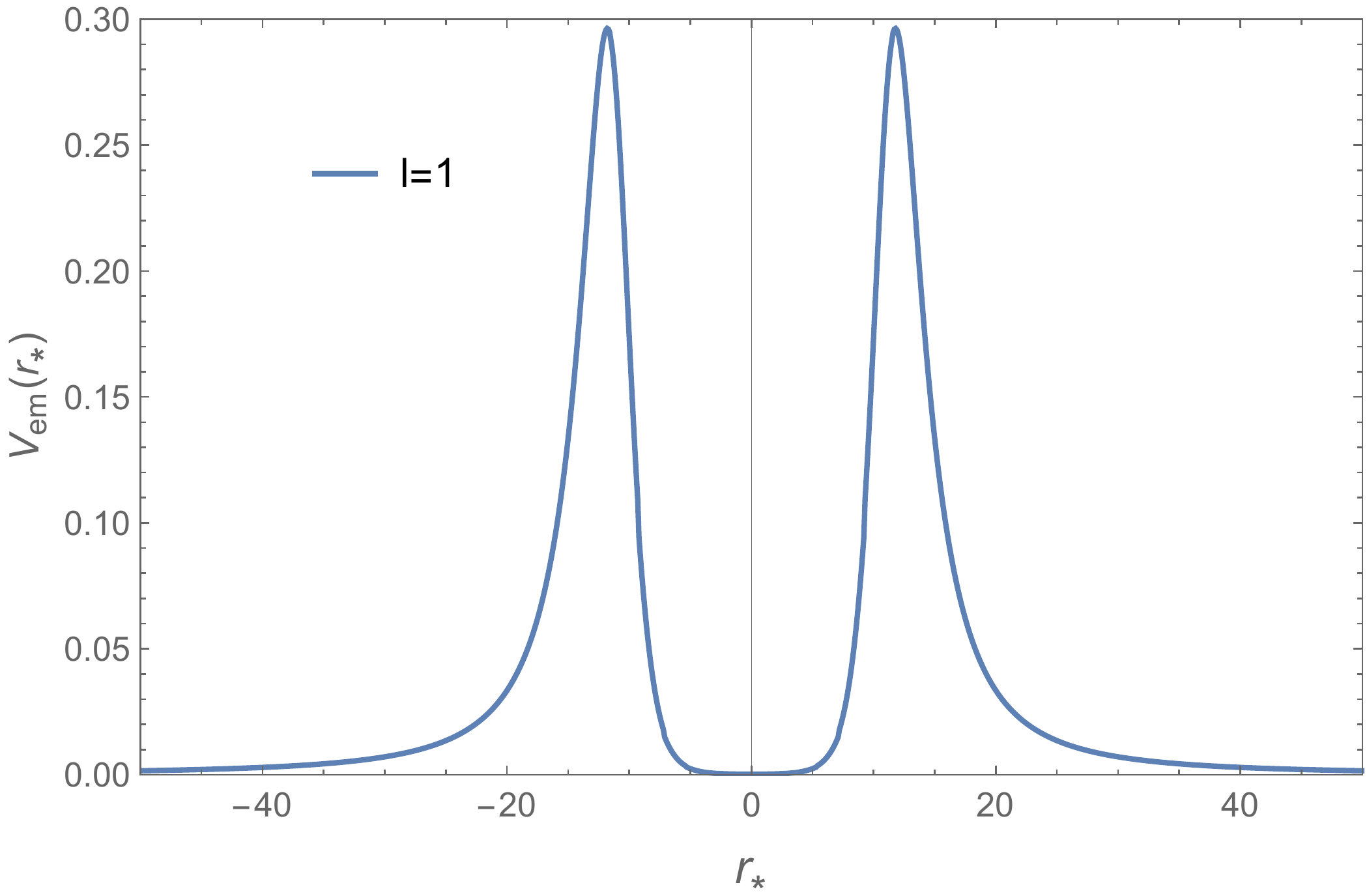}
\includegraphics[height=2.in,width=3.in]{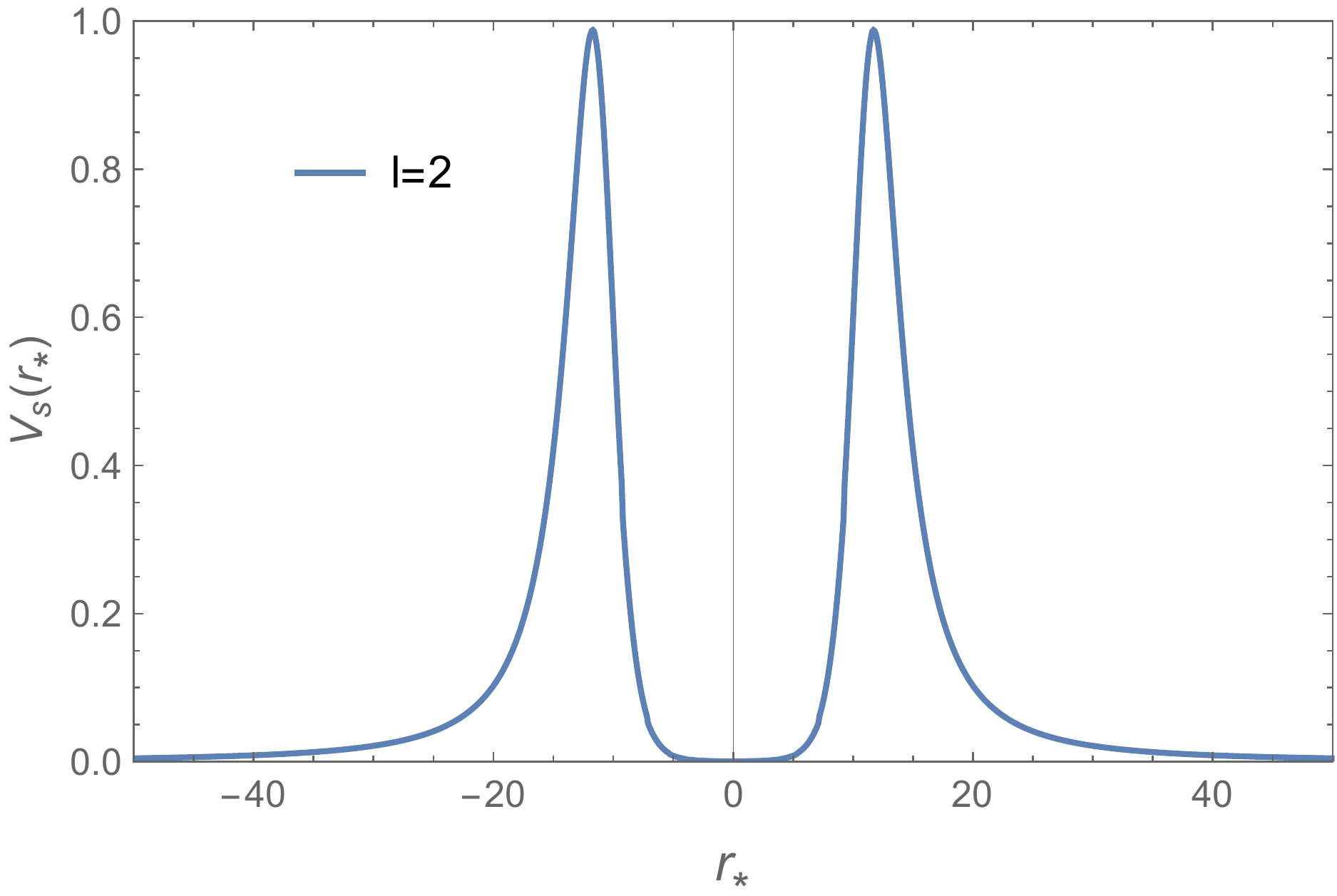}
\includegraphics[height=2.in,width=3.in]{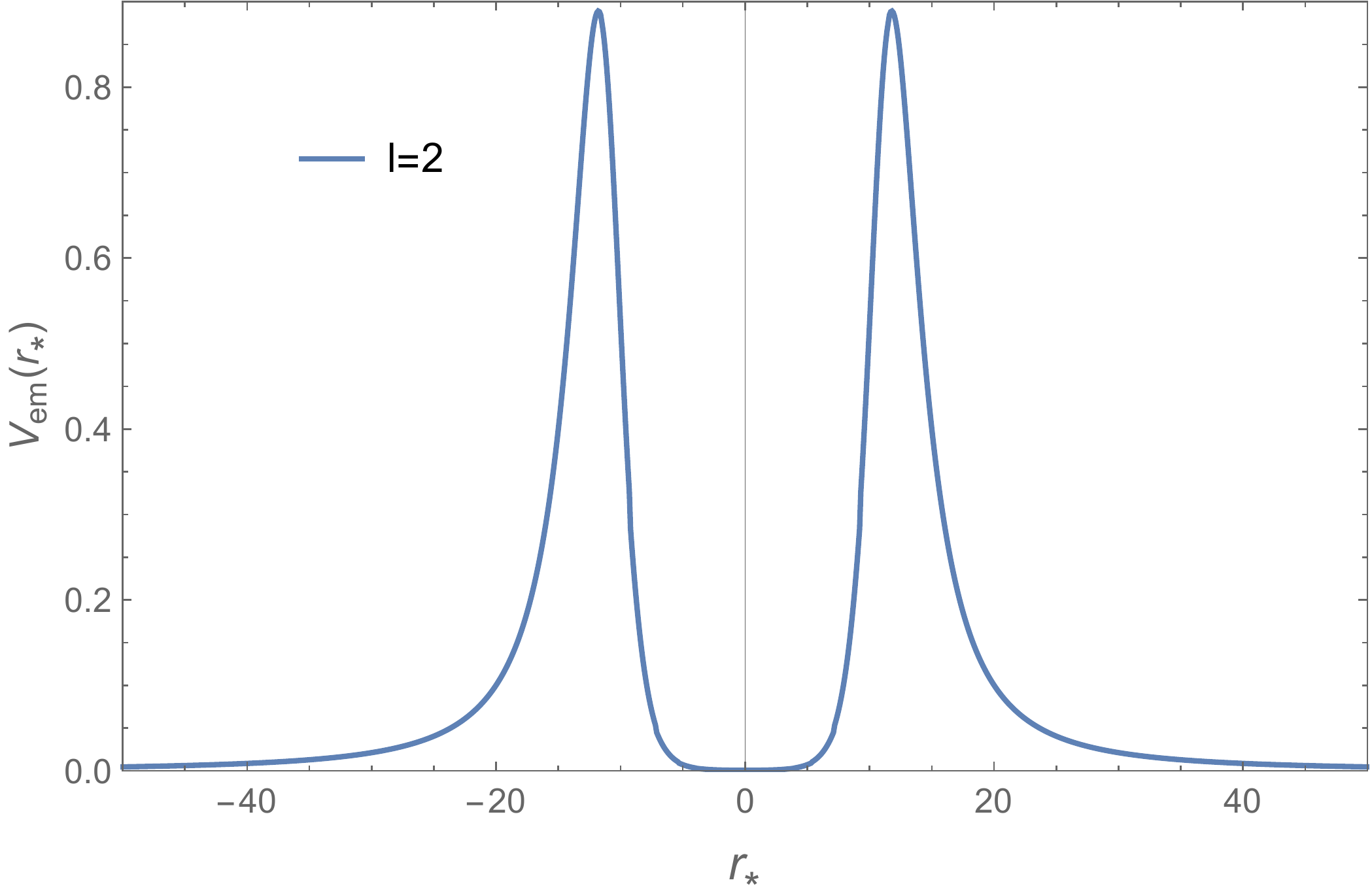}
\includegraphics[height=2.in,width=3.in]{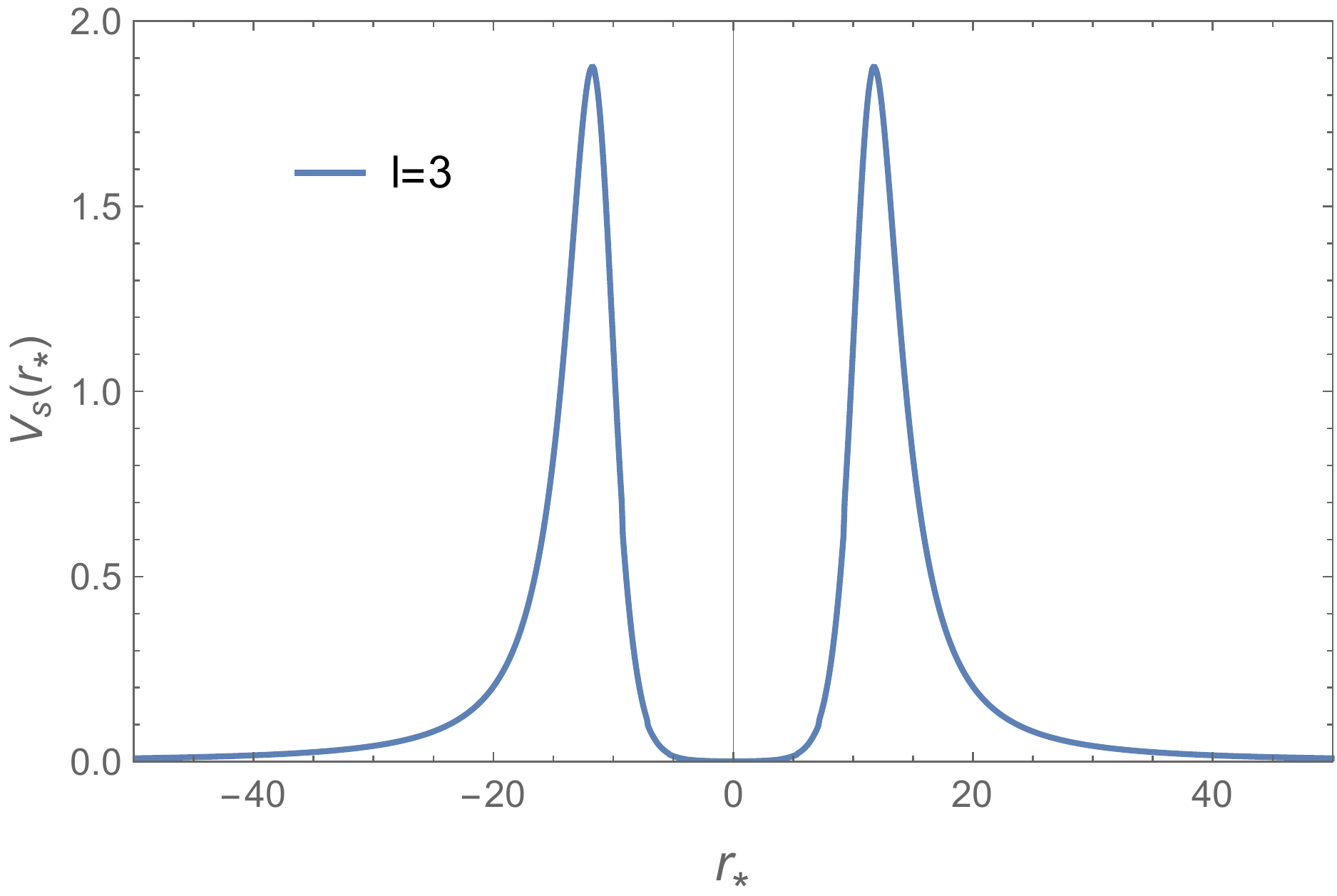}
\includegraphics[height=2.in,width=3.in]{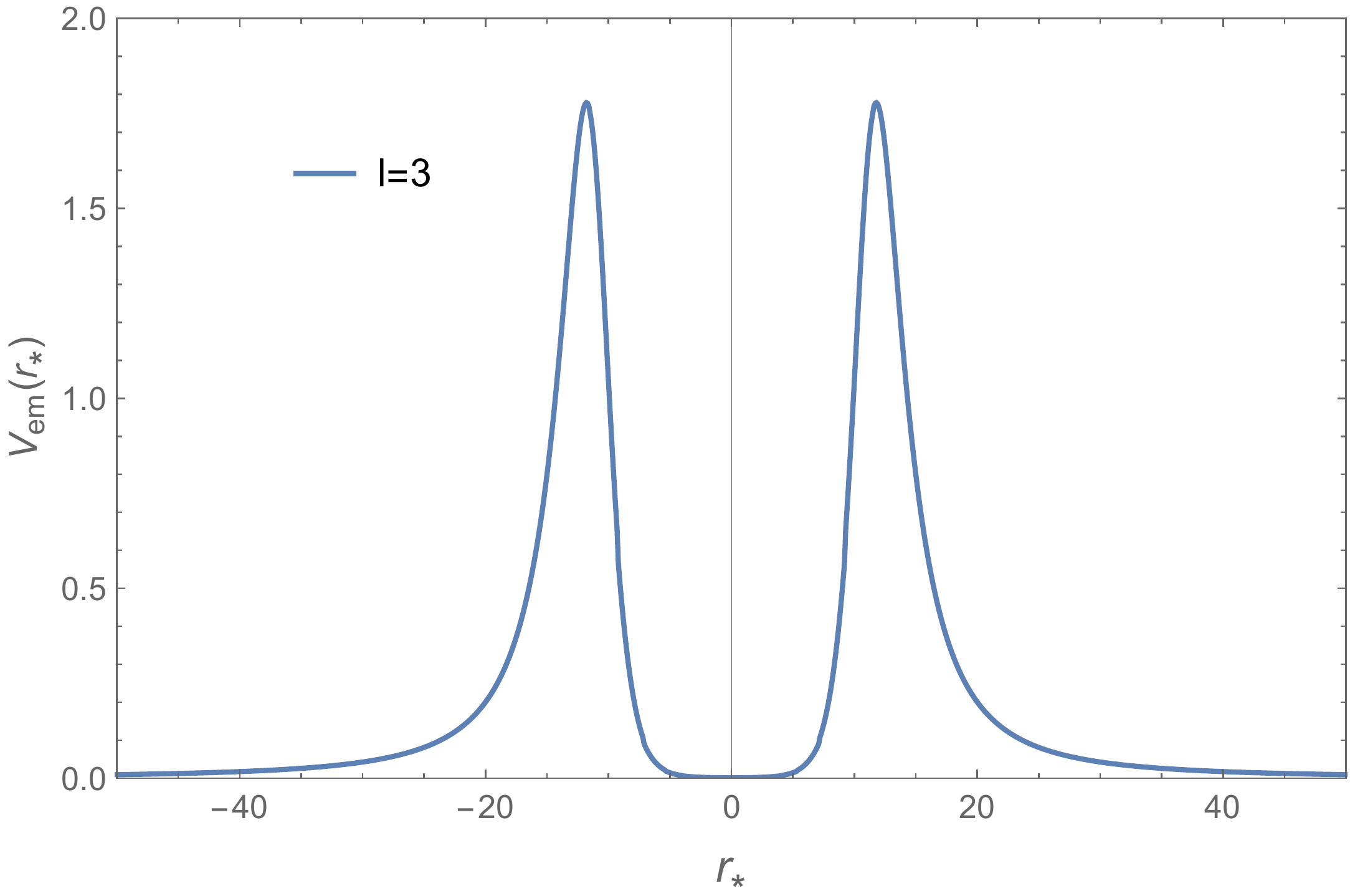}
\includegraphics[height=2.in,width=3.in]{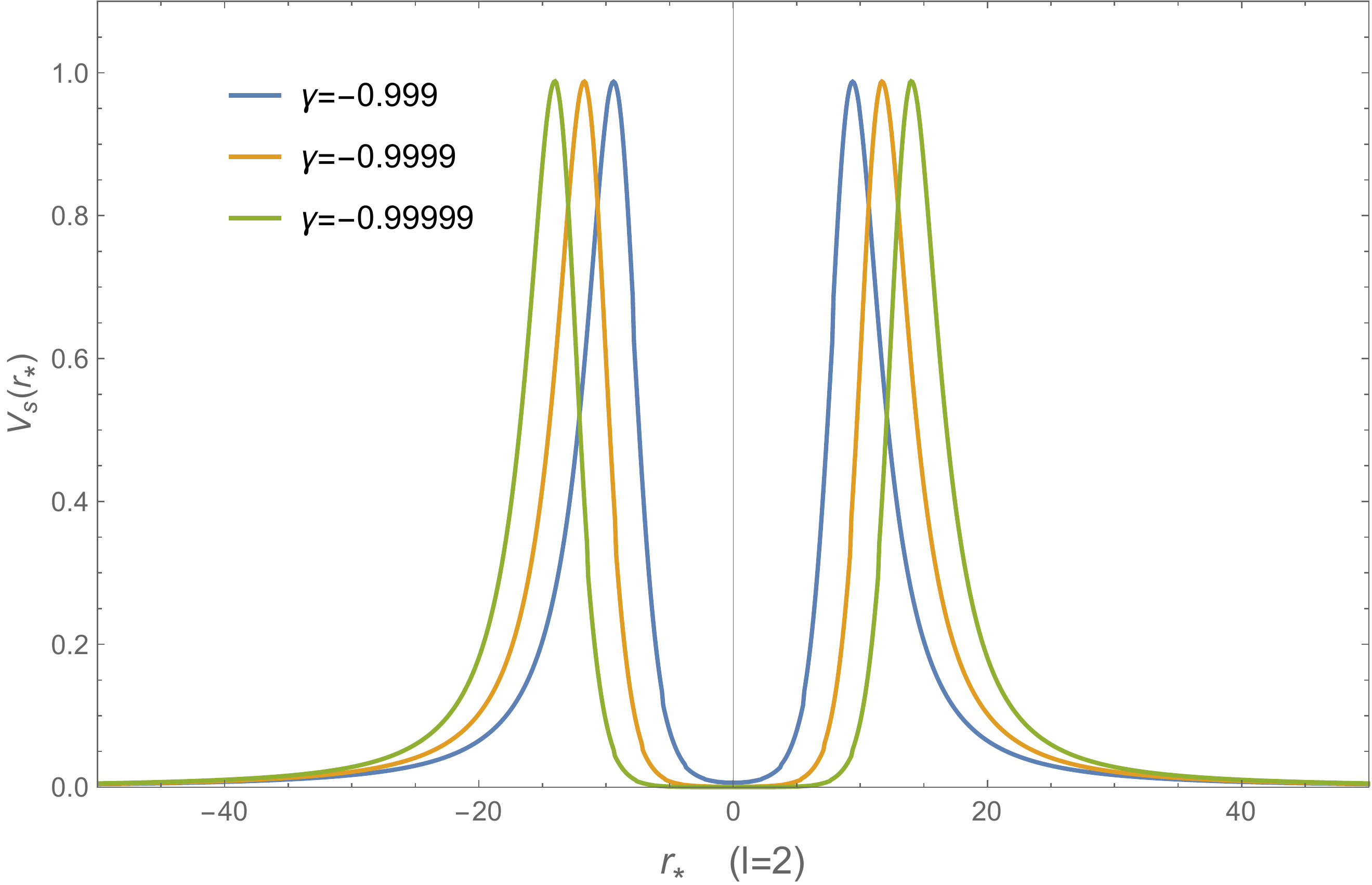}
\includegraphics[height=2in,width=3in]{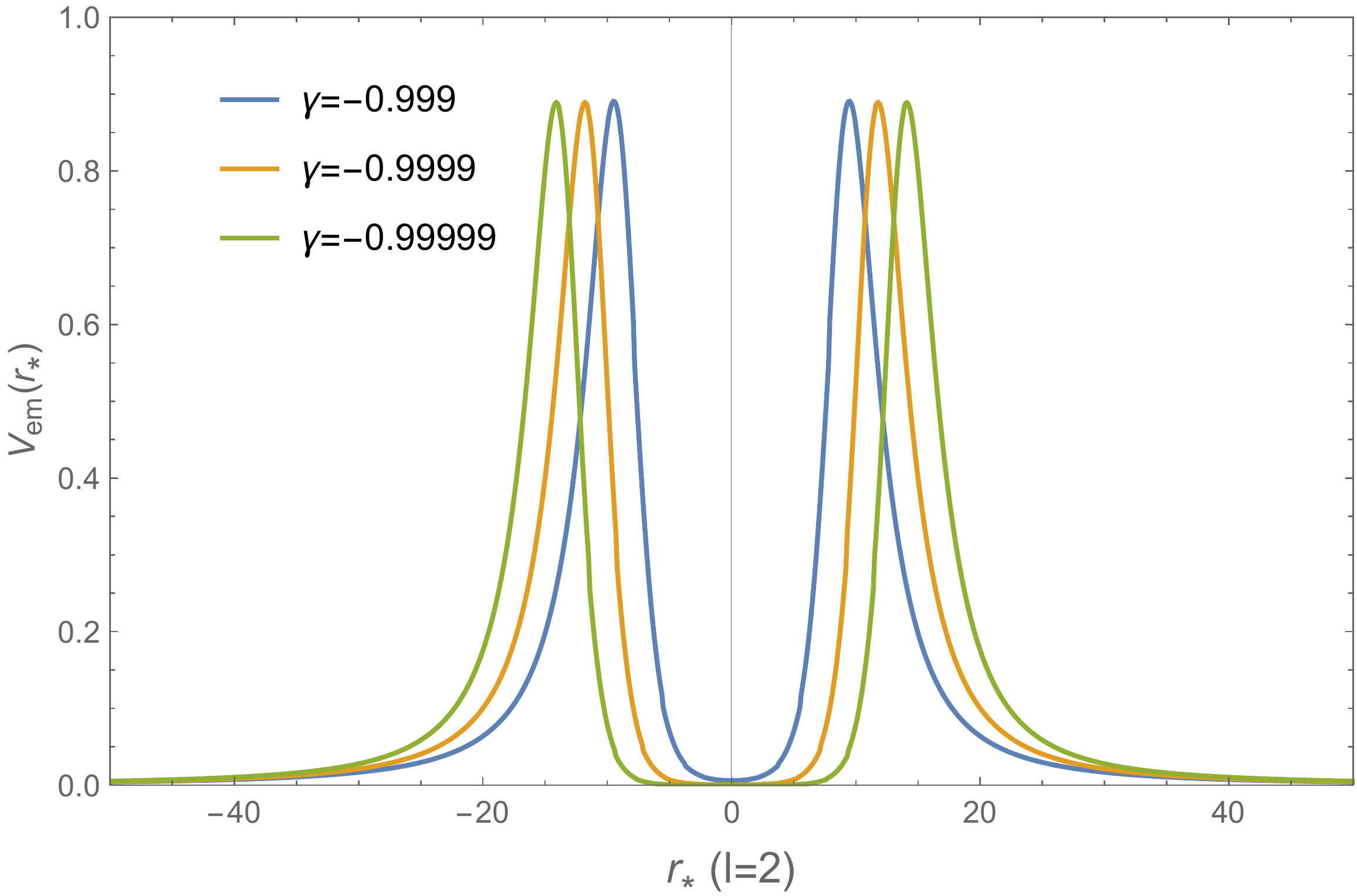}
\caption{The effective potential in coordinate $r_\ast$ for model II (left column is for the scalar field and right column is for the electromagnetic field).
Here we take $\gamma=-0.9999$ and $\omega=-2$.}\label{fig7}
\end{figure}

As revealed in the above section, there is only very slightly difference between the potentials of scalar field and electromagnetic field.
The difference of  the effective potential formula for scalar and electromagnetic field is that an additional term appears for scalar field as shown in  Eq. \eqref{eqsc} and Eq. \eqref{eqem}, and this additional term only contributes little to the potential in wormhole model I studied above.
The almost same potentials also result in the almost the same properties of the signal of echoes.
Before starting this section, we also show the potentials of scalar field and electromagnetic field over Wormhole Model  \uppercase\expandafter{\romannumeral 2}
in Fig. \ref{fig7}. From this figure, we clearly see that the characteristics of effective potential of electromagnetic field are also just slightly different from that of scalar field in this wormhole model
and so we infer that the signals of echo from scalar field and electromagnetic field should be similar.
Based on this point, we only study the time-evolution of scalar field in the configuration of  Wormhole Model  \uppercase\expandafter{\romannumeral 2} in this section.

\begin{figure}
\center{
\includegraphics[scale=0.2]{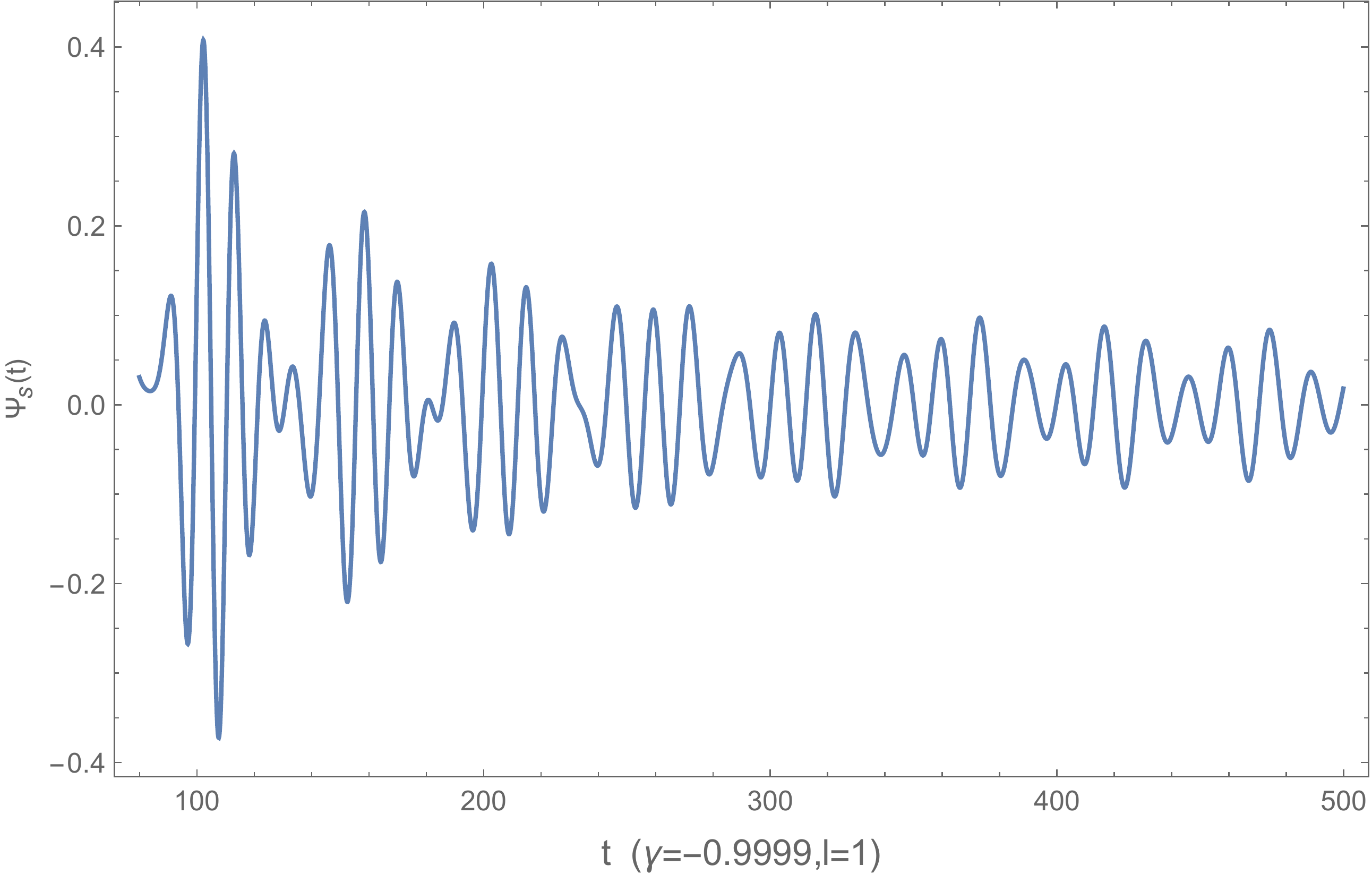}\ \hspace{0.1cm}
\includegraphics[scale=0.2]{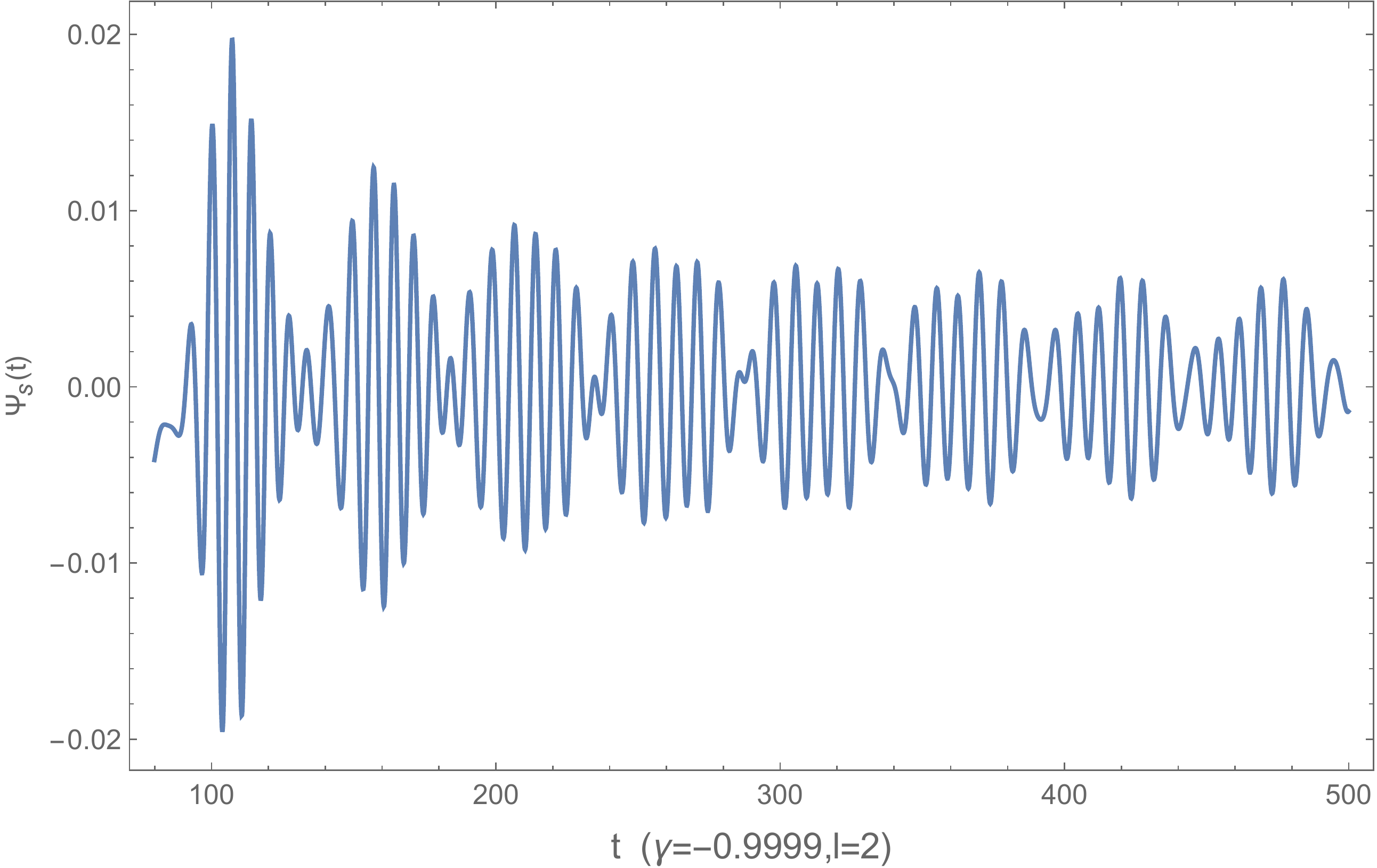}\ \hspace{0.1cm}
\includegraphics[scale=0.2]{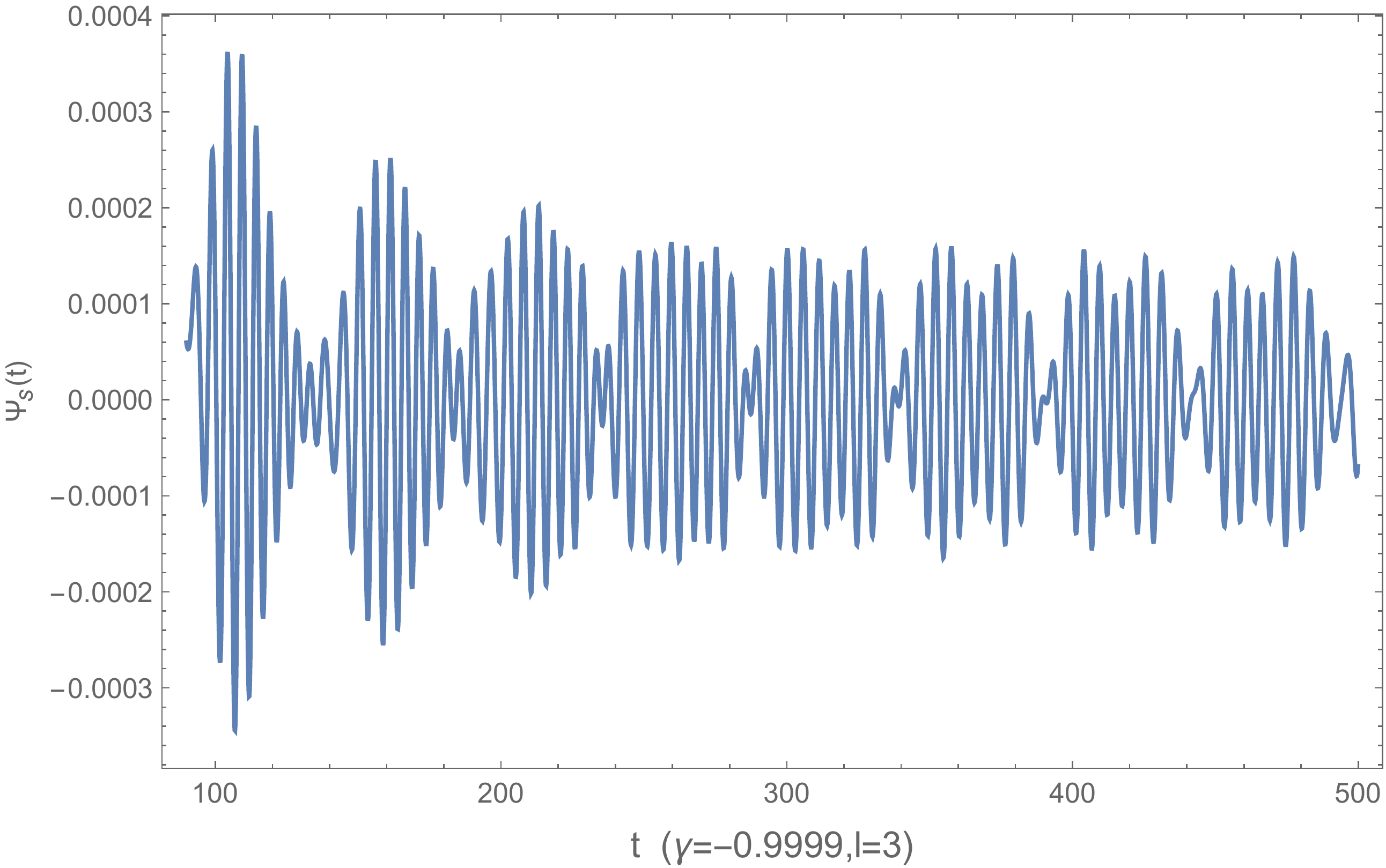}\\
\caption{\label{fig71} The time-evolution of scalar field for different angular number $l$. Here we we take $\gamma=-0.9999$ and $\omega=-2$.
}}
\end{figure}

As we have illuminated above, we shall set $r_0=1$ and $\alpha=1$ and leave the parameters $\gamma$ and $\omega$ free, such that we mainly focus on the effect of the two parameters on the echoes.
Note that to make sure the existence of the wormhole and the potential well, we restrict $\gamma$ to be the region of $\gamma>-1$ but very close to $-1$. Since the state parameter $\omega$ is not involved in the redshift function $A(r)$, we can take any values of $\omega$ in the region $\omega<-1$, which is contrary to the situation in Model I where we must take $\omega\lesssim-1$ to ensure echoes produced.

We also would like to exhibit the properties of the time-evolution profile of scalar field over model II for different angular number $l$ (see Fig. \ref{fig71}).
From this figure, we again confirm that the effects of the angular number $l$ on the signals of echo studied in model I, i.e.,
\begin{itemize}
  \item Higher angular number $l$ leads to echo signals with smaller amplitudes.
  \item The time delay between two successive echoes depend very weakly on the angular number $l$.
  \item The waves with higher $l$  oscillate more rapidly.
\end{itemize}
Therefore, we conclude that the above properties of the dependence of the echoes on the angular number $l$ are universal.
In addition, we note that the amplitude of echo is usually smaller for model II than that for model I. It can attribute to a deeper potential well for model II, for which it is more difficult for the trapped waves to escape.

Next, we turn to study the effect of the parameter $\gamma$ of the phantom wormhole.
The effective potential and the time evolution of scalar field for different $\gamma$ are exhibited in the bottom of  Fig. \ref{fig7}.
It is shown that the maximum of the potential is almost not changed under different $\gamma$, but the width of the potential well is apparently extended with the increase of $\gamma$.
Correspondingly, we find in Fig.\ref{fig71} that the amplitudes of the scalar wave for different $\gamma$ almost remain unchanged, but the time delay of echo signals becomes longer with larger $|\gamma|$.
What's more, we would like to point out that the oscillation frequencies of echoes for different $\gamma$ are in the same order of magnitude. This observation can also attributed to the height of the potential,
which is almost unchanged for different $\gamma$.
\begin{figure}
\center{
\includegraphics[scale=0.185]{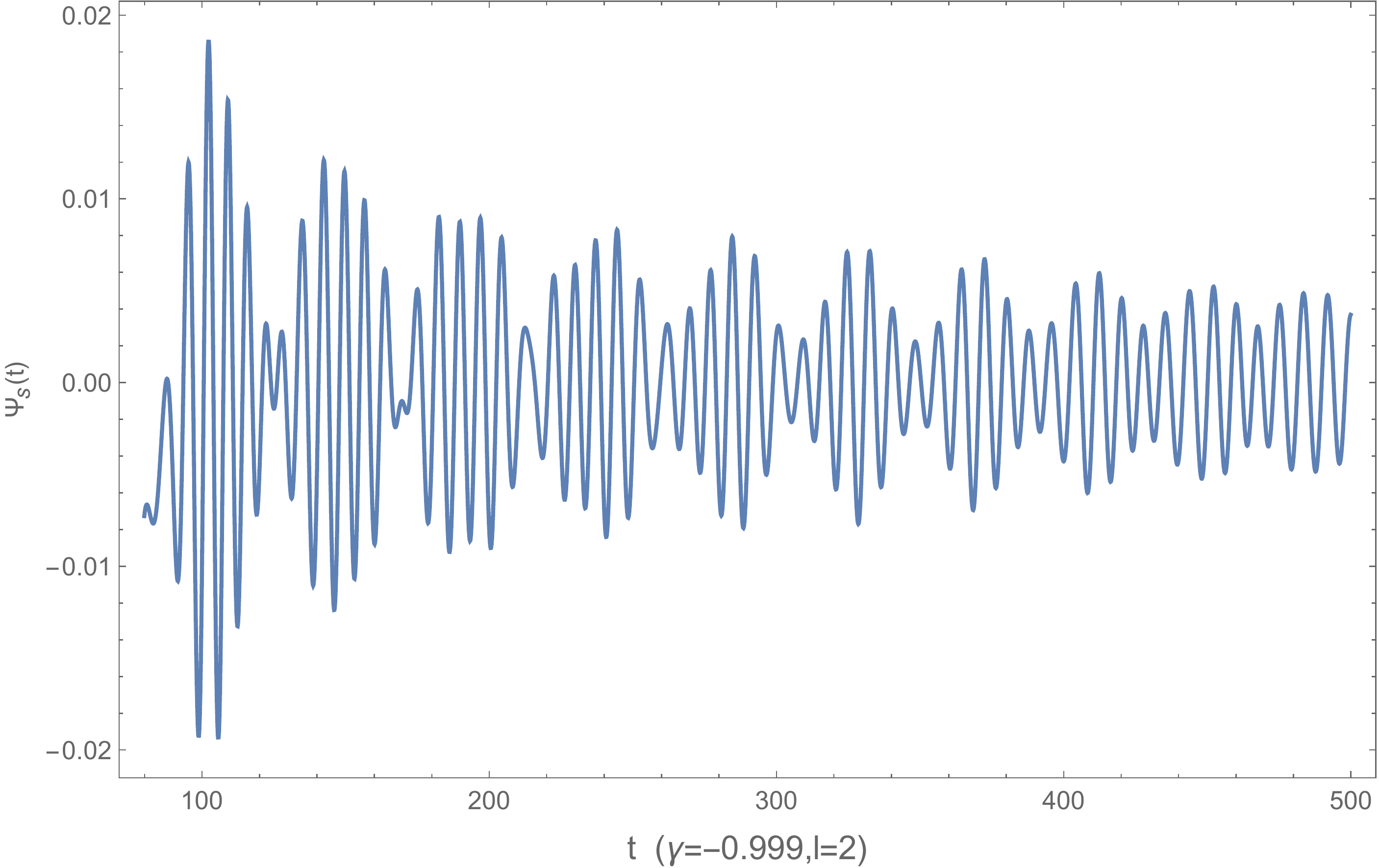}\ \hspace{0.1cm}
\includegraphics[scale=0.2]{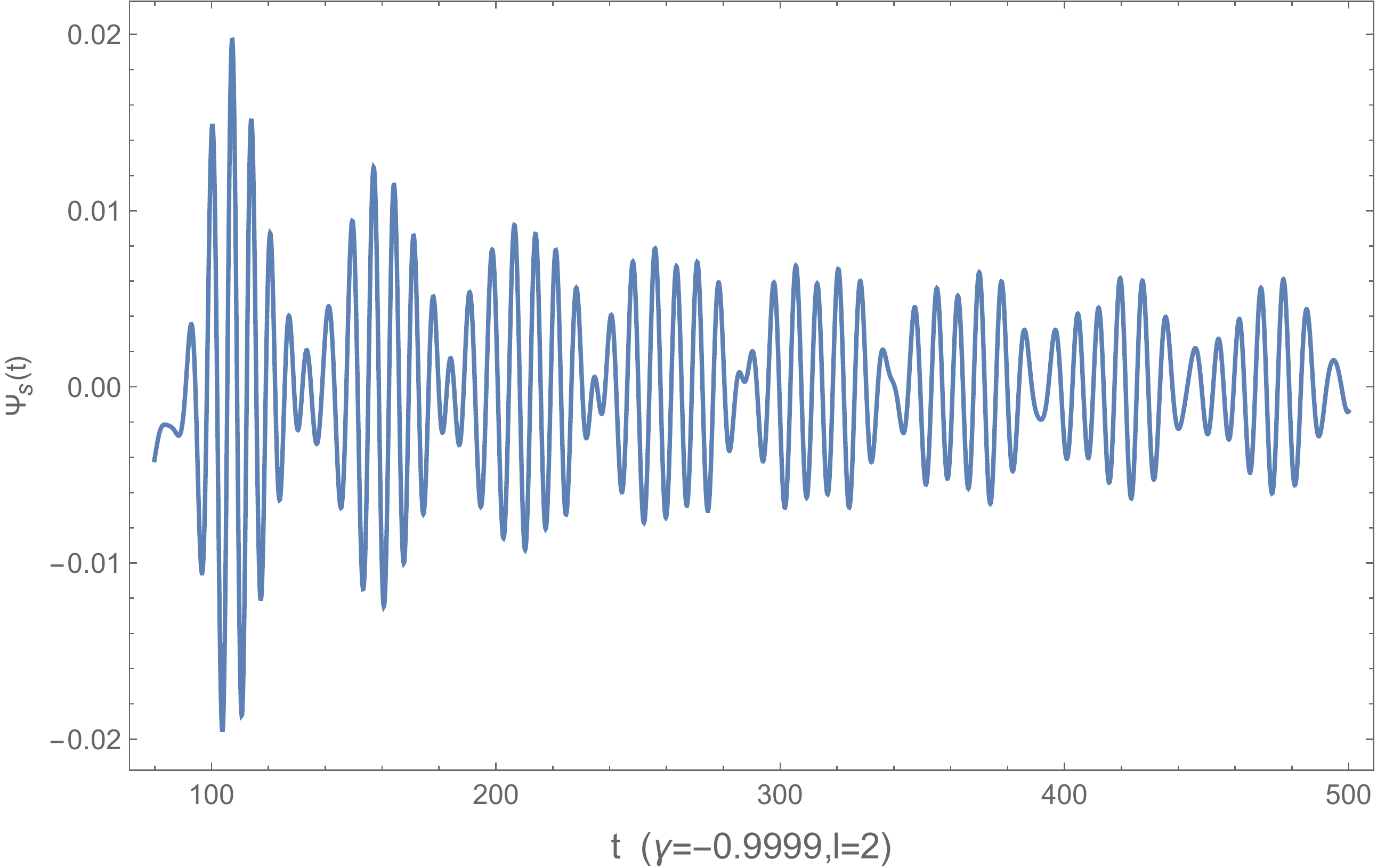}\ \hspace{0.1cm}
\includegraphics[scale=0.2]{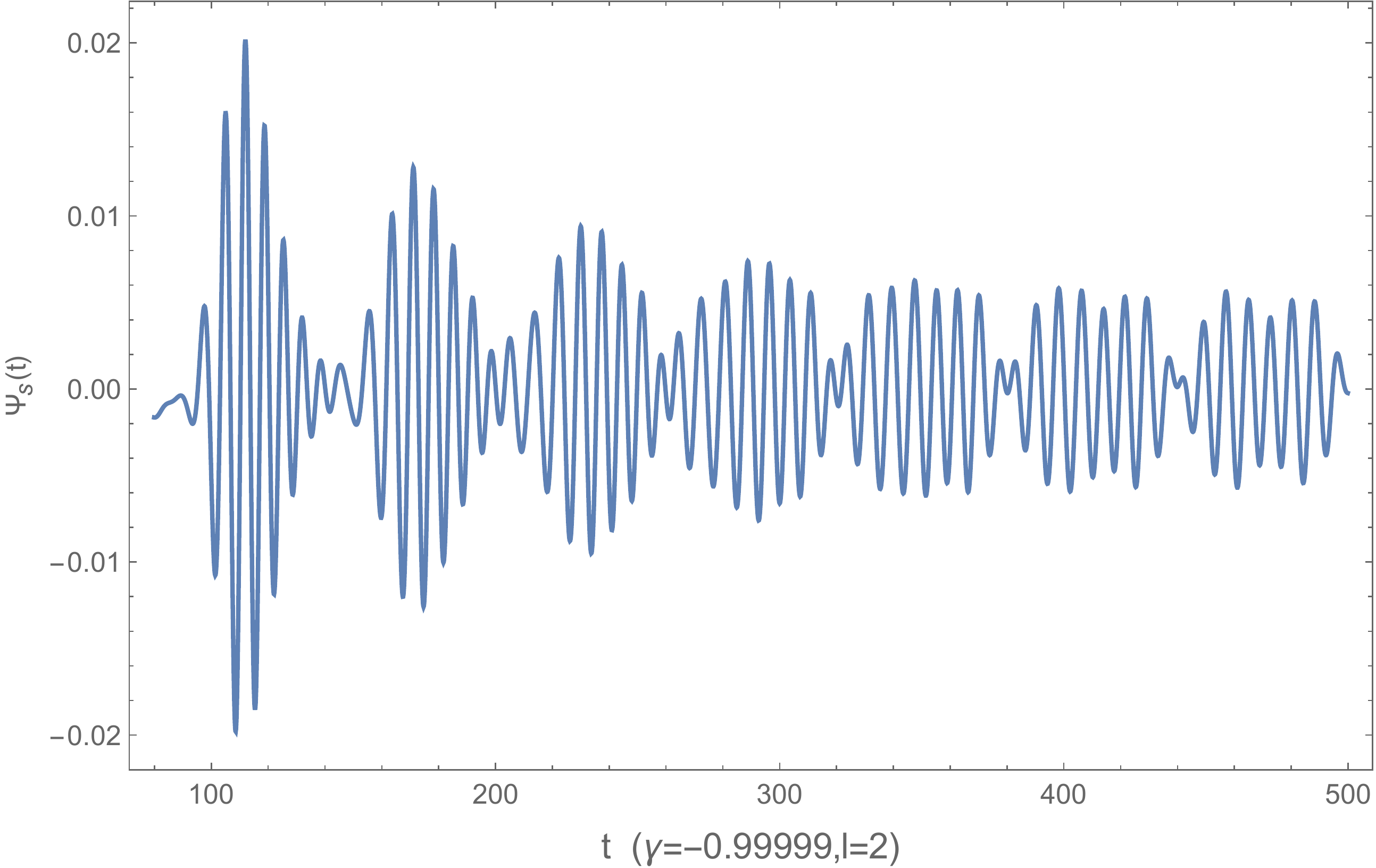}\\
\caption{\label{fig6} The time evolution of scalar field for different parameter $\gamma$. Here we take $\omega=-2$ and $l=2$.
}}
\end{figure}

The state parameter $\omega$ is an important quantity characterized the property of phantom energy. Therefore,
it is necessary to investigate the effects of the state parameter on the property of the evolution of the field perturbations exerted on the wormhole spacetime supported by this phantom matter. As the behaviors of the field perturbations could be affected by different $\omega$, thus it is possible to determine the sate parameter by analysing the characteristics of the perturbations, or echoes as we will demonstrate.
Here, we shall explore carefully the effect of $\omega$ on the signals of echo in model II.
\begin{figure}
\centering
\includegraphics[height=3in,width=4.5in]{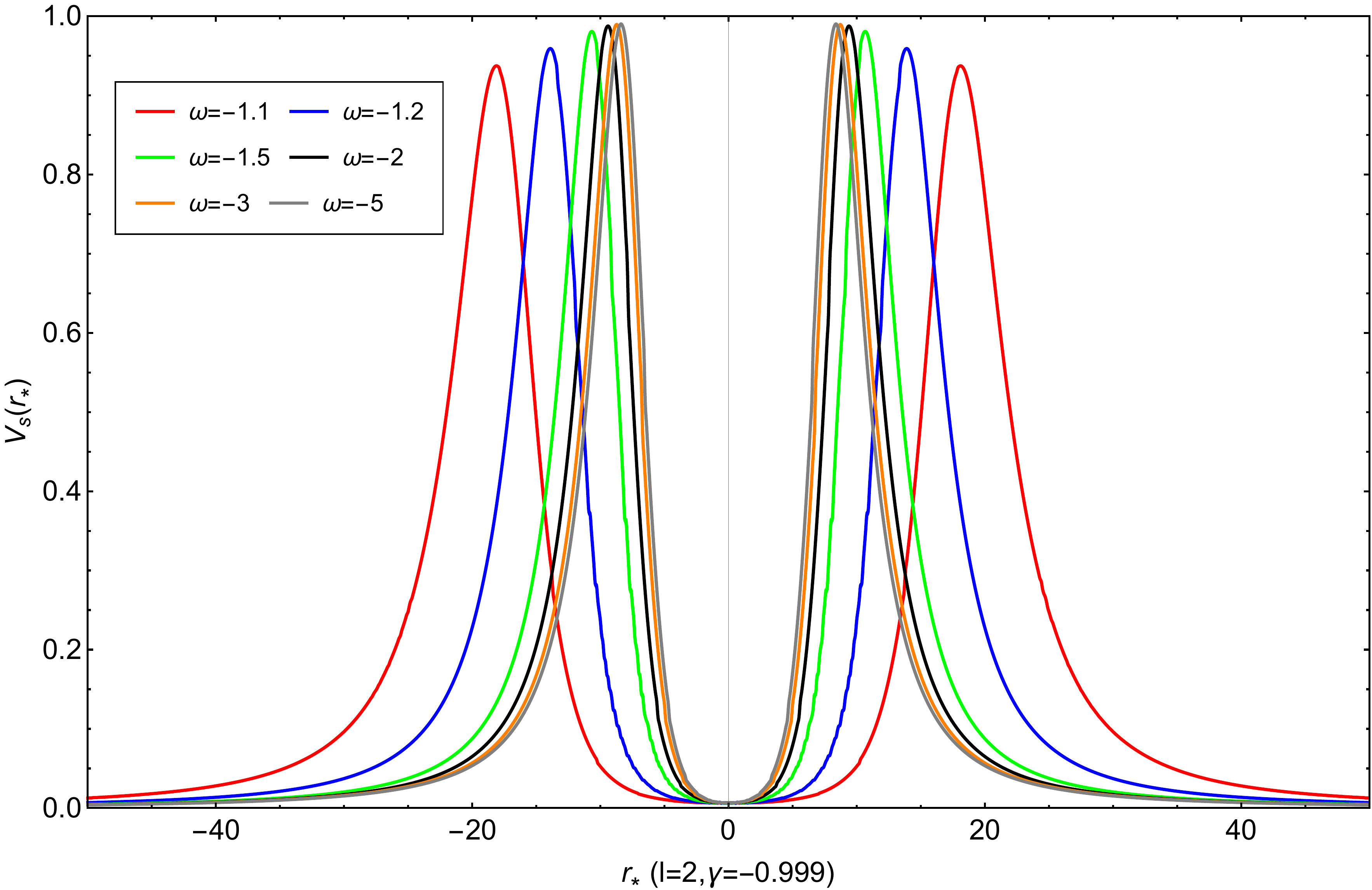}
\caption{The effective potential  of scalar field for different state parameter  $\omega$, and we take  $\gamma=-0.999,l=2$ in this figure.}\label{fig10}
\end{figure}
\begin{figure}
\centering
\includegraphics[height=2.in,width=3.in]{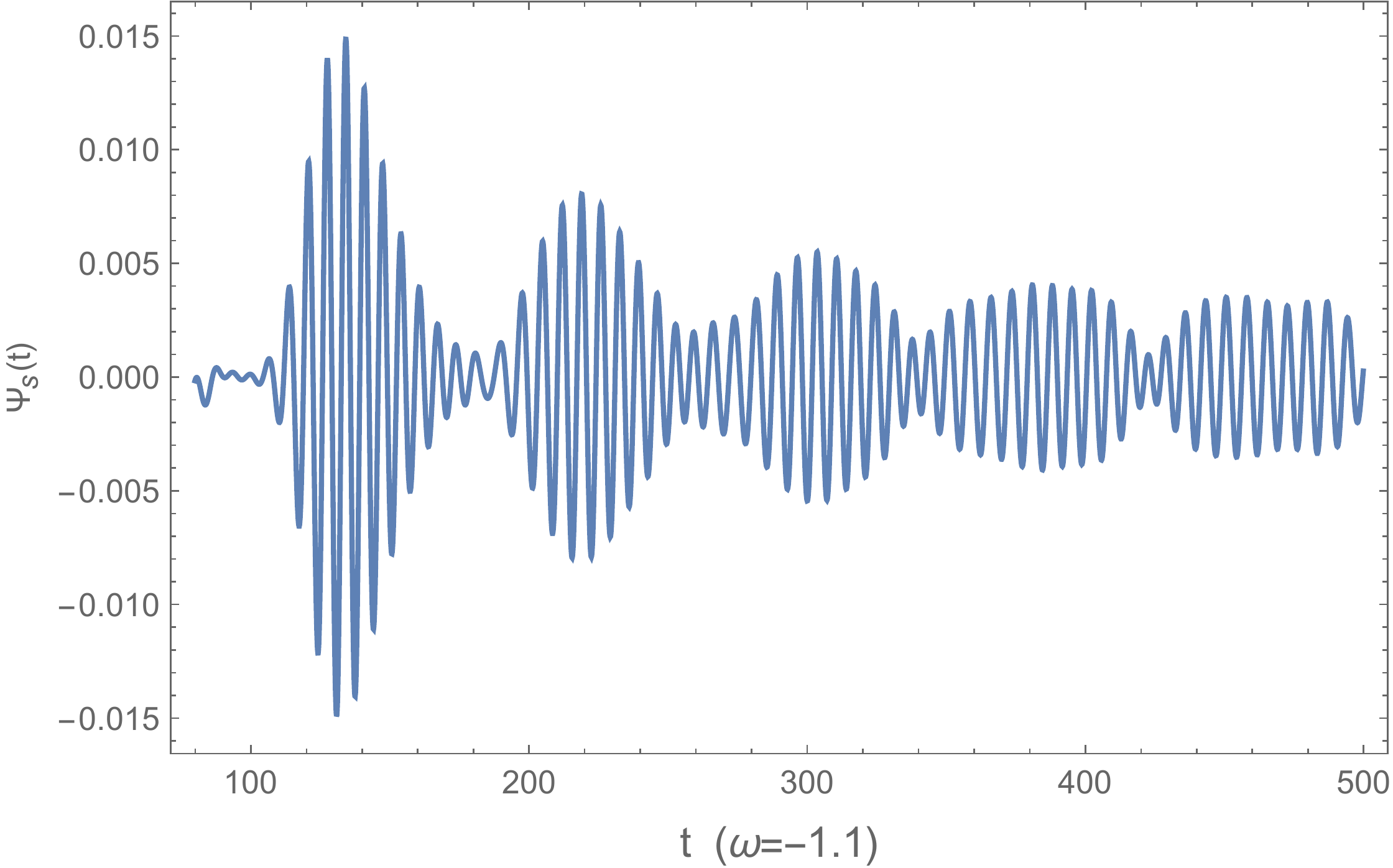}~~
\includegraphics[height=2.in,width=3.in]{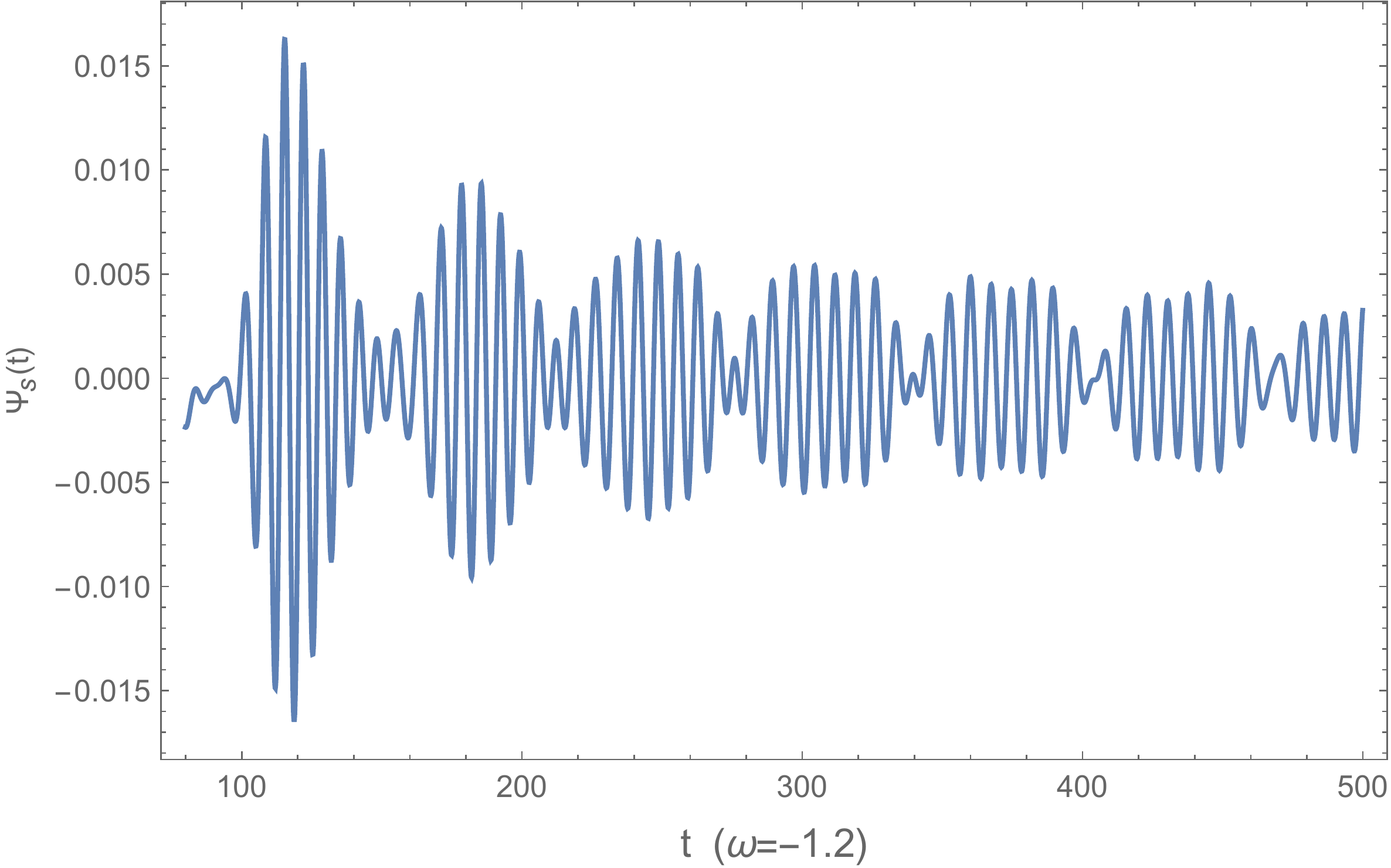}
\includegraphics[height=2.in,width=3.in]{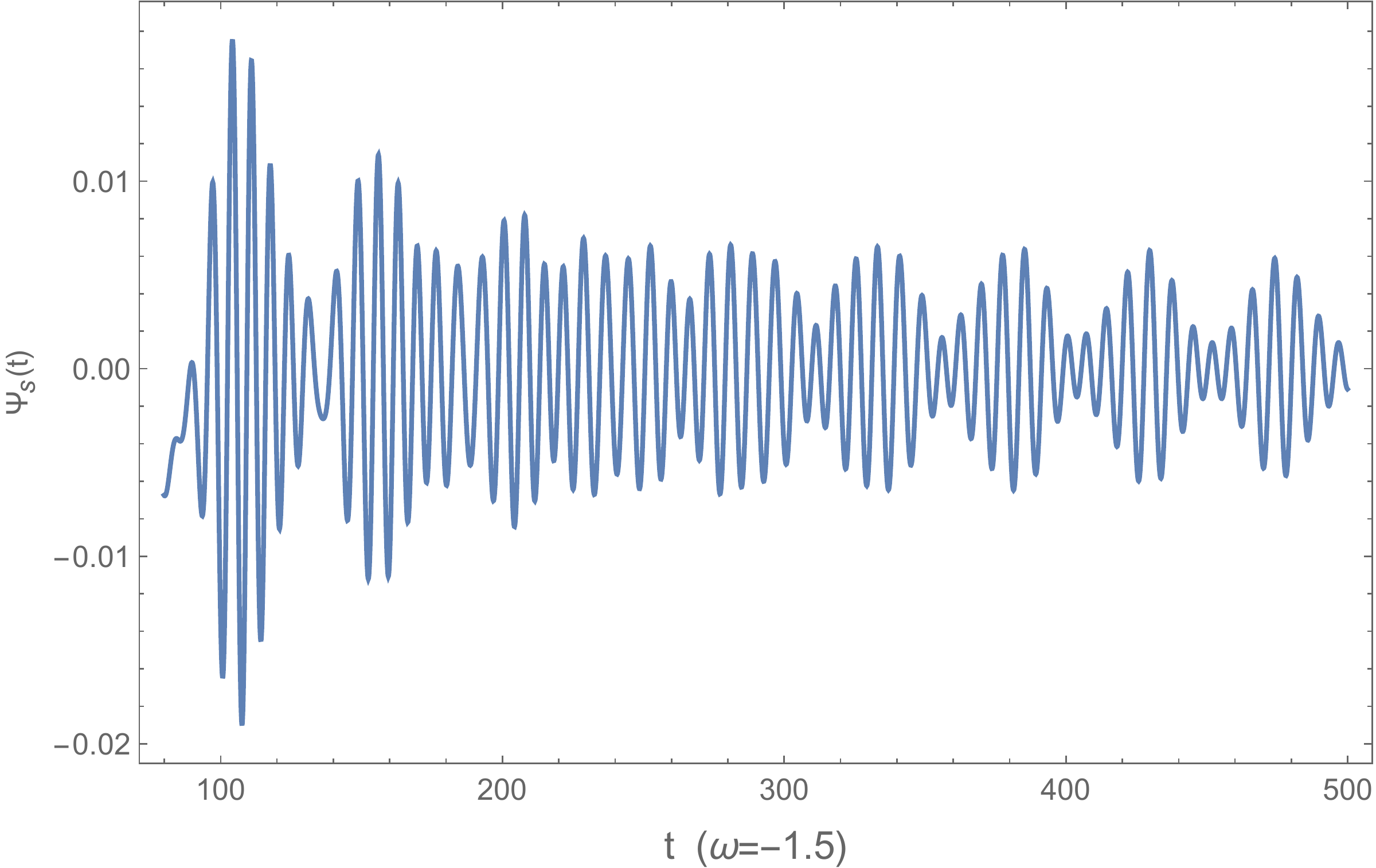}
\includegraphics[height=2.in,width=3.in]{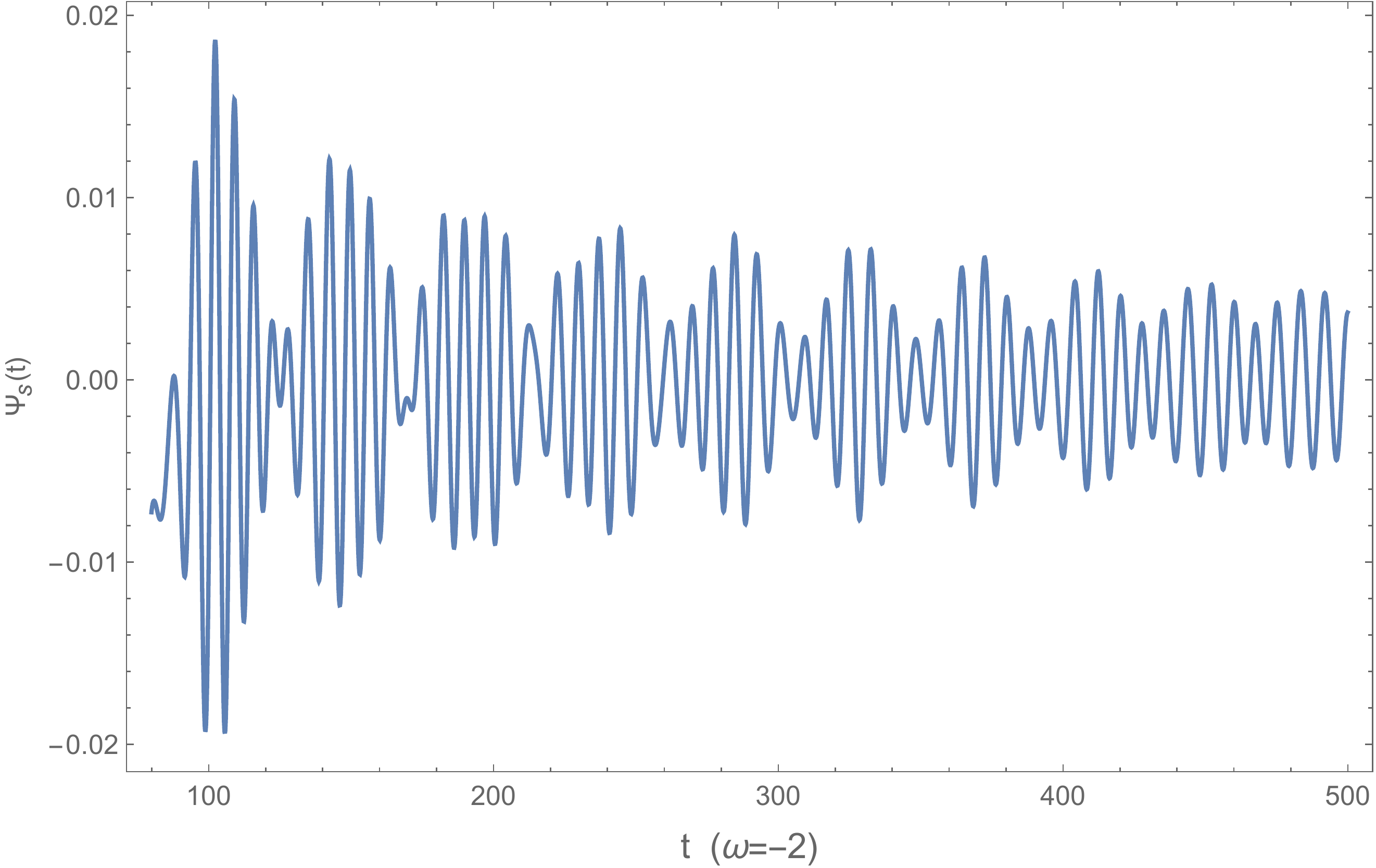}
\includegraphics[height=2.in,width=3.in]{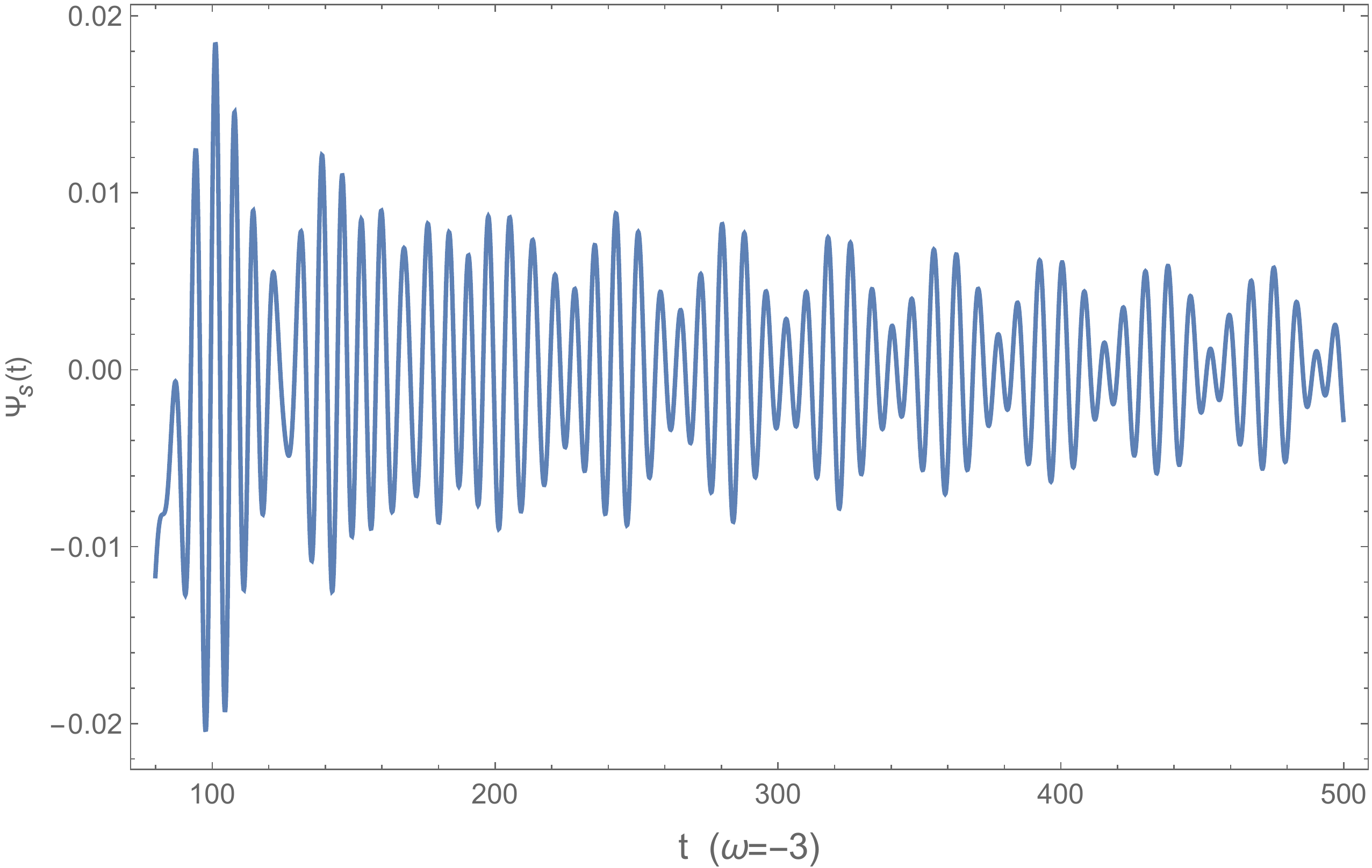}
\includegraphics[height=2.in,width=3.in]{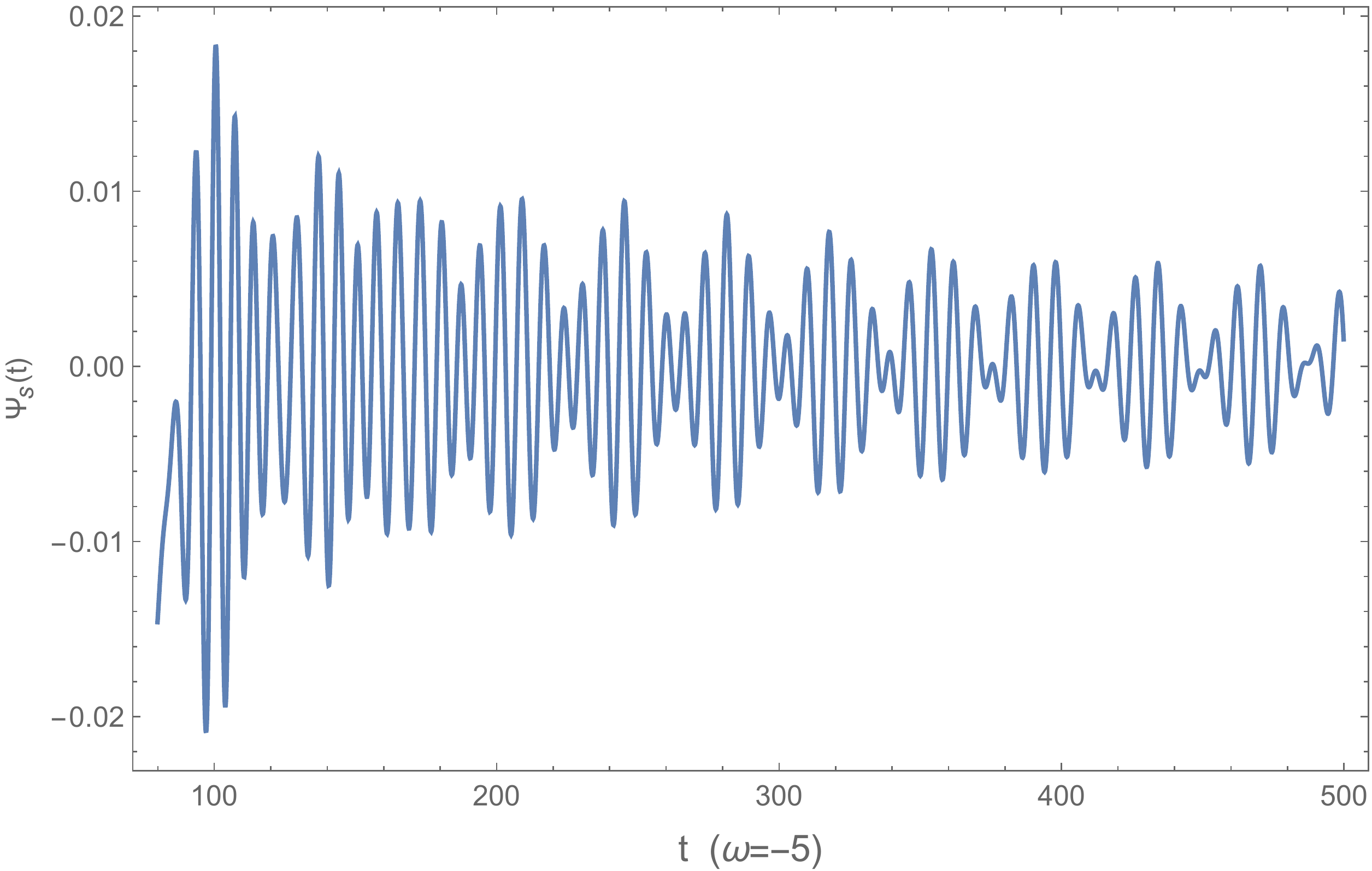}
\caption{The  time-domain profile  of scalar field for different parameter  $\omega$, and we take  $\gamma=-0.999,l=2$ in this figure.}\label{fig11}
\end{figure}

The effective potential and the time-evolution profile of scalar field for different $\omega$ are shown in Fig. \ref{fig10} and Fig. \ref{fig11}, respectively. From Fig. \ref{fig10}, we can see that the height of the potential is only slightly influenced by $\omega$. Correspondingly, the amplitudes of the echo only slightly change as $\omega$ varies (see Fig. \ref{fig11}), which is consistent with the role $\omega$ plays in model I.

Now, we mainly focus on the effect of the state parameter $\omega$ on the delay time of echoes, which is an important testable quantity in future GW observations. First, from Fig. \ref{fig10}, one can see that the width of the potential well is decreased with the increase of $|\omega|$. In particular, as we increase $|\omega|$ from 1 to 2 width of the potential well is quickly shortened. But for $|\omega|>2$, even we further turn up $|\omega|$, the width of the potential well is just slightly changed. As a result, we see that the delay time of echoes becomes longer when $\omega$ tends to $\omega=-1$ from below. As dark energy becomes more phantom, the delay time becomes shorter and shorter. And then for $\omega<-2$, the delay time scale  tends to a constant even when $\omega$ further decreases. Such an obvious characteristic can be easily detected in future GW observations.

\section{Quasinormal Modes of wormholes}
\label{section5}
It would be instructive to compute the quasinormal modes (QNMs) to construct a bridge between the echoes and the
wormhole's modes.
In this section, we only focus on the QNMs frequencies of scalar perturbation for the two wormholes models. For the QNMs of electromagnetic field, some similar conclusions can be found.

Quasinormal ringing belongs to the ringdown  stage of the external perturbation against the compact objects in spacetime and this kind of ringing is characterized by complex frequencies which only depends on the parameters describing the spacetime geometry under consideration, for this reason the QNMs are considered as the characteristic `sound' of the compact objects, such as black holes, neutron stars and wormholes. One can refer to \cite{Nollert1999,Berti:2009kk,Konoplya:2011qq} for a nice review on this topic.
However, the echoes can not be represented by a single dominant QNM. To extract QNMs frequencies from  the ringdown signals  of perturbation, we are supposed to consider wormhole parameters with the values that present the existence of echoes.

In our calculation of QNMs frequencies of wormholes, we would like to employ Prony method by which the oscillation frequencies and the damping rate of the QNMs can be extracted from the time-domain profile of the perturbation. The time-domain profile data can be fitted by superposition of $p$ complex exponentials as \cite{Zhu:2014sya,Berti:2007dg}
\begin{equation}
\Psi(r,t)\simeq \sum_{j=1}^{p}C_je^{-i\omega_j(t-t_0)},
\end{equation}
where $t_0$ is the beginning  time of the time period from $t_0$ to $t=N\Delta t+t_0$ in our consideration, and $N$ is an integer which satisfies  $N\geq 2p-1$. Then we have
\begin{equation}
x_n\equiv \Psi(r,n\Delta t+t_0)=\sum_{j=1}^{p}C_je^{-in\omega_j\Delta t}=\sum_{j=1}^{p}C_jz_j^n.
\end{equation}
Since $\Delta t$ is known  and $x_n$ is the  profile data we have obtained, Prony method allows us to calculate $z_j$
in terms of $x_n$ and thus the QNMs frequencies $\omega_i$ can be found.

\subsection{Quasinormal Modes of Wormhole Model I}
\begin{figure}
\centering
\includegraphics[height=2.2in,width=3.2in]{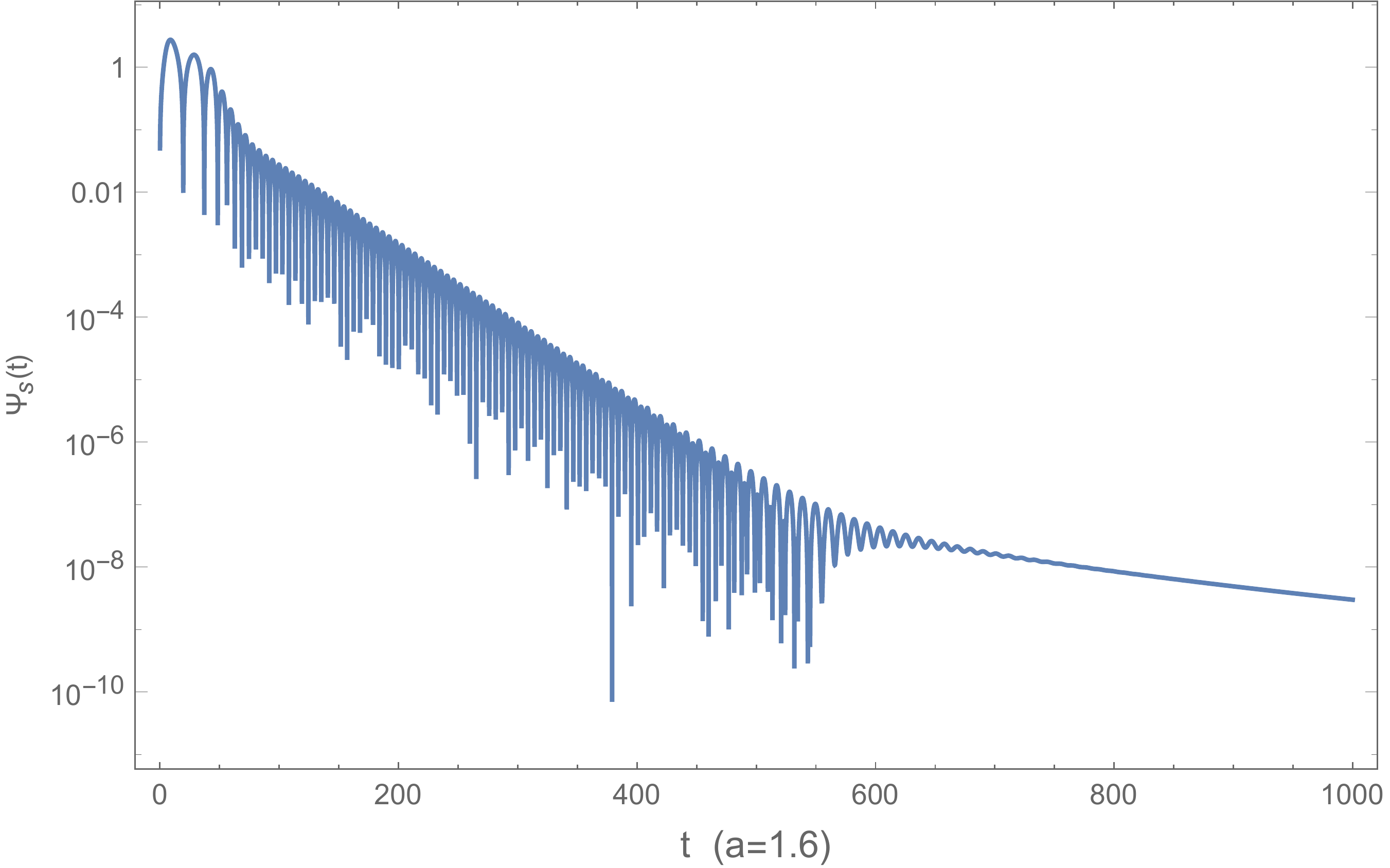}
\includegraphics[height=2.2in,width=3.2in]{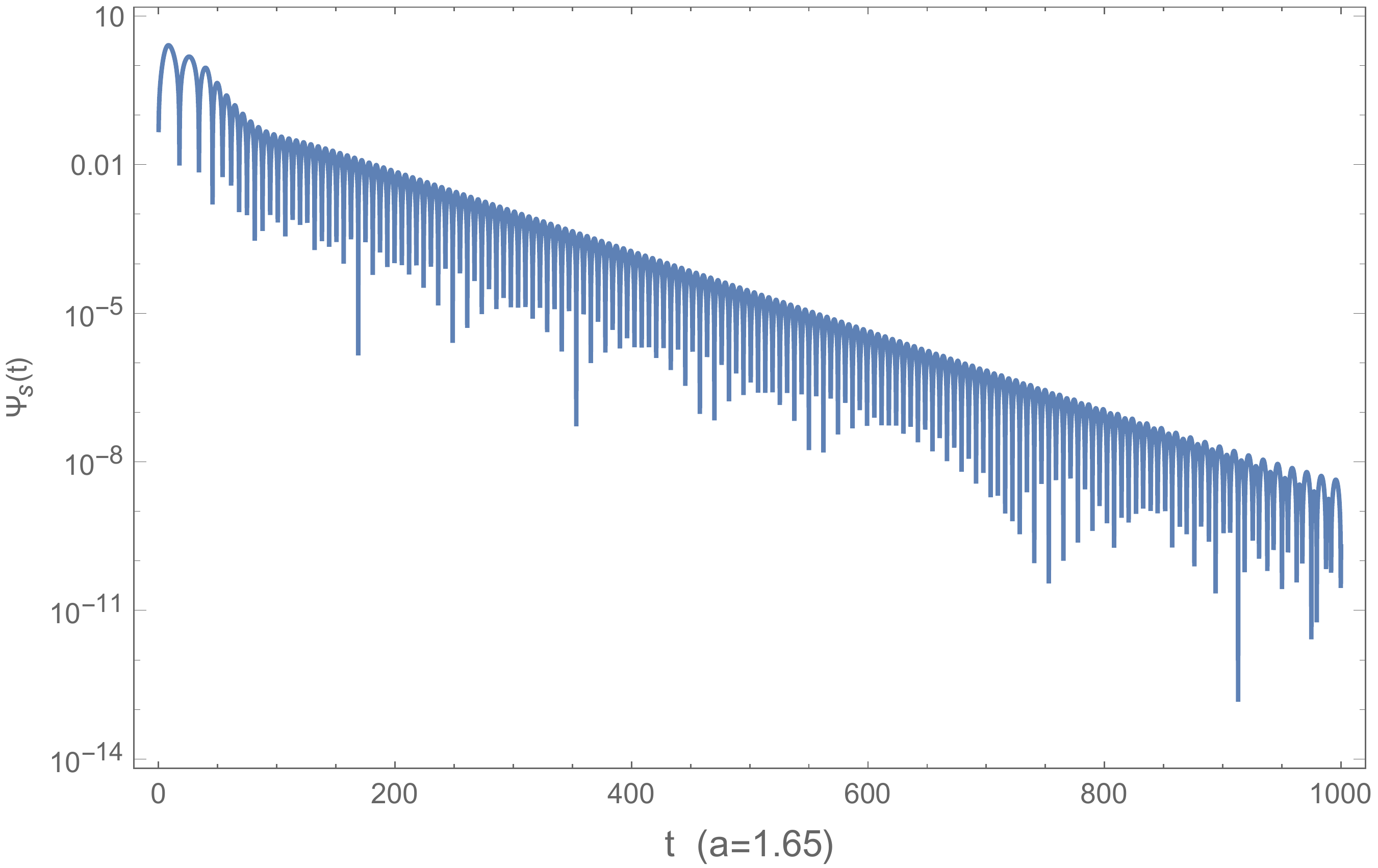}
\includegraphics[height=2.2in,width=3.2in]{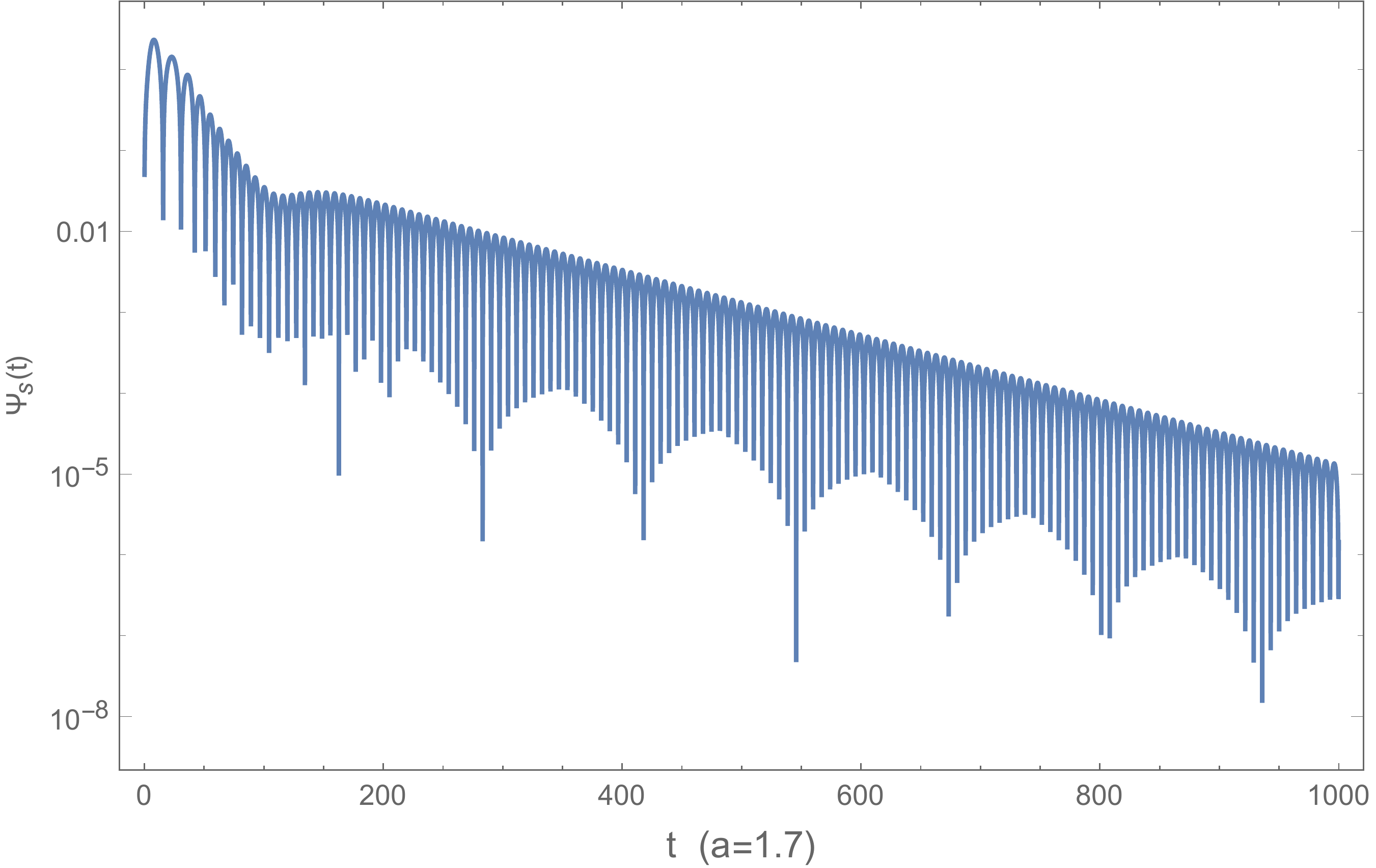}
\includegraphics[height=2.2in,width=3.2in]{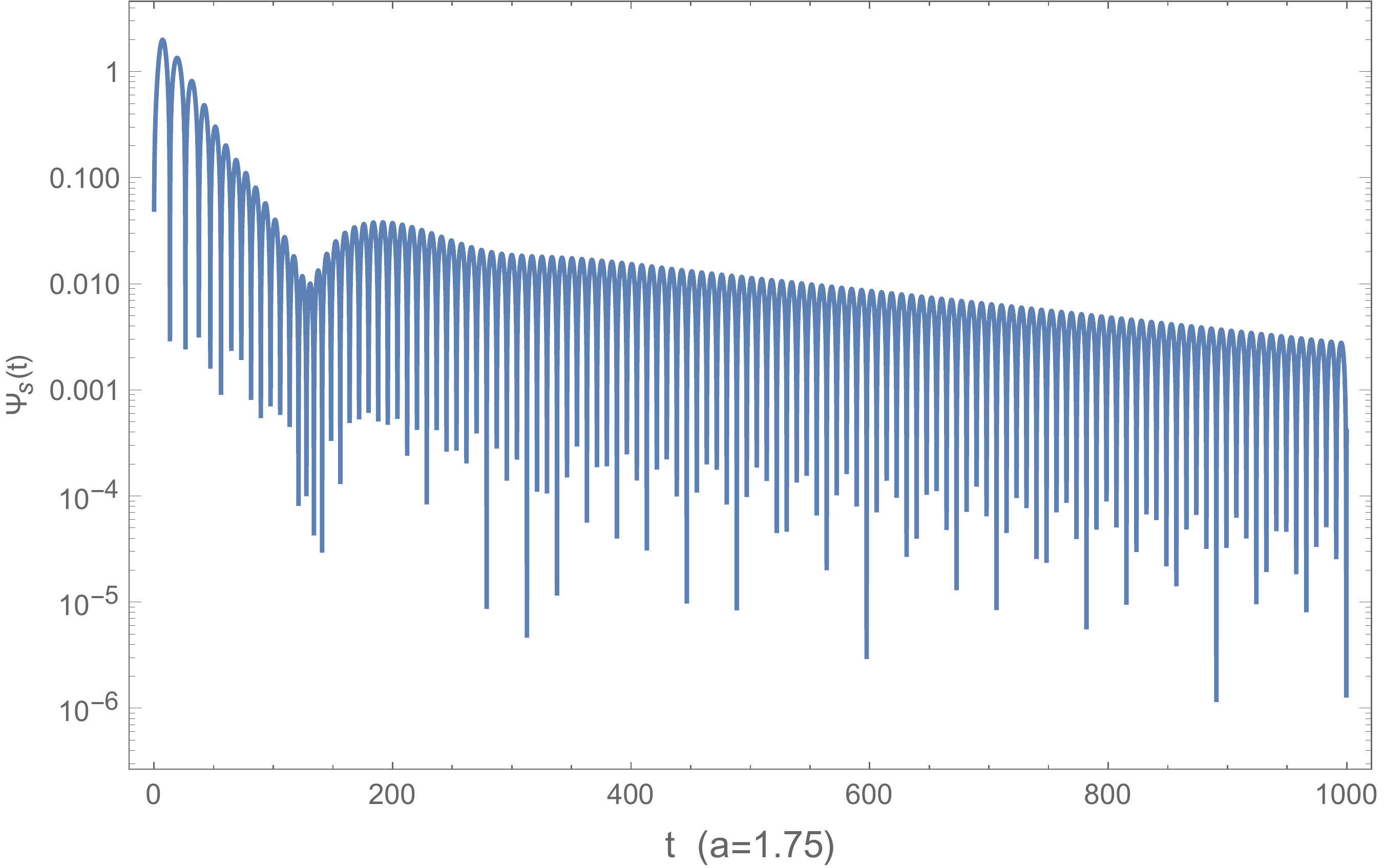}
\caption{The semi-logarithmic plots of the time-domain profile for perturbation of scalar field with angular number $l=1$ with four different value of parameter $a$ for Wormhole Model I.}\label{fig12}
\end{figure}

We show the pictures of evolution of scalar perturbation with angular number $l=1$ in Fig. \ref{fig12} from which the characteristic QNMs ringing can be observed, and  the decaying behavior of the perturbed scalar field indicates that this wormhole spacetime is stable under scalar field perturbation. The perturbation at different values of wormhole parameter $a$ shows a different damping rate. For larger value of $a$, the perturbation decays more slowly, but the change of oscillation frequencies is not as apparent as the change of damping rate. To have a quantitive  understanding of this ringdown stage of perturbation, we are required to work out the QNMs frequencies.

We calculate scalar QNMs frequencies for different angular number $l$ and wormhole parameter $a$ and the numerical results are listed in Table. \ref{table1}. The `echoes' appearing in the table means that under the corresponding values of $(a,l)$ the echoes are observed in the evolution picture of perturbation. From the data listed in Tale. \ref{table1}, it is easy to see how the QNMs frequencies change with the parameter value of $a$ and angular number $l$. By fixing $l$ and increase $a$ from $a=1.6$ to $a=1.78$, both the magnitude of real part and the imaginary part of the QNMs frequencies decrease implying that the oscillation frequencies and  damping rate are reduced which agrees with the behavior of time-domain profiles we demonstrate in Fig. \ref{fig12}. By fixing parameter $a$ and increase angular number $l$, as expected, we can see that the real part of the frequencies grows, while the magnitude of the imaginary part decreases.

\begin{table}[!htbp]
\centering
        \begin{tabular}{ccccccccccc}
    \hline\hline
    $a$ & $l=1$ & $l=2$& $l=3$ \\
    \hline
    $1.60$ & $0.580105-0.029412i$ & $0.978997-0.023685i$ & $1.370779-0.017403i$\\
    $1.62$ & $0.524104-0.024919i$ & $0.931959-0.019711i$ & $1.305067-0.014297i$\\
    $1.65$ & $0.511072-0.018521i$ & $0.861576-0.013897i$ & $1.206465-0.009623i$\\
    $1.68$ & $0.470023-0.012693i$ & $0.791572-0.008510i$ & $1.108091-0.005237i$\\
    $1.70$ & $0.442819-0.009246i$ & $0.745156-0.005402i$ & $1.042695-0.002821i$\\
    $1.72$ & $0.415707-0.006258i$ & $0.698859-0.002921i$ & $0.977260-0.001142i$\\
    $1.75$ & $0.375022-0.002852i$ & $0.629082-0.000724i$ &        echoes       \\
    $1.78$ & $0.333841-0.000899i$ &        echoes        &        echoes       \\

    \hline\hline
\end{tabular}
\caption{The dominant QNMs frequency $\omega$ of the scalar field for Wormhole Model I.}\label{table1}
\end{table}

\subsection{Quasinormal Modes of Wormhole Model II}

\begin{figure}
\centering
\includegraphics[height=2.2in,width=3.2in]{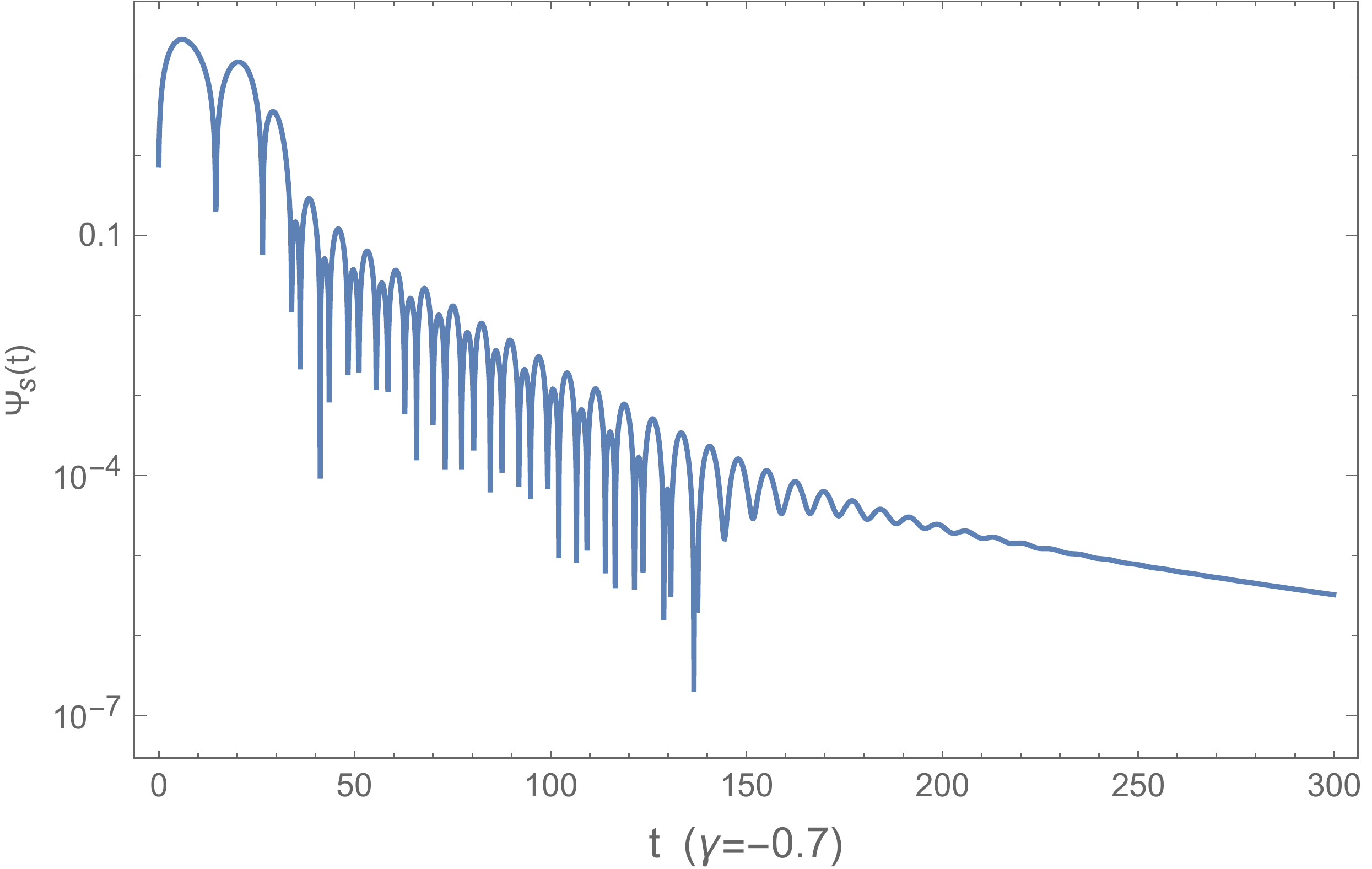}
\includegraphics[height=2.2in,width=3.2in]{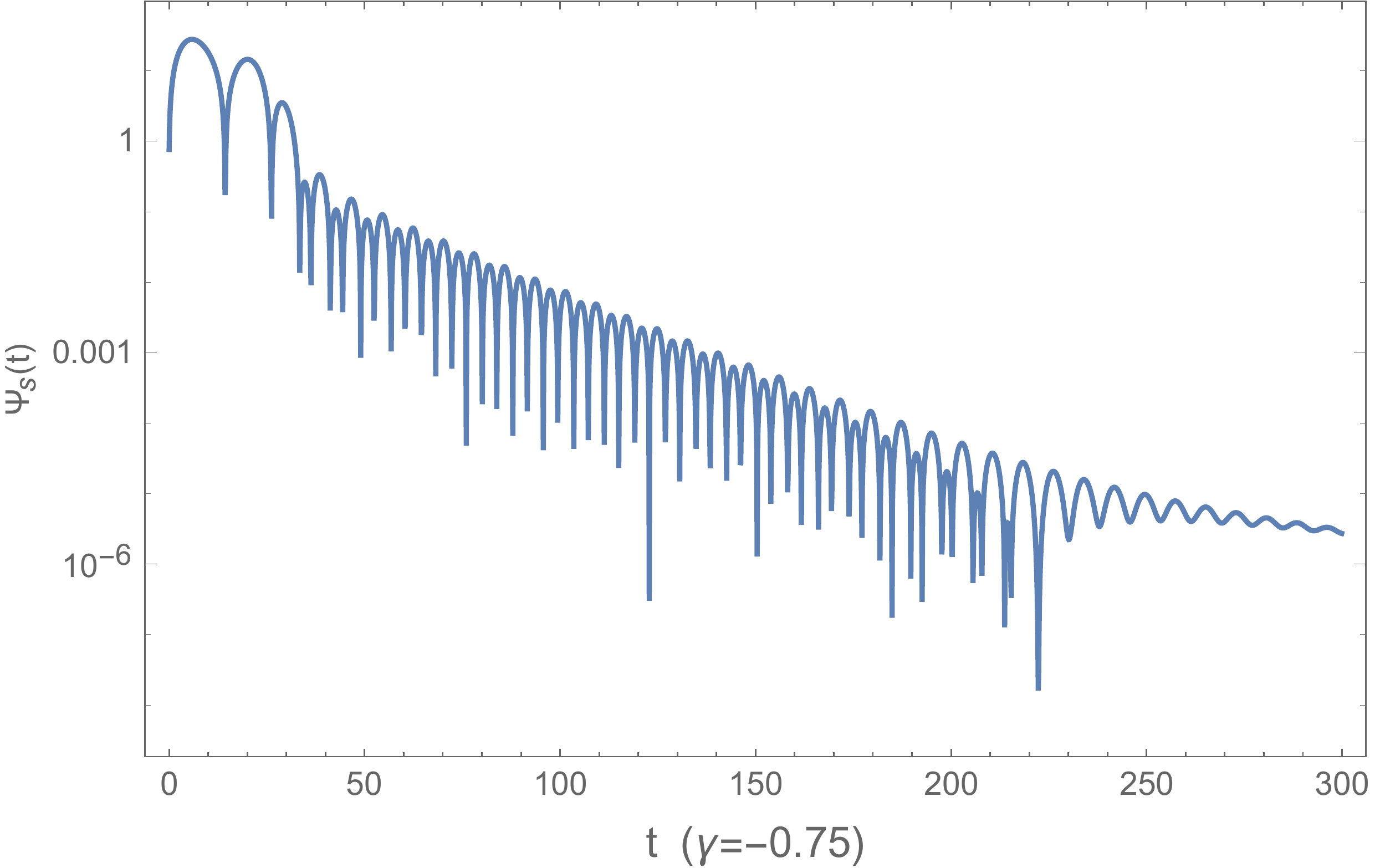}
\includegraphics[height=2.2in,width=3.2in]{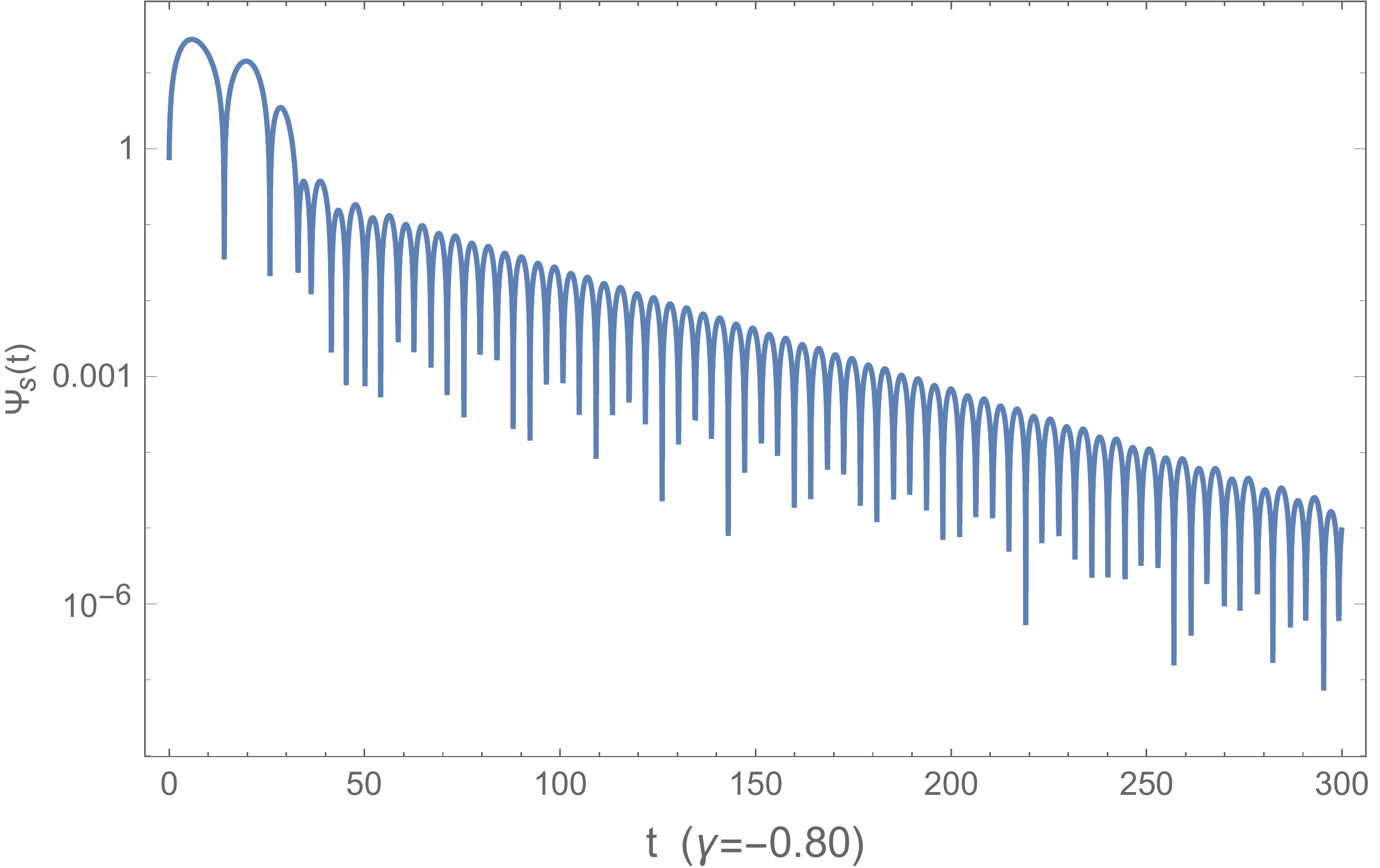}
\includegraphics[height=2.2in,width=3.2in]{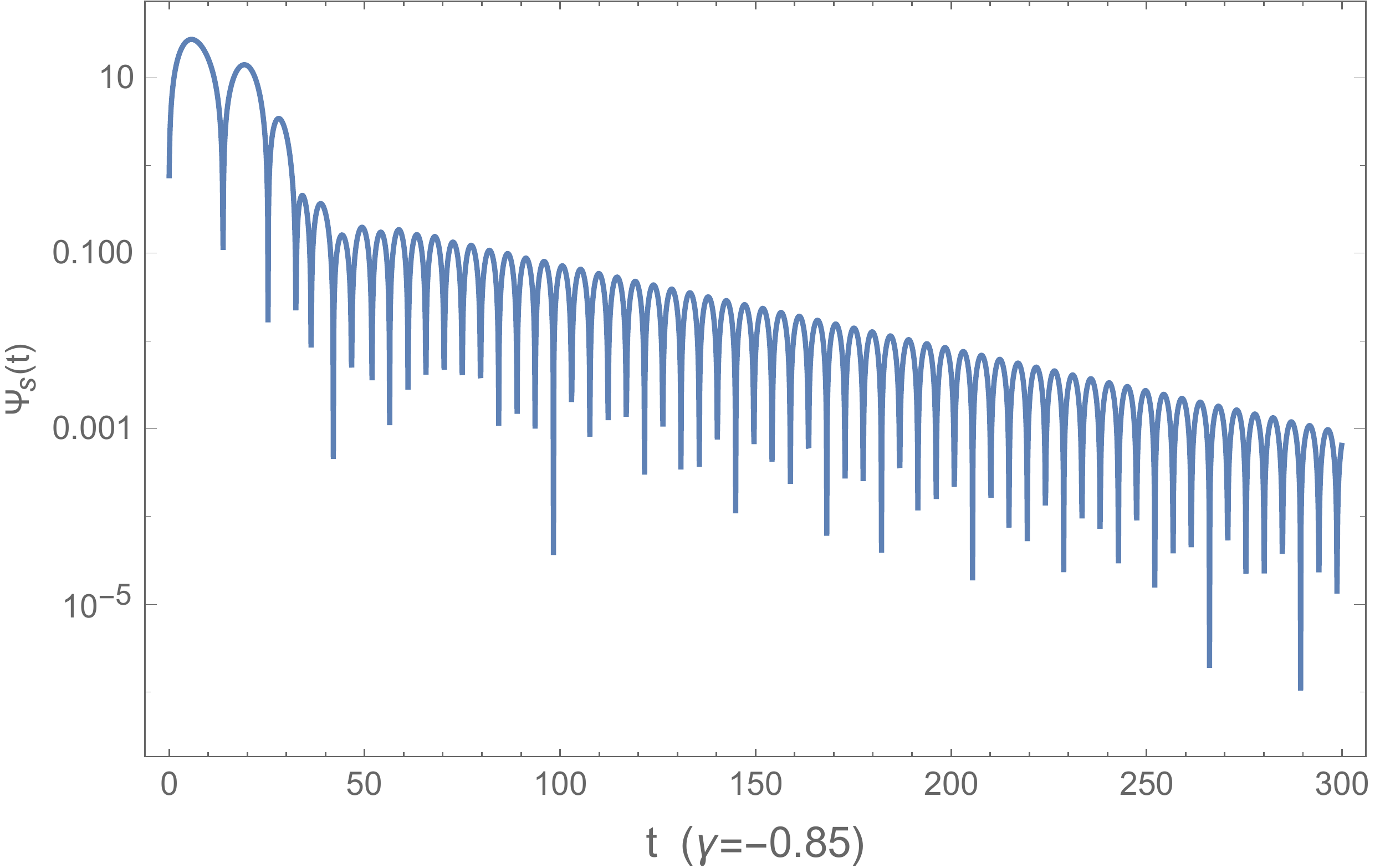}
\caption{The semi-logarithmic plots of the time-domain profile for perturbation of scalar field with angular number $l=1$ with four different value of  parameter $\gamma$ for Wormhole Model II.}\label{fig13}
\end{figure}

In this subsection we deal with the QNMs of scalar field for  Wormhole Model II. The time evolution of the scalar perturbation is displayed in Fig. \ref{fig13}. The features of the QNMs ringing reflected by this figure are qualitatively similar to that in Fig. \ref{fig12}, as the decaying perturbation implies a stable wormhole spacetime under scalar perturbation, and the change of parameter from $\gamma=-0.7$ to $\gamma=-0.85$
leads to a smaller damping rate.

In Table. \ref{table2} we list our numerical results of QNMs frequencies. Interestingly, these complex frequencies behave similarly to what we have revealed for QNMs data in Table. \ref{table1} under the change of wormhole parameter and angular number. Improving  $l$ will increase the real part of the QNMs frequencies which meets our expectation, while the magnitude of imaginary part of frequencies will be decreased. On the other hand, when increasing the magnitude of parameter $\gamma$, it is found that both the real and the magnitude of imaginary part of the QNMs frequencies become smaller.

\begin{table}[!htbp]
\centering
        \begin{tabular}{ccccccccccc}
    \hline\hline
    $\gamma$ & $l=1$ & $l=2$& $l=3$ \\
    \hline

    $-0.70$ & $0.862639-0.069110i$ & $1.410393-0.055482i$ & $1.959254-0.046957i$\\
    $-0.72$ & $0.840143-0.062443i$ & $1.369947-0.047802i$ & $1.900899-0.038571i$\\
    $-0.75$ & $0.805425-0.052499i$ & $1.307465-0.036512i$ & $1.810667-0.026595i$\\
    $-0.78$ & $0.768769-0.042793i$ & $1.241783-0.026130i$ & $1.715816-0.016364i$\\
    $-0.80$ & $0.743357-0.036509i$ & $1.196103-0.019830i$ & $1.649544-0.010745i$\\
    $-0.82$ & $0.716594-0.030449i$ & $1.147950-0.014268i$ & $1.579318-0.006354i$\\
    $-0.85$ & $0.673975-0.021988i$ & $1.070724-0.007633i$ & $1.465840-0.002299i$\\

    \hline\hline
\end{tabular}
\caption{The dominant QNMs frequency $\omega$ of the scalar field for Wormhole Model II.}  \label{table2}
\end{table}

\section{conclusions and discussions}
\label{section5}

In this paper, we have investigated the properties of signals of echo from testing scalar field and electromagnetic field perturbations around phantom wormhole configurations.  The phenomena of echo behaviors root deeply in physical properties of the effective potential well of the corresponding spacetimes. Here  we summarize the main properties of signals of echo from phantom wormhole configurations and their relations to effective potential properties. When the effective potential becomes deeper, for example because of the increase of the angular index, the perturbation will be more difficult to escape from the potential well which results in a smaller signal amplitude of the echo.  When the effective potential well becomes wider due to the change of parameters $a$ or $\gamma$ for example, there appears a time delay in the echo spectrum.

Here we examined the echoes of scalar field and electromagnetic field perturbations. Regardless of an additional term in the scalar potential, echo behavior in the scalar perturbation is found very similar to that in the electromagnetic field perturbation. We expect that the properties of echoes we obtained in scalar field and electromagnetic field perturbations also will persist in the gravitational perturbation,  which we will examine in the future. As described in  \cite{Cardoso:2016rao,Cardoso:2016oxy}, the waveform of echo is composed of two parts. The first part is the initial ringdown signals from the  waves scattered from the potential wall.  The second part is the signal of echo arising from the trapped waves leakage through the potential barrier. This suggests that each echo is a low-frequency filtered version of the previous one. The original shape of the mode gets quickly washed out after a few echoes, such that at later times the lower frequencies of echoes will be observed. This description of echoes also holds in our discussion of the phantom wormhole configurations.

It is of great interest that in phantom wormhole configurations, the dark energy equation of state can influence the echo spectrum. In the wormhole model I, there are two entangled parameters influence the echo spectrum. In this model dark energy equation of state does not show apparently in the parameter space, it is tuned by changing the parameter $a$ through $\omega=-2/a$. Increasing the model parameter $a$ will increase the equation of state of dark energy from $\omega<-1$. When $a$ reaches the maximum allowed value $2$, $\omega=-1$. This maximum value of $a$ is required to keep the configuration. In the process of the increase of $a$, the time delay in the echo becomes longer. In Model II,  there are two independent parameters $\gamma$ and $\omega$ appear in the metric and affect the echo spectrum separately. The parameter $\gamma$ dominates the impacts on echoes, and the influences of $\omega$ is subdominant. The increase of the absolute value $|\gamma|$ will result in the time delay of the echoes, however the oscillation frequency is not sensitive to the change of $\gamma$. The effect of the dark energy equation of state in Model II also modifies the time delay in echoes. For $\omega<-2$, its influence on echo time delay is negligible. However when $\omega$ approaches to phantom divide from $-2$, the delay time in echoes becomes longer.   The influence of the dark energy equation of state shown in the echo spectrum is interesting. Once the echo is detected, the signature of the dark energy equation of state in the echo can serve as a local measurement of the dark energy.

Besides the echoes, the QNMs can be regarded as the `characteristic sound' of wormholes in our consideration and thus it can also serve as a probe of wormhole in the future detection. We have calculated the QNMs frequencies for both wormhole models by Prony method. Interestingly, it is found that QNMs in two wormholes backgrounds behave similarly, as a higher magnitude of wormhole parameter will lead to a lower oscillation frequency and damping rate, while  higher angular number will naturally cause a higher oscillation frequency but smaller damping rate.

In previous literatures \cite{Cardoso:2016rao,Cardoso:2016oxy,Foit:2016uxn,Bueno:2017hyj,Wang:2018mlp,Cardoso:2019apo,Wang:2019rcf,Wang:2019szm,Dey:2020lhq}, the time delay $\Delta t$ of the echoes are usually simply evaluated by calculating the time the null geodesic traveling from one potential peak to another one and comeback to the first one. Therefore, the delay time is given by $\Delta t\approx2L$, where $L$ is the width of the potential well obtained by $r_{\ast(right peak)}-r_{\ast(left peak)}$. We would like to point out that this formula is only valid for a very sharp potential, as in the case of our wormhole model II and the existing literatures \cite{Cardoso:2016rao,Cardoso:2016oxy,Foit:2016uxn,Bueno:2017hyj, Wang:2018mlp,Cardoso:2019apo,Wang:2019rcf,Wang:2019szm,Dey:2020lhq}. However, this formula is not valid if the potential is ``fat" since in this situation we can not determine the locations where the waves are reflected by the potential, as in our wormhole model I. One may note that the null-like geodesics are connected to higher $l$ limit so that this formula may be valid for echoes with large $l$. In fact, as we have demonstrated in Fig. \ref{fig2} from which we can find that when increasing $l$ from 1 to 6, the shape of potentials remain almost unchanged, and just the values of potentials are improved, which implies that formula is still invalid to calculate the period of echoes.
This suggests that the period in echoes are needed to be studied more carefully, especially if we want to employ it as a dark energy probe.

\begin{acknowledgments}

This work is supported by the Natural Science Foundation
of China under Grants Nos. 11775036, 11905083, 11847055, and Fok Ying Tung Education Foundation
under Grant No.171006. Jian-Pin Wu is also supported by Top
Talent Support Program from Yangzhou University.

\end{acknowledgments}

\bibliographystyle{JHEP}

\bibliography{Echoes}

\providecommand{\href}[2]{#2}\begingroup\raggedright\begin{thebibliography}{10}

\bibitem{Akiyama:2019bqs}
{\scshape Event Horizon Telescope} collaboration, K.~Akiyama et~al.,
  \emph{{First M87 Event Horizon Telescope Results. IV. Imaging the Central
  Supermassive Black Hole}},
  \href{http://dx.doi.org/10.3847/2041-8213/ab0e85}{\emph{Astrophys. J. Lett.}
  {\bf 875} (2019) L4}, [\href{http://arxiv.org/abs/1906.11241}{{\tt
  1906.11241}}].

\bibitem{Abramowicz:2002vt}
M.~A. Abramowicz, W.~Kluzniak and J.-P. Lasota, \emph{{No observational proof
  of the black hole event-horizon}},
  \href{http://dx.doi.org/10.1051/0004-6361:20021645}{\emph{Astron. Astrophys.}
  {\bf 396} (2002) L31--L34},
  [\href{http://arxiv.org/abs/astro-ph/0207270}{{\tt astro-ph/0207270}}].

\bibitem{Cardoso:2016rao}
V.~Cardoso, E.~Franzin and P.~Pani, \emph{{Is the gravitational-wave ringdown a
  probe of the event horizon?}},
  \href{http://dx.doi.org/10.1103/PhysRevLett.116.171101}{\emph{Phys.\ Rev.\
  Lett.} {\bf 116} (2016) 171101}, [\href{http://arxiv.org/abs/1602.07309}{{\tt
  1602.07309}}].

\bibitem{Almheiri:2013hfa}
A.~Almheiri, D.~Marolf, J.~Polchinski, D.~Stanford and J.~Sully, \emph{{An
  Apologia for Firewalls}},
  \href{http://dx.doi.org/10.1007/JHEP09(2013)018}{\emph{JHEP} {\bf 09} (2013)
  018}, [\href{http://arxiv.org/abs/1304.6483}{{\tt 1304.6483}}].

\bibitem{Cardoso:2019rvt}
V.~Cardoso and P.~Pani, \emph{{Testing the nature of dark compact objects: a
  status report}},
  \href{http://dx.doi.org/10.1007/s41114-019-0020-4}{\emph{Living Rev. Rel.}
  {\bf 22} (2019) 4}, [\href{http://arxiv.org/abs/1904.05363}{{\tt
  1904.05363}}].

\bibitem{Bueno:2017hyj}
P.~Bueno, P.~A. Cano, F.~Goelen, T.~Hertog and B.~Vercnocke, \emph{{Echoes of
  Kerr-like wormholes}},
  \href{http://dx.doi.org/10.1103/PhysRevD.97.024040}{\emph{Phys.\ Rev.\ D}
  {\bf 97} (2018) 024040}, [\href{http://arxiv.org/abs/1711.00391}{{\tt
  1711.00391}}].

\bibitem{Mazur:2004fk}
P.~O. Mazur and E.~Mottola, \emph{{Gravitational vacuum condensate stars}},
  \href{http://dx.doi.org/10.1073/pnas.0402717101}{\emph{Proc. Nat. Acad. Sci.}
  {\bf 101} (2004) 9545--9550}, [\href{http://arxiv.org/abs/gr-qc/0407075}{{\tt
  gr-qc/0407075}}].

\bibitem{Visser:2003ge}
M.~Visser and D.~L. Wiltshire, \emph{{Stable gravastars: An Alternative to
  black holes?}},
  \href{http://dx.doi.org/10.1088/0264-9381/21/4/027}{\emph{Class. Quant.
  Grav.} {\bf 21} (2004) 1135--1152},
  [\href{http://arxiv.org/abs/gr-qc/0310107}{{\tt gr-qc/0310107}}].

\bibitem{Wang:2018cum}
Y.-T. Wang, J.~Zhang and Y.-S. Piao, \emph{{Primordial gravastar from
  inflation}},
  \href{http://dx.doi.org/10.1016/j.physletb.2019.06.036}{\emph{Phys. Lett. B}
  {\bf 795} (2019) 314--318}, [\href{http://arxiv.org/abs/1810.04885}{{\tt
  1810.04885}}].

\bibitem{Schunck:2003kk}
F.~E. Schunck and E.~W. Mielke, \emph{{General relativistic boson stars}},
  \href{http://dx.doi.org/10.1088/0264-9381/20/20/201}{\emph{Class. Quant.
  Grav.} {\bf 20} (2003) R301--R356},
  [\href{http://arxiv.org/abs/0801.0307}{{\tt 0801.0307}}].

\bibitem{Carloni:2017bck}
S.~Carloni and D.~Vernieri, \emph{{Covariant Tolman-Oppenheimer-Volkoff
  equations. II. The anisotropic case}},
  \href{http://dx.doi.org/10.1103/PhysRevD.97.124057}{\emph{Phys. Rev. D} {\bf
  97} (2018) 124057}, [\href{http://arxiv.org/abs/1709.03996}{{\tt
  1709.03996}}].

\bibitem{Isayev:2018hqx}
A.~Isayev, \emph{{Comment on ``Covariant Tolman-Oppenheimer-Volkoff equations.
  II. The anisotropic case''}},
  \href{http://dx.doi.org/10.1103/PhysRevD.98.088503}{\emph{Phys. Rev. D} {\bf
  98} (2018) 088503}, [\href{http://arxiv.org/abs/1808.05699}{{\tt
  1808.05699}}].

\bibitem{Raposo:2018rjn}
G.~Raposo, P.~Pani, M.~Bezares, C.~Palenzuela and V.~Cardoso,
  \emph{{Anisotropic stars as ultracompact objects in General Relativity}},
  \href{http://dx.doi.org/10.1103/PhysRevD.99.104072}{\emph{Phys. Rev. D} {\bf
  99} (2019) 104072}, [\href{http://arxiv.org/abs/1811.07917}{{\tt
  1811.07917}}].

\bibitem{Lemos:2003gx}
J.~P. Lemos and E.~J. Weinberg, \emph{{Quasiblack holes from extremal charged
  dust}}, \href{http://dx.doi.org/10.1103/PhysRevD.69.104004}{\emph{Phys. Rev.
  D} {\bf 69} (2004) 104004}, [\href{http://arxiv.org/abs/gr-qc/0311051}{{\tt
  gr-qc/0311051}}].

\bibitem{Lemos:2008cv}
J.~P. Lemos and O.~B. Zaslavskii, \emph{{Black hole mimickers: Regular versus
  singular behavior}},
  \href{http://dx.doi.org/10.1103/PhysRevD.78.024040}{\emph{Phys. Rev. D} {\bf
  78} (2008) 024040}, [\href{http://arxiv.org/abs/0806.0845}{{\tt 0806.0845}}].

\bibitem{Kawai:2013mda}
H.~Kawai, Y.~Matsuo and Y.~Yokokura, \emph{{A Self-consistent Model of the
  Black Hole Evaporation}},
  \href{http://dx.doi.org/10.1142/S0217751X13500504}{\emph{Int. J. Mod. Phys.
  A} {\bf 28} (2013) 1350050}, [\href{http://arxiv.org/abs/1302.4733}{{\tt
  1302.4733}}].

\bibitem{Baccetti:2017oas}
V.~Baccetti, R.~B. Mann and D.~R. Terno, \emph{{Do event horizons exist?}},
  \href{http://dx.doi.org/10.1142/S0218271817170088}{\emph{Int. J. Mod. Phys.
  D} {\bf 26} (2017) 1743008}, [\href{http://arxiv.org/abs/1706.01180}{{\tt
  1706.01180}}].

\bibitem{Baccetti:2016lsb}
V.~Baccetti, R.~B. Mann and D.~R. Terno, \emph{{Role of evaporation in
  gravitational collapse}},
  \href{http://dx.doi.org/10.1088/1361-6382/aad70e}{\emph{Class. Quant. Grav.}
  {\bf 35} (2018) 185005}, [\href{http://arxiv.org/abs/1610.07839}{{\tt
  1610.07839}}].

\bibitem{Baccetti:2018qrp}
V.~Baccetti, S.~Murk and D.~R. Terno, \emph{{Black hole evaporation and
  semiclassical thin shell collapse}},
  \href{http://dx.doi.org/10.1103/PhysRevD.100.064054}{\emph{Phys. Rev. D} {\bf
  100} (2019) 064054}, [\href{http://arxiv.org/abs/1812.07727}{{\tt
  1812.07727}}].

\bibitem{Giddings:1992hh}
S.~B. Giddings, \emph{{Black holes and massive remnants}},
  \href{http://dx.doi.org/10.1103/PhysRevD.46.1347}{\emph{Phys. Rev. D} {\bf
  46} (1992) 1347--1352}, [\href{http://arxiv.org/abs/hep-th/9203059}{{\tt
  hep-th/9203059}}].

\bibitem{Unruh:2017uaw}
W.~G. Unruh and R.~M. Wald, \emph{{Information Loss}},
  \href{http://dx.doi.org/10.1088/1361-6633/aa778e}{\emph{Rept. Prog. Phys.}
  {\bf 80} (2017) 092002}, [\href{http://arxiv.org/abs/1703.02140}{{\tt
  1703.02140}}].

\bibitem{Lunin:2001jy}
O.~Lunin and S.~D. Mathur, \emph{{AdS / CFT duality and the black hole
  information paradox}},
  \href{http://dx.doi.org/10.1016/S0550-3213(01)00620-4}{\emph{Nucl. Phys. B}
  {\bf 623} (2002) 342--394}, [\href{http://arxiv.org/abs/hep-th/0109154}{{\tt
  hep-th/0109154}}].

\bibitem{Lunin:2002qf}
O.~Lunin and S.~D. Mathur, \emph{{Statistical interpretation of Bekenstein
  entropy for systems with a stretched horizon}},
  \href{http://dx.doi.org/10.1103/PhysRevLett.88.211303}{\emph{Phys. Rev.
  Lett.} {\bf 88} (2002) 211303},
  [\href{http://arxiv.org/abs/hep-th/0202072}{{\tt hep-th/0202072}}].

\bibitem{Mathur:2005zp}
S.~D. Mathur, \emph{{The Fuzzball proposal for black holes: An Elementary
  review}}, \href{http://dx.doi.org/10.1002/prop.200410203}{\emph{Fortsch.
  Phys.} {\bf 53} (2005) 793--827},
  [\href{http://arxiv.org/abs/hep-th/0502050}{{\tt hep-th/0502050}}].

\bibitem{Mathur:2008nj}
S.~D. Mathur, \emph{{Fuzzballs and the information paradox: A Summary and
  conjectures}},  \href{http://arxiv.org/abs/0810.4525}{{\tt 0810.4525}}.

\bibitem{Mathur:2012jk}
S.~D. Mathur and D.~Turton, \emph{{Comments on black holes I: The possibility
  of complementarity}},
  \href{http://dx.doi.org/10.1007/JHEP01(2014)034}{\emph{JHEP} {\bf 01} (2014)
  034}, [\href{http://arxiv.org/abs/1208.2005}{{\tt 1208.2005}}].

\bibitem{Almheiri:2012rt}
A.~Almheiri, D.~Marolf, J.~Polchinski and J.~Sully, \emph{{Black Holes:
  Complementarity or Firewalls?}},
  \href{http://dx.doi.org/10.1007/JHEP02(2013)062}{\emph{JHEP} {\bf 02} (2013)
  062}, [\href{http://arxiv.org/abs/1207.3123}{{\tt 1207.3123}}].

\bibitem{Oshita:2018fqu}
N.~Oshita and N.~Afshordi, \emph{{Probing microstructure of black hole
  spacetimes with gravitational wave echoes}},
  \href{http://dx.doi.org/10.1103/PhysRevD.99.044002}{\emph{Phys. Rev. D} {\bf
  99} (2019) 044002}, [\href{http://arxiv.org/abs/1807.10287}{{\tt
  1807.10287}}].

\bibitem{Oshita:2019sat}
N.~Oshita, Q.~Wang and N.~Afshordi, \emph{{On Reflectivity of Quantum Black
  Hole Horizons}},
  \href{http://dx.doi.org/10.1088/1475-7516/2020/04/016}{\emph{JCAP} {\bf 04}
  (2020) 016}, [\href{http://arxiv.org/abs/1905.00464}{{\tt 1905.00464}}].

\bibitem{Abbott:2016blz}
{\scshape LIGO Scientific, Virgo} collaboration, B.~Abbott et~al.,
  \emph{{Observation of Gravitational Waves from a Binary Black Hole Merger}},
  \href{http://dx.doi.org/10.1103/PhysRevLett.116.061102}{\emph{Phys. Rev.
  Lett.} {\bf 116} (2016) 061102}, [\href{http://arxiv.org/abs/1602.03837}{{\tt
  1602.03837}}].

\bibitem{TheLIGOScientific:2017qsa}
{\scshape LIGO Scientific, Virgo} collaboration, B.~Abbott et~al.,
  \emph{{GW170817: Observation of Gravitational Waves from a Binary Neutron
  Star Inspiral}},
  \href{http://dx.doi.org/10.1103/PhysRevLett.119.161101}{\emph{Phys. Rev.
  Lett.} {\bf 119} (2017) 161101}, [\href{http://arxiv.org/abs/1710.05832}{{\tt
  1710.05832}}].

\bibitem{Giudice:2016zpa}
G.~F. Giudice, M.~McCullough and A.~Urbano, \emph{{Hunting for Dark Particles
  with Gravitational Waves}},
  \href{http://dx.doi.org/10.1088/1475-7516/2016/10/001}{\emph{JCAP} {\bf 10}
  (2016) 001}, [\href{http://arxiv.org/abs/1605.01209}{{\tt 1605.01209}}].

\bibitem{Macedo:2013jja}
C.~F. Macedo, P.~Pani, V.~Cardoso and L.~C.~B. Crispino, \emph{{Astrophysical
  signatures of boson stars: quasinormal modes and inspiral resonances}},
  \href{http://dx.doi.org/10.1103/PhysRevD.88.064046}{\emph{Phys. Rev. D} {\bf
  88} (2013) 064046}, [\href{http://arxiv.org/abs/1307.4812}{{\tt 1307.4812}}].

\bibitem{Cardoso:2016oxy}
V.~Cardoso, S.~Hopper, C.~F.~B. Macedo, C.~Palenzuela and P.~Pani,
  \emph{{Gravitational-wave signatures of exotic compact objects and of quantum
  corrections at the horizon scale}},
  \href{http://dx.doi.org/10.1103/PhysRevD.94.084031}{\emph{Phys.\ Rev.\ D}
  {\bf 94} (2016) 084031}, [\href{http://arxiv.org/abs/1608.08637}{{\tt
  1608.08637}}].

\bibitem{Abedi:2020ujo}
J.~Abedi, N.~Afshordi, N.~Oshita and Q.~Wang, \emph{{Quantum Black Holes in the
  Sky}}, \href{http://dx.doi.org/10.3390/universe6030043}{\emph{Universe} {\bf
  6} (2020) 43}, [\href{http://arxiv.org/abs/2001.09553}{{\tt 2001.09553}}].

\bibitem{Foit:2016uxn}
V.~F. Foit and M.~Kleban, \emph{{Testing Quantum Black Holes with Gravitational
  Waves}}, \href{http://dx.doi.org/10.1088/1361-6382/aafcba}{\emph{Class.\
  Quant.\ Grav.} {\bf 36} (2019) 035006},
  [\href{http://arxiv.org/abs/1611.07009}{{\tt 1611.07009}}].

\bibitem{Price:2017cjr}
R.~H. Price and G.~Khanna, \emph{{Gravitational wave sources: reflections and
  echoes}}, \href{http://dx.doi.org/10.1088/1361-6382/aa8f29}{\emph{Class.\
  Quant.\ Grav.} {\bf 34} (2017) 225005},
  [\href{http://arxiv.org/abs/1702.04833}{{\tt 1702.04833}}].

\bibitem{Nakano:2017fvh}
H.~Nakano, N.~Sago, H.~Tagoshi and T.~Tanaka, \emph{{Black hole ringdown echoes
  and howls}}, \href{http://dx.doi.org/10.1093/ptep/ptx093}{\emph{PTEP} {\bf
  2017} (2017) 071E01}, [\href{http://arxiv.org/abs/1704.07175}{{\tt
  1704.07175}}].

\bibitem{Mark:2017dnq}
Z.~Mark, A.~Zimmerman, S.~M. Du and Y.~Chen, \emph{{A recipe for echoes from
  exotic compact objects}},
  \href{http://dx.doi.org/10.1103/PhysRevD.96.084002}{\emph{Phys.\ Rev.\ D}
  {\bf 96} (2017) 084002}, [\href{http://arxiv.org/abs/1706.06155}{{\tt
  1706.06155}}].

\bibitem{Burgess:2018pmm}
C.~Burgess, R.~Plestid and M.~Rummel, \emph{{Effective Field Theory of Black
  Hole Echoes}}, \href{http://dx.doi.org/10.1007/JHEP09(2018)113}{\emph{JHEP}
  {\bf 09} (2018) 113}, [\href{http://arxiv.org/abs/1808.00847}{{\tt
  1808.00847}}].

\bibitem{Konoplya:2018yrp}
R.~Konoplya, Z.~Stuchlík and A.~Zhidenko, \emph{{Echoes of compact objects:
  new physics near the surface and matter at a distance}},
  \href{http://dx.doi.org/10.1103/PhysRevD.99.024007}{\emph{Phys.\ Rev.\ D}
  {\bf 99} (2019) 024007}, [\href{http://arxiv.org/abs/1810.01295}{{\tt
  1810.01295}}].

\bibitem{Testa:2018bzd}
A.~Testa and P.~Pani, \emph{{Analytical template for gravitational-wave echoes:
  signal characterization and prospects of detection with current and future
  interferometers}},
  \href{http://dx.doi.org/10.1103/PhysRevD.98.044018}{\emph{Phys.\ Rev.\ D}
  {\bf 98} (2018) 044018}, [\href{http://arxiv.org/abs/1806.04253}{{\tt
  1806.04253}}].

\bibitem{Wang:2018mlp}
Y.-T. Wang, Z.-P. Li, J.~Zhang, S.-Y. Zhou and Y.-S. Piao, \emph{{Are
  gravitational wave ringdown echoes always equal-interval?}},
  \href{http://dx.doi.org/10.1140/epjc/s10052-018-5974-y}{\emph{Eur.\ Phys.\
  J.\ C} {\bf 78} (2018) 482}, [\href{http://arxiv.org/abs/1802.02003}{{\tt
  1802.02003}}].

\bibitem{Cardoso:2019apo}
V.~Cardoso, V.~F. Foit and M.~Kleban, \emph{{Gravitational wave echoes from
  black hole area quantization}},
  \href{http://dx.doi.org/10.1088/1475-7516/2019/08/006}{\emph{JCAP} {\bf 08}
  (2019) 006}, [\href{http://arxiv.org/abs/1902.10164}{{\tt 1902.10164}}].

\bibitem{Wang:2019rcf}
Q.~Wang, N.~Oshita and N.~Afshordi, \emph{{Echoes from Quantum Black Holes}},
  \href{http://dx.doi.org/10.1103/PhysRevD.101.024031}{\emph{Phys.\ Rev.\ D}
  {\bf 101} (2020) 024031}, [\href{http://arxiv.org/abs/1905.00446}{{\tt
  1905.00446}}].

\bibitem{Ghersi:2019trn}
J.~T. Gálvez~Ghersi, A.~V. Frolov and D.~A. Dobre, \emph{{Echoes from the
  scattering of wavepackets on wormholes}},
  \href{http://dx.doi.org/10.1088/1361-6382/ab23c8}{\emph{Class.\ Quant.\
  Grav.} {\bf 36} (2019) 135006}, [\href{http://arxiv.org/abs/1901.06625}{{\tt
  1901.06625}}].

\bibitem{Li:2019kwa}
Z.-P. Li and Y.-S. Piao, \emph{{Mixing of gravitational wave echoes}},
  \href{http://dx.doi.org/10.1103/PhysRevD.100.044023}{\emph{Phys.\ Rev.\ D}
  {\bf 100} (2019) 044023}, [\href{http://arxiv.org/abs/1904.05652}{{\tt
  1904.05652}}].

\bibitem{Wang:2019szm}
Y.-T. Wang, J.~Zhang, S.-Y. Zhou and Y.-S. Piao, \emph{{On echo intervals in
  gravitational wave echo analysis}},
  \href{http://dx.doi.org/10.1140/epjc/s10052-019-7234-1}{\emph{Eur.\ Phys.\
  J.\ C} {\bf 79} (2019) 726}, [\href{http://arxiv.org/abs/1904.00212}{{\tt
  1904.00212}}].

\bibitem{Graham:2020gwr}
M.~Graham et~al., \emph{{Candidate Electromagnetic Counterpart to the Binary
  Black Hole Merger Gravitational Wave Event S190521g}},
  \href{http://dx.doi.org/10.1103/PhysRevLett.124.251102}{\emph{Phys. Rev.
  Lett.} {\bf 124} (2020) 251102}, [\href{http://arxiv.org/abs/2006.14122}{{\tt
  2006.14122}}].

\bibitem{Mastrogiovanni:2020gua}
S.~Mastrogiovanni, D.~Steer and M.~Barsuglia, \emph{{Probing modified gravity
  theories and cosmology using gravitational-waves and associated
  electromagnetic counterparts}},
  \href{http://dx.doi.org/10.1103/PhysRevD.102.044009}{\emph{Phys. Rev. D} {\bf
  102} (2020) 044009}, [\href{http://arxiv.org/abs/2004.01632}{{\tt
  2004.01632}}].

\bibitem{Soares-Santos:2017lru}
{\scshape DES, Dark Energy Camera GW-EM} collaboration, M.~Soares-Santos
  et~al., \emph{{The Electromagnetic Counterpart of the Binary Neutron Star
  Merger LIGO/Virgo GW170817. I. Discovery of the Optical Counterpart Using the
  Dark Energy Camera}},
  \href{http://dx.doi.org/10.3847/2041-8213/aa9059}{\emph{Astrophys. J. Lett.}
  {\bf 848} (2017) L16}, [\href{http://arxiv.org/abs/1710.05459}{{\tt
  1710.05459}}].

\bibitem{Valenti2017}
S.~Valenti, J.~S. David, S.~Yang, E.~Cappellaro, L.~Tartaglia, A.~Corsi et~al.,
  \emph{The discovery of the electromagnetic counterpart of {GW}170817:
  Kilonova {AT} 2017gfo/{DLT}17ck},
  \href{http://dx.doi.org/10.3847/2041-8213/aa8edf}{\emph{The Astrophysical
  Journal} {\bf 848} (oct, 2017) L24}.

\bibitem{Morris:1988cz}
M.~Morris and K.~Thorne, \emph{{Wormholes in space-time and their use for
  interstellar travel: A tool for teaching general relativity}},
  \href{http://dx.doi.org/10.1119/1.15620}{\emph{Am. J. Phys.} {\bf 56} (1988)
  395--412}.

\bibitem{Dai:2019mse}
D.-C. Dai and D.~Stojkovic, \emph{{Observing a Wormhole}},
  \href{http://dx.doi.org/10.1103/PhysRevD.100.083513}{\emph{Phys. Rev. D} {\bf
  100} (2019) 083513}, [\href{http://arxiv.org/abs/1910.00429}{{\tt
  1910.00429}}].

\bibitem{Damour:2007ap}
T.~Damour and S.~N. Solodukhin, \emph{{Wormholes as black hole foils}},
  \href{http://dx.doi.org/10.1103/PhysRevD.76.024016}{\emph{Phys. Rev. D} {\bf
  76} (2007) 024016}, [\href{http://arxiv.org/abs/0704.2667}{{\tt 0704.2667}}].

\bibitem{Volkel:2018hwb}
S.~H. Völkel and K.~D. Kokkotas, \emph{{Wormhole Potentials and Throats from
  Quasi-Normal Modes}},
  \href{http://dx.doi.org/10.1088/1361-6382/aabce6}{\emph{Class. Quant. Grav.}
  {\bf 35} (2018) 105018}, [\href{http://arxiv.org/abs/1802.08525}{{\tt
  1802.08525}}].

\bibitem{Cramer:1994qj}
J.~G. Cramer, R.~L. Forward, M.~S. Morris, M.~Visser, G.~Benford and G.~A.
  Landis, \emph{{Natural wormholes as gravitational lenses}},
  \href{http://dx.doi.org/10.1103/PhysRevD.51.3117}{\emph{Phys. Rev. D} {\bf
  51} (1995) 3117--3120}, [\href{http://arxiv.org/abs/astro-ph/9409051}{{\tt
  astro-ph/9409051}}].

\bibitem{Safonova:2001vz}
M.~Safonova, D.~F. Torres and G.~E. Romero, \emph{{Microlensing by natural
  wormholes: Theory and simulations}},
  \href{http://dx.doi.org/10.1103/PhysRevD.65.023001}{\emph{Phys. Rev. D} {\bf
  65} (2002) 023001}, [\href{http://arxiv.org/abs/gr-qc/0105070}{{\tt
  gr-qc/0105070}}].

\bibitem{Nandi:2006ds}
K.~K. Nandi, Y.-Z. Zhang and A.~V. Zakharov, \emph{{Gravitational lensing by
  wormholes}}, \href{http://dx.doi.org/10.1103/PhysRevD.74.024020}{\emph{Phys.
  Rev. D} {\bf 74} (2006) 024020},
  [\href{http://arxiv.org/abs/gr-qc/0602062}{{\tt gr-qc/0602062}}].

\bibitem{Nakajima:2012pu}
K.~Nakajima and H.~Asada, \emph{{Deflection angle of light in an Ellis wormhole
  geometry}}, \href{http://dx.doi.org/10.1103/PhysRevD.85.107501}{\emph{Phys.
  Rev. D} {\bf 85} (2012) 107501}, [\href{http://arxiv.org/abs/1204.3710}{{\tt
  1204.3710}}].

\bibitem{Dey:2008kn}
T.~K. Dey and S.~Sen, \emph{{Gravitational lensing by wormholes}},
  \href{http://dx.doi.org/10.1142/S0217732308025498}{\emph{Mod. Phys. Lett. A}
  {\bf 23} (2008) 953--962}, [\href{http://arxiv.org/abs/0806.4059}{{\tt
  0806.4059}}].

\bibitem{Bhattacharya:2010zzb}
A.~Bhattacharya and A.~A. Potapov, \emph{{Bending of light in Ellis wormhole
  geometry}}, \href{http://dx.doi.org/10.1142/S0217732310033748}{\emph{Mod.
  Phys. Lett. A} {\bf 25} (2010) 2399--2409}.

\bibitem{Tsukamoto:2012xs}
N.~Tsukamoto, T.~Harada and K.~Yajima, \emph{{Can we distinguish between black
  holes and wormholes by their Einstein ring systems?}},
  \href{http://dx.doi.org/10.1103/PhysRevD.86.104062}{\emph{Phys. Rev. D} {\bf
  86} (2012) 104062}, [\href{http://arxiv.org/abs/1207.0047}{{\tt 1207.0047}}].

\bibitem{Kuhfittig:2013hva}
P.~K. Kuhfittig, \emph{{Gravitational lensing of wormholes in the galactic halo
  region}}, \href{http://dx.doi.org/10.1140/epjc/s10052-014-2818-2}{\emph{Eur.
  Phys. J. C} {\bf 74} (2014) 2818},
  [\href{http://arxiv.org/abs/1311.2274}{{\tt 1311.2274}}].

\bibitem{Tsukamoto:2016qro}
N.~Tsukamoto, \emph{{Strong deflection limit analysis and gravitational lensing
  of an Ellis wormhole}},
  \href{http://dx.doi.org/10.1103/PhysRevD.94.124001}{\emph{Phys. Rev. D} {\bf
  94} (2016) 124001}, [\href{http://arxiv.org/abs/1607.07022}{{\tt
  1607.07022}}].

\bibitem{Jusufi:2017mav}
K.~Jusufi and A.~Övgün, \emph{{Gravitational Lensing by Rotating Wormholes}},
  \href{http://dx.doi.org/10.1103/PhysRevD.97.024042}{\emph{Phys. Rev. D} {\bf
  97} (2018) 024042}, [\href{http://arxiv.org/abs/1708.06725}{{\tt
  1708.06725}}].

\bibitem{Ovgun:2018fnk}
A.~Övgün, \emph{{Light deflection by Damour-Solodukhin wormholes and
  Gauss-Bonnet theorem}},
  \href{http://dx.doi.org/10.1103/PhysRevD.98.044033}{\emph{Phys. Rev. D} {\bf
  98} (2018) 044033}, [\href{http://arxiv.org/abs/1805.06296}{{\tt
  1805.06296}}].

\bibitem{Nedkova:2013msa}
P.~G. Nedkova, V.~K. Tinchev and S.~S. Yazadjiev, \emph{{Shadow of a rotating
  traversable wormhole}},
  \href{http://dx.doi.org/10.1103/PhysRevD.88.124019}{\emph{Phys. Rev. D} {\bf
  88} (2013) 124019}, [\href{http://arxiv.org/abs/1307.7647}{{\tt 1307.7647}}].

\bibitem{Amir:2018szm}
M.~Amir, A.~Banerjee and S.~D. Maharaj, \emph{{Shadow of charged wormholes in
  Einstein--Maxwell--dilaton theory}},
  \href{http://dx.doi.org/10.1016/j.aop.2018.11.004}{\emph{Annals Phys.} {\bf
  400} (2019) 198--207}, [\href{http://arxiv.org/abs/1805.12435}{{\tt
  1805.12435}}].

\bibitem{Bronnikov:2019sbx}
K.~A. Bronnikov and R.~A. Konoplya, \emph{{Echoes in brane worlds: ringing at a
  black hole--wormhole transition}},
  \href{http://dx.doi.org/10.1103/PhysRevD.101.064004}{\emph{Phys. Rev.} {\bf
  D101} (2020) 064004}, [\href{http://arxiv.org/abs/1912.05315}{{\tt
  1912.05315}}].

\bibitem{Churilova:2019cyt}
M.~Churilova and Z.~Stuchlik, \emph{{Ringing of the regular black-hole/wormhole
  transition}}, \href{http://dx.doi.org/10.1088/1361-6382/ab7717}{\emph{Class.
  Quant. Grav.} {\bf 37} (2020) 075014},
  [\href{http://arxiv.org/abs/1911.11823}{{\tt 1911.11823}}].

\bibitem{Micchi:2019yze}
L.~F.~L. Micchi and C.~Chirenti, \emph{{Spicing up the recipe for echoes from
  exotic compact objects: orbital differences and corrections in rotating
  backgrounds}},
  \href{http://dx.doi.org/10.1103/PhysRevD.101.084010}{\emph{Phys. Rev. D} {\bf
  101} (2020) 084010}, [\href{http://arxiv.org/abs/1912.05419}{{\tt
  1912.05419}}].

\bibitem{Lobo:2012qq}
F.~S.~N. Lobo, F.~Parsaei and N.~Riazi, \emph{{New asymptotically flat phantom
  wormhole solutions}},
  \href{http://dx.doi.org/10.1103/PhysRevD.87.084030}{\emph{Phys. Rev.} {\bf
  D87} (2013) 084030}, [\href{http://arxiv.org/abs/1212.5806}{{\tt
  1212.5806}}].

\bibitem{Caldwell:1999ew}
R.~Caldwell, \emph{{A Phantom menace?}},
  \href{http://dx.doi.org/10.1016/S0370-2693(02)02589-3}{\emph{Phys. Lett. B}
  {\bf 545} (2002) 23--29}, [\href{http://arxiv.org/abs/astro-ph/9908168}{{\tt
  astro-ph/9908168}}].

\bibitem{Sushkov:2005kj}
S.~V. Sushkov, \emph{{Wormholes supported by a phantom energy}},
  \href{http://dx.doi.org/10.1103/PhysRevD.71.043520}{\emph{Phys. Rev. D} {\bf
  71} (2005) 043520}, [\href{http://arxiv.org/abs/gr-qc/0502084}{{\tt
  gr-qc/0502084}}].

\bibitem{Lobo:2005us}
F.~S. Lobo, \emph{{Phantom energy traversable wormholes}},
  \href{http://dx.doi.org/10.1103/PhysRevD.71.084011}{\emph{Phys. Rev. D} {\bf
  71} (2005) 084011}, [\href{http://arxiv.org/abs/gr-qc/0502099}{{\tt
  gr-qc/0502099}}].

\bibitem{Lobo:2005yv}
F.~S. Lobo, \emph{{Stability of phantom wormholes}},
  \href{http://dx.doi.org/10.1103/PhysRevD.71.124022}{\emph{Phys. Rev. D} {\bf
  71} (2005) 124022}, [\href{http://arxiv.org/abs/gr-qc/0506001}{{\tt
  gr-qc/0506001}}].

\bibitem{PhysRevD.72.061303}
O.~B. Zaslavskii, \emph{Exactly solvable model of a wormhole supported by
  phantom energy},
  \href{http://dx.doi.org/10.1103/PhysRevD.72.061303}{\emph{Phys. Rev. D} {\bf
  72} (Sep, 2005) 061303}.

\bibitem{Jamil:2009vn}
M.~Jamil, P.~K. Kuhfittig, F.~Rahaman and S.~Rakib, \emph{{Wormholes supported
  by polytropic phantom energy}},
  \href{http://dx.doi.org/10.1140/epjc/s10052-010-1325-3}{\emph{Eur. Phys. J.
  C} {\bf 67} (2010) 513--520}, [\href{http://arxiv.org/abs/0906.2142}{{\tt
  0906.2142}}].

\bibitem{Lemos:2003jb}
J.~P. Lemos, F.~S. Lobo and S.~Quinet~de Oliveira, \emph{{Morris-Thorne
  wormholes with a cosmological constant}},
  \href{http://dx.doi.org/10.1103/PhysRevD.68.064004}{\emph{Phys. Rev. D} {\bf
  68} (2003) 064004}, [\href{http://arxiv.org/abs/gr-qc/0302049}{{\tt
  gr-qc/0302049}}].

\bibitem{Lobo:2004id}
F.~S. Lobo, \emph{{Surface stresses on a thin shell surrounding a traversable
  wormhole}}, \href{http://dx.doi.org/10.1088/0264-9381/21/21/005}{\emph{Class.
  Quant. Grav.} {\bf 21} (2004) 4811--4832},
  [\href{http://arxiv.org/abs/gr-qc/0409018}{{\tt gr-qc/0409018}}].

\bibitem{Lemos:2004vs}
J.~P. Lemos and F.~S. Lobo, \emph{{Plane symmetric traversable wormholes in an
  Anti-de Sitter background}},
  \href{http://dx.doi.org/10.1103/PhysRevD.69.104007}{\emph{Phys. Rev. D} {\bf
  69} (2004) 104007}, [\href{http://arxiv.org/abs/gr-qc/0402099}{{\tt
  gr-qc/0402099}}].

\bibitem{Lobo:2005zu}
F.~S. Lobo and P.~Crawford, \emph{{Stability analysis of dynamic thin shells}},
  \href{http://dx.doi.org/10.1088/0264-9381/22/22/012}{\emph{Class. Quant.
  Grav.} {\bf 22} (2005) 4869--4886},
  [\href{http://arxiv.org/abs/gr-qc/0507063}{{\tt gr-qc/0507063}}].

\bibitem{Lemos:2008aj}
J.~P. Lemos and F.~S. Lobo, \emph{{Plane symmetric thin-shell wormholes:
  Solutions and stability}},
  \href{http://dx.doi.org/10.1103/PhysRevD.78.044030}{\emph{Phys. Rev. D} {\bf
  78} (2008) 044030}, [\href{http://arxiv.org/abs/0806.4459}{{\tt 0806.4459}}].

\bibitem{Garcia:2011aa}
N.~M. Garcia, F.~S. Lobo and M.~Visser, \emph{{Generic spherically symmetric
  dynamic thin-shell traversable wormholes in standard general relativity}},
  \href{http://dx.doi.org/10.1103/PhysRevD.86.044026}{\emph{Phys. Rev. D} {\bf
  86} (2012) 044026}, [\href{http://arxiv.org/abs/1112.2057}{{\tt 1112.2057}}].

\bibitem{He:2009jd}
X.~He, B.~Wang, S.-F. Wu and C.-Y. Lin, \emph{{Quasinormal modes of black holes
  absorbing dark energy}},
  \href{http://dx.doi.org/10.1016/j.physletb.2009.02.002}{\emph{Phys. Lett. B}
  {\bf 673} (2009) 156--160}, [\href{http://arxiv.org/abs/0901.0034}{{\tt
  0901.0034}}].

\bibitem{MersiniHoughton:2008aw}
L.~Mersini-Houghton and A.~Kelleher, \emph{{Probing Dark Energy with Black Hole
  Binaries}},  \href{http://arxiv.org/abs/0808.3419}{{\tt 0808.3419}}.

\bibitem{Enander:2009pq}
J.~Enander and E.~Mortsell, \emph{{On the use of black hole binaries as probes
  of local dark energy properties}},
  \href{http://dx.doi.org/10.1016/j.physletb.2009.11.057}{\emph{Phys. Lett. B}
  {\bf 683} (2010) 7--10}, [\href{http://arxiv.org/abs/0910.2337}{{\tt
  0910.2337}}].

\bibitem{Toshmatov:2018tyo}
B.~Toshmatov, Z.~Stuchl\'\i{}k, J.~Schee and B.~Ahmedov, \emph{{Electromagnetic
  perturbations of black holes in general relativity coupled to nonlinear
  electrodynamics}},
  \href{http://dx.doi.org/10.1103/PhysRevD.97.084058}{\emph{Phys. Rev. D} {\bf
  97} (2018) 084058}, [\href{http://arxiv.org/abs/1805.00240}{{\tt
  1805.00240}}].

\bibitem{Toshmatov:2018ell}
B.~Toshmatov, Z.~Stuchl\'\i{}k and B.~Ahmedov, \emph{{Electromagnetic
  perturbations of black holes in general relativity coupled to nonlinear
  electrodynamics: Polar perturbations}},
  \href{http://dx.doi.org/10.1103/PhysRevD.98.085021}{\emph{Phys. Rev. D} {\bf
  98} (2018) 085021}, [\href{http://arxiv.org/abs/1810.06383}{{\tt
  1810.06383}}].

\bibitem{Zinhailo:2018ska}
A.~F. Zinhailo, \emph{{Quasinormal modes of the four-dimensional black hole in
  Einstein–Weyl gravity}},
  \href{http://dx.doi.org/10.1140/epjc/s10052-018-6467-8}{\emph{Eur. Phys. J.}
  {\bf C78} (2018) 992}, [\href{http://arxiv.org/abs/1809.03913}{{\tt
  1809.03913}}].

\bibitem{Zhu:2014sya}
Z.~Zhu, S.-J. Zhang, C.~E. Pellicer, B.~Wang and E.~Abdalla, \emph{{Stability
  of Reissner-Nordström black hole in de Sitter background under charged
  scalar perturbation}}, \href{http://dx.doi.org/10.1103/PhysRevD.90.044042,
  10.1103/PhysRevD.90.049904}{\emph{Phys. Rev.} {\bf D90} (2014) 044042},
  [\href{http://arxiv.org/abs/1405.4931}{{\tt 1405.4931}}].

\bibitem{Nollert1999}
H.-P. Nollert, \emph{Quasinormal modes: the characteristic `sound' of black
  holes and neutron stars}, {\emph{Classical and Quantum Gravity} {\bf 16}
  (nov, 1999) R159--R216}.

\bibitem{Berti:2009kk}
E.~Berti, V.~Cardoso and A.~O. Starinets, \emph{{Quasinormal modes of black
  holes and black branes}},
  \href{http://dx.doi.org/10.1088/0264-9381/26/16/163001}{\emph{Class. Quant.
  Grav.} {\bf 26} (2009) 163001}, [\href{http://arxiv.org/abs/0905.2975}{{\tt
  0905.2975}}].

\bibitem{Konoplya:2011qq}
R.~Konoplya and A.~Zhidenko, \emph{{Quasinormal modes of black holes: From
  astrophysics to string theory}},
  \href{http://dx.doi.org/10.1103/RevModPhys.83.793}{\emph{Rev. Mod. Phys.}
  {\bf 83} (2011) 793--836}, [\href{http://arxiv.org/abs/1102.4014}{{\tt
  1102.4014}}].

\bibitem{Berti:2007dg}
E.~Berti, V.~Cardoso, J.~A. Gonzalez and U.~Sperhake, \emph{{Mining information
  from binary black hole mergers: A Comparison of estimation methods for
  complex exponentials in noise}},
  \href{http://dx.doi.org/10.1103/PhysRevD.75.124017}{\emph{Phys. Rev. D} {\bf
  75} (2007) 124017}, [\href{http://arxiv.org/abs/gr-qc/0701086}{{\tt
  gr-qc/0701086}}].

\bibitem{Dey:2020lhq}
R.~Dey, S.~Chakraborty and N.~Afshordi, \emph{{Echoes from braneworld black
  holes}}, \href{http://dx.doi.org/10.1103/PhysRevD.101.104014}{\emph{Phys.
  Rev. D} {\bf 101} (2020) 104014},
  [\href{http://arxiv.org/abs/2001.01301}{{\tt 2001.01301}}].

\end{thebibliography}\endgroup

\end{document}